\documentclass[12pt]{article}
\usepackage{epsfig}
\usepackage{epsfig}
\usepackage{axodraw}
\usepackage{color}
\usepackage{a4}
\usepackage{latexsym}
\usepackage{cite}
\usepackage{bm}
\usepackage{amssymb}

\usepackage{hyperref}

\usepackage{pslatex}
\usepackage[latin1]{inputenc}
\usepackage[T1]{fontenc}

\usepackage{color,colordvi}
\def\colour4colour#1{\Blue{#1}}
\def\colour2colour#1{\Red{#1}}
\newcommand{\colourBlue}[1]{{\color{blue}{#1}}}

\newcommand{\binomial}[2]{{#1\choose #2}}

\renewcommand{\theequation}{\thesection.\arabic{equation}}

\newcommand{\hspm}{{\hspace{-2cm}}}
\newcommand{\hspn}{{\hspace{-4mm}}}
\newcommand{\hspp}{{\hspace{4mm}}}
\newcommand{\hspq}{{\hspace{8mm}}}
\newcommand{\beq}{\begin{equation}}
\newcommand{\eeq}{\end{equation}}
\newcommand{\bea}{\begin{eqnarray}}
\newcommand{\eea}{\end{eqnarray}}
\newcommand{\nn}{\nonumber}
\newcommand{\MSb}{$\overline{\mbox{MS}}$}
\newcommand{\ra}{\rightarrow}

\def\b#1{{\beta_{#1}}}
\def\B(#1,#2){{\beta_{#1}^{\,#2}}}
\newcommand{\als}{\alpha_{\rm s}}
\newcommand{\ars}{a_{\rm s}}
\newcommand{\muS}{\mu^{\:\!2}}
\newcommand{\ep}{\varepsilon}

\begin{document}
\setlength{\parskip}{0.2cm}
\setlength{\baselineskip}{0.53cm}

\def\nc{{n_c}}
\def\ncs{{n_{c}^{\,2}}}
\def\nct{{n_{c}^{\,3}}}
\def\ncf{{n_{c}^{\,4}}}

\def\mus{{\mu^{\,2}}}

\def\z#1{{\zeta_{#1}^{}}}
\def\zz(#1,#2){{\zeta_{#1}^{\,#2}}}
\def\zss{\zeta_2^{\,2}}
\def\zts{\zeta_3^{\,2}}

\def\ca{{C^{}_A}}
\def\cas{{C^{\,2}_A}}
\def\cat{{C^{\,3}_A}}
\def\caf{{C^{\,4}_A}}
\def\cai{{C^{\,5}_A}}
\def\cf{{C^{}_F}}
\def\cfs{{C^{\, 2}_F}}
\def\cft{{C^{\, 3}_F}}
\def\cff{{C^{\, 4}_F}}
\def\cfi{{C^{\, 5}_F}}
\def\nfz{{n^{\,0}_{\! f}}}
\def\nfo{{n^{\,1}_{\! f}}}
\def\nf{{n^{}_{\! f}}}
\def\nfs{{n^{\,2}_{\! f}}}
\def\nft{{n^{\,3}_{\! f}}}
\def\nff{{n^{\,4}_{\! f}}}

\def\dabc2n{{d^{\:\!abc\!}d_{abc}}/{n_c}}

\def\dfRAnr{{ {d_{\,F}^{\,abcd} d_{\,A}^{\,abcd} \over n_c} }}
\def\dfRRnr{{ {d_{\,F}^{\,abcd} d_{\,F}^{\,abcd} \over n_c} }}
\def\dabctnr{{ {d^{abc}d_{abc}}\over{n_c} }}

\def\dfAAna{{ {d_{\,A}^{\,abcd}d_{\,A}^{\,abcd} \over n_a} }}
\def\dfFAna{{ {d_{\,F}^{\,abcd}d_{\,A}^{\,abcd} \over n_a} }}
\def\dfFFna{{ {d_{\,F}^{\,abcd}d_{\,F}^{\,abcd} \over n_a} }}

\def\dfRAnA{{ {d_{F}^{\,abcd}d_{A}^{\,abcd}\! / n_a} }}
\def\dfRRnA{{ {d_{F}^{\,abcd}d_{F}^{\,abcd}\! / n_a} }}
\def\dfRAnc{{ {d_{F}^{\,abcd}d_{A}^{\,abcd}\! / n_c} }}
\def\dfRRnc{{ {d_{F}^{\,abcd}d_{F}^{\,abcd}\! / n_c} }}

\def\dfAA{d_A^{\,abcd}d_A^{\,abcd}}
\def\dfFA{d_F^{\,abcd}d_A^{\,abcd}}
\def\dfFF{d_F^{\,abcd}d_F^{\,abcd}}

\def\nF{{n_{c}^{}}}
\def\nA{{n_{a}^{}}}

\def\cI{{C^{}_{\!I}}}
\def\nI{{n_{\!I}^{}}}
\def\dfIAnI{{\frac{d_I^{\,abcd}d_A^{\,abcd}}{n_I^{} }}}
\def\dfIFnI{{\frac{d_I^{\,abcd}d_F^{\,abcd}}{n_I^{} }}}

\def\als{{\alpha_{\rm s}}}
\def\as(#1){{\alpha_{\rm s}^{\:#1}}}
\def\ar(#1){{a_{\rm s}^{\:#1}}}
\def\ars{{a_{\rm s}}}

\def\frkt#1#2{\mbox{\large{$\frac{#1}{#2}$}}}
\def\frct#1#2{\mbox{\small{$\displaystyle\frac{#1}{#2}$}}}

\def\xm{(1\!-\!x)}
\def\LntO{\ln(1\!-\!x)}
\def\Lnt(#1){\ln^{\,#1}(1\!-\!x)}

\def\S(#1){{{S}_{#1}}}
\def\BS(#1){{{\mathbb S}_{#1}}}
\def\BSss(#1,#2,#3){{{\mathbb S}_{#1,#2,#3}}}
\def\BSsss(#1,#2,#3,#4){{{\mathbb S}_{#1,#2,#3,#4}}}
\def\BSssss(#1,#2,#3,#4,#5){{{\mathbb S}_{#1,#2,#3,#4,#5}}}
\def\BSsssss(#1,#2,#3,#4,#5,#6){{{\mathbb S}_{#1,#2,#3,#4,#5,#6}}}

\def\Ge{{\gamma_{\:\! \rm e}}}
\def\lnN{{\ln \widetilde N}}

\def\pqq(#1){p_{\rm{qq}}(#1)}


\begin{titlepage}
\noindent
%
%
PUBDB-2026-01424 \hfill arXiv:2605.03889v2\\    
LTH 1416 \\
\vspace{1.5cm}
\begin{center}
{\LARGE \bf Properties and implications of the\\[1ex] 
 four-loop non-singlet splitting functions in QCD}\\
\vspace{2.5cm}
\Large
S. Moch$^{\, a}$ and A. Vogt$^{\, b\!,\,a\,\ast}$\\
\vspace{1.0cm}
\normalsize
{\it $^a$II.~Institute for Theoretical Physics, Hamburg University\\[0.5mm]
Luruper Chaussee 149, D-22761 Hamburg, Germany}\\
\vspace{5mm}
{\it $^b$Department of Mathematical Sciences, University of Liverpool\\[0.5mm]
Peach Street, Liverpool L69 3BX, United Kingdom}\\
\vspace{2.5cm}
\vfill
{\large \bf Abstract}
\vspace{-0.2cm}
\end{center}
We have studied the recently completed analytic all-$N$ expressions for the 
four-loop anomalous dimensions corresponding to the next-to-next-to-next-to-%
leading order splitting functions for the non-singlet quark distribution in 
perturbative QCD.
The results agree with fixed-$N$ values beyond those published so far.
Their structural consistency with theoretical requirements is established. 
\linebreak
They are used to cast the four-loop gluon virtual anomalous dimension and 
the next-to-next-to-next-to-next-to-leading logarithmic threshold-resummation
coefficients for lepton-pair and Higgs production in hadron-hadron collisions 
and deep-inelastic scattering into their final analytical forms.
Further properties and consequences of the new results are addressed,
in particular a new structure seen most clearly in the small-$x$ logarithms 
occurring in a quartic-Casimir contribution.
 
\vspace{1cm}
\noindent
$\ast$ {\it Supported by the Alexander-von-Humboldt Foundation}
\end{titlepage}
 
%
\section{Introduction}
\label{sec:intro}
%

Recently, the penultimate milestone has been reached of research projects 
spanning more than a decade: the exact expressions have been completed of the 
fourth-order contributions to the splitting functions for the scale dependence 
(evolution) of the non-singlet quark distributions of hadrons 
\cite{Gehrmann:2026qbl}. 
Until then only the fermionic parts, which depend on the number $\nf$ of 
light quark flavours, were known exactly \cite{Gracey:1994nn,Davies:2016jie,%
Basdew-Sharma:2022vya,Gehrmann:2023iah,Kniehl:2025ttz}. 
The non-fermionic parts had hitherto been completed only in the (planar) limit 
of a large number of colours $\nc$ \cite{Moch:2017uml}, and approximate 
expressions had been published for the remaining contributions from their
lowest eight moments supplemented by end-point constraints.
The four-loop singlet splitting functions are not yet fully known, for their 
status see refs.~%
\cite{Davies:2016jie,Moch:2021qrk,Moch:2023tdj,Gehrmann:2023cqm,%
Falcioni:2023tzp,Falcioni:2023luc,Falcioni:2023vqq,Falcioni:2024xyt,%
Falcioni:2024qpd,Falcioni:2025hfz,Kniehl:2026eij}.

The new expressions for the non-singlet splitting functions are the result of 
a computational tour de force. 
They deserve checks of their correctness, and studies of their properties and 
consequences beyond those included in ref.~\cite{Gehrmann:2026qbl}. 
The present article addresses some of these issues in, hopefully, a 
sufficiently detailed manner to be useful for further research in this field.
 
After recalling the notations and some basic properties of the non-singlet 
splitting functions in section 2, we provide some numerical and structural 
checks in section 3. 
Some implications of the results of ref.~\cite{Gehrmann:2026qbl} on other 
four-loop quantities in perturbative QCD are addressed in section~4, before 
we briefly turn to a few predictions for five-loop splitting functions in 
section~5. Our~findings are summarized in section 6.
Some extra material can be found in appendices and ancillary files.

%
\section{Notations and basic features}
\label{sec:basics}
\setcounter{equation}{0}
%

The general structure of the (anti-)$\,$quark (anti-)$\,$quark splitting
functions, constrained by charge conjugation invariance and flavour
symmetry, is given by
\bea
\label{eq:Pqiqk}
  P_{{\rm q}_{\,\rm i}^{}{\rm q}_{\,\rm k}^{}} \: = \:
  P_{\,\bar{\rm q}_{\,\rm i}^{}\bar{\rm q}_{\,\rm k}^{}} 
  \: = \: & \delta_{\,\rm ik}^{}\, P_{\rm qq}^{\,\rm v}
        + P_{\rm qq}^{\,\rm s} 
\quad , \qquad
  P_{{\rm q}_{\,\rm i}^{}\bar{\rm q}_{\,\rm k}^{}} \: = \:
  P_{\,\bar{\rm q}_{\,\rm i}^{}{\rm q}_{\,\rm k}^{}} 
  \: = \: & \delta_{\,\rm ik}\, P_{{\rm q}\bar{{\rm q}}}^{\,\rm v}
        + P_{{\rm q}\bar{{\rm q}}}^{\,\rm s}
  \:\: .
\eea
The flavour-diagonal (`valence') quantity $P_{\rm qq}^{\,\rm v}$ starts at 
the first order in the strong coupling $\als$, while 
$P_{{\rm q}\bar{{\rm q}}}^{\,\rm v}$ 
and the flavour-independent (`sea') contributions 
$P_{{\rm qq}}^{\,\rm s}$ and $P_{{\rm q}\bar{{\rm q}}}^{\,\rm s}$ 
are of order $\as(2)$.  A non-vanishing difference 
$P_{{\rm qq}}^{\,\rm s} - P_{{\rm q}\bar{{\rm q}}}^{\,\rm s}$ occurs at 
the third order for the first time, see fig.~1. 
This difference it is proportional to the cubic group invariant $\dabc2n$ 
which does not enter the other contributions.

This structure leads to three type of combinations of quark distributions
that decouple from the gluon distribution. 
The two types of flavour asymmetries and the total valence distribution,
\beq
\label{eq:q-pmv}
  q_{{\rm ns},ik}^{\,\pm} \: = \: q_i^{} \pm
  \bar{q}_i^{} - (q_k^{} \pm \bar{q}_k^{})
\qquad \mbox{ and } \qquad
  q_{\rm ns}^{\rm v} \: = \: \sum_{r=1}^{\nf} (q_r^{}
  - \bar{q}_r^{}) 
\:\: ,
\eeq
evolve with
\beq
\label{eq:P-pmv}
  P_{\rm ns}^{\,\pm} \: = \: P_{{\rm q}{\rm q}}^{\,\rm v}
  \pm P_{{\rm q}\bar{{\rm q}}}^{\,\rm v} 
\qquad \mbox{ and } \qquad
  P_{\rm ns}^{\,\rm v} \: = \: P_{\rm qq}^{\,\rm v}
  - P_{{\rm q}\bar{{\rm q}}}^{\,\rm v} + \nf (P_{\rm qq}^{\,\rm s}
  - P_{{\rm q}\bar{{\rm q}}}^{\,\rm s}) \: \equiv \:
  P_{\rm ns}^{\, -} + P_{\rm ns}^{\,\rm s} 
\:\: ,
\eeq
where $P_{{\rm q}\bar{{\rm q}}}^{\,\rm v}$ vanishes in the large-$\nc$ limit.
The first moments of $P_{\rm ns}^{\,-}$ and $P_{\rm ns}^{\,\rm v}$ are zero,
since the first moments of $q_{\rm ns}^-$ and $q_{\rm ns}^{\rm v}$ reflect 
conserved additive quantum numbers.

\begin{figure}[tbh]
\label{fig:PqqDiags}

\vspace{4mm}
\hspace*{1.5cm}
\begin{picture}(310,50)(-10,0)
\scalebox{1.2}{


\SetWidth{1.1}
\SetColor{OliveGreen}
\Photon(0,50)(10,40){1.5}{2}
\Photon(60,50)(50,40){1.5}{2}
\SetColor{Red}
\ArrowLine(0,0)(10,10)
\ArrowLine(10,10)(10,40)
\SetColor{OliveGreen}
\ArrowLine(10,40)(50,40)
\ArrowLine(50,40)(50,10)
\ArrowLine(50,10)(60,0)
\Gluon(10,10)(50,10){2.5}{6}

\Vertex(10,10){1.5}
\Vertex(50,10){1.5}
\Vertex(10,40){1.5}
\Vertex(50,40){1.5}

\SetWidth{0.8}
\SetColor{Black}
\DashLine(30,0)(30,50){5}


\SetWidth{1.1}
\SetColor{OliveGreen}
\Photon(160,50)(170,40){1.5}{2}
\Photon(220,50)(210,40){1.5}{2}
\SetColor{Red}
\ArrowLine(160,0)(170,10)
\Line(170,27)(170,40)
\SetColor{OliveGreen}
\ArrowLine(170,10)(210,10)
\ArrowLine(210,10)(220,0)
\Gluon(170,10)(170,27){2}{2}
\Gluon(210,10)(210,27){-2}{2}
\Line(170,40)(210,40)
\Line(210,27)(170,27)
\Line(210,40)(210,27)

\Vertex(170,10){1.5}
\Vertex(210,10){1.5}
\Vertex(170,40){1.5}
\Vertex(210,40){1.5}
\Vertex(170,27){1.5}
\Vertex(210,27){1.5}

\SetWidth{0.8}
\SetColor{Black}
\DashLine(110,0)(110,50){5}


\SetWidth{1.1}
\SetColor{OliveGreen}
\Photon(80,50)(90,40){1.5}{2}
\Photon(140,50)(130,40){1.5}{2}
\SetColor{Red}
\ArrowLine(80,0)(90,10)
\ArrowLine(90,40)(90,25)
\SetColor{OliveGreen}
\ArrowLine(90,10)(110,17.5)
\Line(110,17.5)(130,25)
\ArrowLine(130,10)(140,0)
\Gluon(90,10)(90,25){2}{2}
\Gluon(130,10)(130,25){-2}{2}
\ArrowLine(130,40)(90,40)
\Line(90,25)(107,18.5)
\ArrowLine(113,16.5)(130,10)
\ArrowLine(130,25)(130,40)

\Vertex(90,10){1.5}
\Vertex(130,10){1.5}
\Vertex(90,40){1.5}
\Vertex(130,40){1.5}
\Vertex(90,25){1.5}
\Vertex(130,25){1.5}

\SetWidth{0.8}
\SetColor{Black}
\DashLine(190,0)(190,50){5}


\SetWidth{1.1}
\SetColor{OliveGreen}
\Photon(240,50)(250,40){1.5}{2}
\Photon(300,50)(290,40){1.5}{2}
\SetColor{Red}
\ArrowLine(240,0)(250,10)
\Line(250,27)(250,40)
\SetColor{OliveGreen}
\Line(250,10)(262,10)
\ArrowLine(262,10)(290,10)
\ArrowLine(290,10)(300,0)
\Gluon(250,10)(250,27){2}{2}
\Gluon(262,10)(278,27){2}{3}
\Gluon(290,10)(290,27){-2}{2}
\Line(250,40)(290,40)
\Line(290,27)(250,27)
\Line(290,40)(290,27)

\Vertex(250,10){1.5}
\Vertex(290,10){1.5}
\Vertex(250,40){1.5}
\Vertex(290,40){1.5}
\Vertex(250,27){1.5}
\Vertex(290,27){1.5}
\Vertex(262,10){1.5}
\Vertex(278,27){1.5}

\SetWidth{0.8}
\SetColor{Black}
\DashLine(270,0)(270,50){5}

}
\end{picture}
\vspace*{-1mm}

\caption{Representative forward-Compton diagrams for the four cases
on the r.h.s of eq.~(\ref{eq:Pqiqk}).}
\vspace*{-1mm}
\end{figure}
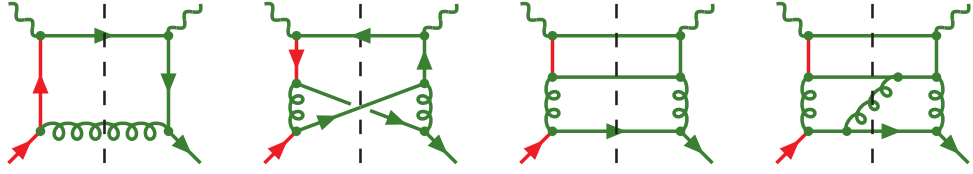

The evolution equations for the momentum distributions 
$\,q_{\rm ns}^{\, a}(x,\muS)$, $a=\pm,{\rm v},$ can be written as
\beq
\label{eq:evol}
  \frac{d}{d \ln \muS} \: q_{\rm ns}^{\, a}(x,\muS)\: =\:
 \sum_{n=0}\: \ar(n+1) P_{\rm ns}^{\,(n)a} (x) \,\otimes\, 
 q_{\rm ns}^{\, a}(x,\muS)
\quad \mbox{ with } \quad
 \ars \,\equiv\, \als(\muS) / (4\pi) 
\;\; ,
\eeq
where $\otimes$ represents the standard Mellin convolution.
Throughout this article, we work in the \MSb~scheme. 
We identify the second argument of the $q_{\rm ns}^{\, a}$, the factorization 
scale, and the argument of $\als$, the renormalization scale. 
The additional terms for the general case are provided by lower-order 
quantities. Hence they are not relevant in the present context.

The even- or odd-$N$ moments of the splitting functions are related to the
anomalous dimensions $\gamma_{\,\rm ns}^{}$ of twist-two operators in the 
light-cone expansion,
\beq
\label{eq:PvsGam}
  \gamma_{\,\rm ns}^{\:(n)a}(N) 
  \: = \: - P_{\,\rm ns}^{\,(n)a}(N)    
  \: = \: - \int_0^1 \!dx\:\, x^{\,N-1}\, P_{\,\rm ns}^{\,(n)a}(x)
\eeq
where the relative sign is a standard convention. 
In general, eq.~(\ref{eq:PvsGam}) holds at even $N$ for $a = +$ (and the
singlet cases) and at odd $N$ for $a = -,{\rm v}$. 
In the large-$\nc$ limit $P_{\rm ns}^{\,+}$ and $P_{\rm ns}^{\,-}$ 
coincide, see eq.~(\ref{eq:P-pmv}), hence there the above relation is 
applicable at all integer values of $N$. 
This was crucial for the all-$N$ reconstruction in ref.~\cite{Moch:2017uml}.

All four-loop ($n=3$ in eq.~(\ref{eq:evol})) computations for refs.~%
\cite{Gehrmann:2026qbl,Gracey:1994nn,Davies:2016jie,Basdew-Sharma:2022vya,
Gehrmann:2023iah,Kniehl:2025ttz,Moch:2017uml,Moch:2021qrk,Moch:2023tdj,
Gehrmann:2023cqm,Falcioni:2023tzp,Falcioni:2023luc,Falcioni:2023vqq,
Falcioni:2024xyt,Falcioni:2024qpd,Falcioni:2025hfz},
as well as the much earlier determination of the three-loop splitting functions
in refs.~\cite{Moch:2002sn,Moch:2004pa,Vogt:2004mw}, have been performed in 
\mbox{$N$-space}, either via operator matrix elements or via forward Compton
 amplitudes for inclusive deep-inelastic scattering (DIS). 
Therefore they all obtain $\gamma_{\,\rm ns}^{}$ at even or odd $N$, from 
which the splitting functions $P_{\rm ns}^{}(x)$ are determined by an 
inverse Mellin transformation. 
The anomalous dimensions can be expressed in terms of harmonic sums 
$S_{\vec{w}}(N)$ \cite{Vermaseren:1998uu,Blumlein:1998if}, see below. 
Hence this transformation can be carried out by a completely algebraic
procedure~\cite{Remiddi:1999ew,Moch:1999eb} based on the fact that harmonic 
sums occur as coefficients of the Taylor expansion of harmonic polylogarithms.

The anomalous dimensions $\gamma_{\,\rm ns}^{\:\pm}(N)$ to now (at least) 
four loops \cite{Gehrmann:2026qbl} can be cast in the form 
\beq
\label{eq:gamStr}
  \gamma_{\,\rm ns}^{\:(\ell)\pm}(N) \;=\;
  \sum_{n\,=\,0}^{2\ell-1} \;
  \sum_{a\,=\,0,1} 
  \sum_{k\,=\,0}^{2\ell+1-n\,} \;
  \sum_{\vec{w},\,w\,=\,0}^{2\ell+1-n-k} 
  c_{akw}^{(\ell,n)} \: 
  \zeta_{n}^{} \: \frac{1}{(N\!+\!a)^{k}} \: S_{\vec{w}}(N) \; .
\eeq
Here $\zeta_{0}^{} \equiv 1$, and the sum over $n$ excludes $n=1,2$.
The sum over $\vec{w}$, where $w$ denotes the weight of the sums, 
excludes any sums with an index -1. 

For the $d^{\:\!abc\!}d_{abc}$ anomalous dimension 
$\gamma_{\,\rm ns}^{\:\rm s}(N)$ in eq.~(\ref{eq:P-pmv}), additional 
denominators $1/(N-1)$ and $1/(N+2)$ enter already at its lowest order, 
$\ell=2$ \cite{Moch:2004pa}, see eq.~(2.14) of ref.~\cite{Davies:2016jie}.
For these functions, the maximal weight of the sums is lower than in 
eq.~(\ref{eq:gamStr}) with $w=3$ at three loops and $w=5$ at four loops
\cite{Gehrmann:2026qbl}. The latter result includes a new feature, see 
below: $\gamma_{\,\rm ns}^{\:(3)\rm s}(N)$ includes terms with $k=0$, 
unlike $\gamma_{\,\rm ns}^{\:(2)\rm s}(N)$ and the third-order 
pure-singlet anomalous dimension \cite{Vogt:2004mw}.
 
%
\section{Checks and properties}
\label{sec:props}
\setcounter{equation}{0}
%

The fixed-$N$ moments of the splitting functions $P_{\,\rm ns}^{\,(3)+}(x)$ 
and $P_{\,\rm ns}^{\,(3)-,\,\rm s}(x)$ were published for even $N \leq 16$ 
and odd $N \leq 15$, respectively, in ref.~\cite{Moch:2017uml} 
(for $N \leq 4$ see also 
refs.~\cite{Velizhanin:2011es,Velizhanin:2014fua,Baikov:2015tea}). 
Beyond these results, which were available for checks in 
ref.~\cite{Gehrmann:2026qbl}, three more moments were computed by the autumn 
of 2023.
A consequence of these additional constraints was included in 
ref.~\cite{Moch:2023tdj}, see eq.~(\ref{eq:b4FAappr}) below.
More recently, we extended $\gamma_{\,\rm ns}^{\:(3)\rm s}(N)$ to $N=29$, 
in an attempt to gather enough information for a construction of its all-$N$ 
form analogous to refs.~\cite{Davies:2016jie,Moch:2017uml,Kniehl:2025ttz}.
These computations were performed using an efficiency-improved in-house
version of the {\sc Forcer} \cite{Ruijl:2017cxj}, a package for the parametric 
reduction of four-loop self-energy integrals programmed in {\sc Form}
\cite{Vermaseren:2000nd,Kuipers:2012rf,Ruijl:2017dtg}.

A comparison of these fixed-$N$ results with those of 
ref.~\cite{Gehrmann:2026qbl} provides a check of their computations:
we find perfect agreement in all cases. 
A sample of our results can be found in appendix A, where we present the
respective highest moments of $P_{\,\rm ns}^{\,(3)\pm,\rm s}(x)$ computed
using {\sc Forcer}.

The authors of ref.~\cite{Gehrmann:2026qbl} determined a huge number of values 
of $\gamma_{\,\rm ns}^{\:(3)\pm,\rm s}(N)$ from which they inferred the
functional forms without taking into account some structural and parametric
constraints on these functions. 
Consequently, these constraints can now serve as further checks of their 
results. By far the most important feature seems to arise from conformal
symmetry of QCD for $D = 4 -2\,\ep$ dimensions at a particular value of the
coupling constant, which implies the relation 
\bea
\label{eq:gamma-u}
  \gamma_{\,\rm ns}^{}(N) &\!=\!& \gamma_{\,\rm u}^{}
  \left( N + \sigma \,\gamma_{\,\rm ns}^{}(N)-\beta(\ars)/\ars ) \right)
\eea
first written down in refs.~%
\cite{Dokshitzer:2005bf,Dokshitzer:2006nm,Basso:2006nk}, for a detailed
discussion see, e.g., ref.~\cite{Strohmaier:2018tjo}.
Here $\,\sigma = -1\,$ for the present initial-state (space-like) 
splitting functions and $\,\sigma = 1\,$ for their final-state (time-like) 
counterparts governing the evolution of fragmentation functions.
The crucial point is that the non-singlet universal evolution kernel 
$\gamma_{\rm u}^{}$ is reciprocity respecting (RR), i.e., its Mellin
inverse fulfils $P_{\rm u}^{}(x) = - x P_{\rm u}^{}(1/x)$ corresponding to
invariance under $N \ra -\,N-1$ in $N$-space. Eq.~(\ref{eq:gamma-u}) leads to
\bea
\label{eq:gu-exp}
  \gamma_{\,\rm u}^{} \:=\: 
         \sum_{n=0}\: \ar(n+1)\, \gamma_{\,\rm u}^{\,(n)} 
   &\!=\!&
         \ars \,\* \gamma_0^{}
   \:+\: \ar(2) \* \left( \,
         \overline{\gamma}_1^{}
      \,-\, \b0 \,\* d_N^{} \* \gamma_0^{}
       \right)
\nn \\[-2mm] & & \mbox{\hspn}
   \:+\: \ar(3) \* \Big(
         \overline{\gamma}_2^{}
      \,-\, \frct{1}{6}\, \* d_N^{\,2\,} \* \gamma_0^{\:3}
      \,-\, \b1 \,\* d_N^{} \* \gamma_0^{}
      \,-\, \b0 \,\* d_N^{} \* \overline{\gamma}_1^{}
      \,+\, \frct{1}{2}\, \* \B(0,2) \,\* d_N^{\,2\,} \* \gamma_0^{}
      \Big)
\nn \\ & & \mbox{\hspn}
   \:+\: \ar(4) \* \Big(
         \overline{\gamma}_3^{}
      \,-\, \frct{1}{2}\, \* d_N^{\,2\,} \*
         ( \gamma_0^{\:2}\,\overline{\gamma}_1^{} )
      \,-\, \b2\,\* d_N^{} \* \gamma_0^{}
      \,-\, \b1\,\* d_N^{} \* \overline{\gamma}_1^{}
      \,-\, \b0\,\* d_N^{} \* \overline{\gamma}_2^{}
    \\ & & \mbox{\hspp}
      \,+\, \frct{1}{6}\, \* \b0 \,\* d_N^{\,3\,} \* \gamma_0^{\:3}
      \,+\, \b0\,\* \b1\,\* d_N^{\,2\,} \* \gamma_0^{}
      \,+\, \frct{1}{2}\, \* \B(0,2) \,\* d_N^{\,2\:} \* \overline{\gamma}_1^{}
      \,-\, \frct{1}{6}\, \* \B(0,3) \,\* d_N^{\,3\:} \* \gamma_0^{}
      \Big)
  \:+\: {\cal O}(\ar(5))
\nn
\:\: .
\eea
$\overline{\gamma}_n^{}$ denotes the average of the space-like and 
time-like expansion coefficients, and we have suppressed all arguments $N$.
{The derivatives $d_N^{\,m} \equiv d^{\,m}/dN^{\,m}$ can 
be taken via inverse Mellin transforms to $x$-space, where they correspond
to products with $\ln^{\,m\!}x$,} and Mellin transforms of these
products. The publicly available tools for these manipulations in {\sc Form}
have been described in~ref.~\cite{Vermaseren:2000nd}.

Since the difference between the time-like and space-like anomalous dimensions 
is known to four loops \cite{Mitov:2006ic,Moch:2017uml}, the new result of 
ref.~\cite{Gehrmann:2026qbl} can be employed to completely determine the 
four-loop coefficient $\gamma_{\,\rm u}^{\,(3)}$ in eq.~(\ref{eq:gu-exp}).
So far only all $\nf$ contributions \cite{Davies:2016jie,Kniehl:2025ttz}
and the $\nfz$ $\zeta$-function terms were known 
\cite{Moch:2017uml,Davies:2017hyl,Vogt:2023unp,Kniehl:2025jfs} besides the
complete large-$\nc$ limit \cite{Moch:2017uml}. We find that -- and this is
a noteworthy check of ref.~\cite{Gehrmann:2026qbl} -- all three functions 
$\gamma_{\,\rm u}^{\,(n)a}(N)$, $a = \pm,\rm s$ are indeed reciprocity 
respecting.

The complete $N$-dependence of $\,\gamma_{\,\rm u}^{\; a}$, $\,a= \pm,\rm s\,$ 
to four loops, expressed in a basis of RR sums, see appendix B, is much more 
compact than the corresponding space-like expressions for the \MSb\ anomalous 
dimensions, but still rather lengthy, especially for $a = \pm$. 
It is therefore deferred to an ancillary file. 
The quartic-Casimir parts are given in appendix C.
Here we write down the most complicated (or~simplest, depending on the
point of view) part, the terms with $\,w=7$ in eq.~(\ref{eq:gamStr}): 
\bea
\label{eq:gns3W7}
 {\lefteqn{ \gamma_{\,\rm u}^{\:(3)+}(N)\big|_{w=7} \;=\; \quad \quad
}} 
\nn 
\\[0.5mm]  && \nn \mbox{\hspn} \phantom{+}
      16\,\* {\colourBlue{\cf\,\*\cat}} \,\* \left\{
          - 4\,\* \BS(1,1,1,2,2)
          + 6\,\* \BS(1,1,1,3,1)
          + 4\,\* \BS(1,1,1,4)
          + 2\,\* \BS(1,1,2,1,2)
          - 8\,\* \BS(1,1,2,2,1)
          - 6\,\* \BS(1,1,2,3)
\right.
\\ && \nn \mbox{}
          + 4\,\* \BS(1,1,3,1,1)
          - 2\,\* \BS(1,1,3,2)
          + 4\,\* \BS(1,1,4,1)
          + 2\,\* \BS(1,2,1,1,2)
          - 2\,\* \BS(1,2,1,3)
          - 4\,\* \BS(1,2,2,1,1)
          + 9\,\* \BS(1,2,2,2)
\\ && \nn \mbox{}
          - 9\,\* \BS(1,2,3,1)
          - 2\,\* \BS(1,2,4)
          + 2\,\* \BS(1,3,1,1,1)
          + 3\,\* \BS(1,3,1,2)
          - 5\,\* \BS(1,3,2,1)
          + 4\,\* \BS(1,4,1,1)
          + 2\,\* \BS(1,5,1)
\\ && \nn \mbox{}
          + 2\,\* \BS(\,2,1,1,1,2)
          - 2\,\* \BS(\,2,1,1,3)
          + 5\,\* \BS(\,2,1,2,2)
          - 5\,\* \BS(\,2,1,3,1)
          - 2\,\* \BS(\,2,2,1,1,1)
          + 8\,\* \BS(\,2,2,2,1)
          + 4\,\* \BS(\,2,2,3)
\\ && \nn \mbox{}
          - 6\,\* \BS(\,2,3,1,1)
          + \BS(\,2,3,2)
          - 5\,\* \BS(\,2,4,1)
          + 3\,\* \BS(\,3,1,1,2)
          - \BS(\,3,1,2,1)
          + 2\,\* \BS(\,3,1,3)
          - 4\,\* \BS(\,3,2,1,1)
          - 3\,\* \BS(\,3,2,2)
\\ && \nn \mbox{}
\left.
          - \BS(3,3,1)
          + 2\,\* \BS(\,4,1,1,1)
          + 4\,\* \BS(\,4,1,2)
          - 4\,\* \BS(\,4,2,1)
          + 2\,\* \BS(\,5,1,1)
          \right\}
\\[1mm] && \nn \mbox{\hspn}
   + 64\* \left( {\colourBlue{\dfRAnc - 1/24\:\* \cf \* \cat}} \right) 
          \* \left\{
            4\,\* \BS(1,1,1,2,2)
          - 12\,\* \BS(1,1,1,3,1)
          + 8\,\* \BS(1,1,1,4)
          + 4\,\* \BS(1,1,2,1,2)
\right.
\\ && \nn \mbox{}
          - 4\,\* \BS(1,1,2,2,1)
          + 4\,\* \BS(1,1,2,3)
          + 8\,\* \BS(1,1,3,1,1)
          - 8\,\* \BS(1,1,3,2)
          - 4\,\* \BS(1,1,4,1)
          + 8\,\* \BS(1,2,1,1,2)
          + 2\,\* \BS(1,2,1,2,1)
\\ && \nn \mbox{}
          - \BS(1,2,1,3)
          - 14\,\* \BS(1,2,2,1,1)
          + 3\,\* \BS(1,2,2,2)
          + 6\,\* \BS(1,2,3,1)
          - 4\,\* \BS(1,2,4)
          + 4\,\* \BS(1,3,1,1,1)
          - 14\,\* \BS(1,3,1,2)
\\ && \nn \mbox{}
          - 3\,\* \BS(1,3,2,1)
          - 2\,\* \BS(1,3,3)
          + 9\,\* \BS(1,4,1,1)
          + 2\,\* \BS(1,4,2)
          + 4\,\* \BS(1,5,1)
          + 4\,\* \BS(2,1,1,1,2)
          + 4\,\* \BS(2,1,1,2,1)
          - 3\,\* \BS(2,1,1,3)
\\ && \nn \mbox{}
          - 4\,\* \BS(2,1,2,1,1)
          - \BS(2,1,2,2)
          - 4\,\* \BS(2,2,1,1,1)
          + \BS(2,2,2,1)
          - 6\,\* \BS(2,2,3)
          + 3\,\* \BS(2,3,1,1)
          + 2\,\* \BS(2,3,2)
          + 4\,\* \BS(2,4,1)
\\ && \nn \mbox{}
          - 8\,\* \BS(3,1,1,2)
          - 8\,\* \BS(3,1,2,1)
          + 12\,\* \BS(3,1,3)
          + 12\,\* \BS(3,2,1,1)
          - 6\,\* \BS(3,2,2)
          - 2\,\* \BS(3,3,1)
          + 4\,\* \BS(4,1,1,1)
          + 2\,\* \BS(4,1,2)
\\ && \nn \left. \mbox{}
          + 6\,\* \BS(4,2,1)
          - 2\,\* \BS(4,3)
          - 12\,\* \BS(5,1,1)
          + 2\,\* \BS(5,2)
          \right\}
\\[1mm] && \nn \mbox{\hspn}
   + 16\,\* {\colourBlue{ \cf\*\cas \* ( \ca - \cf ) }} \* \left\{
            28\,\*\BS(1,1,1,2,2)
          - 60\,\*\BS(1,1,1,3,1)
          - 28\,\*\BS(1,1,1,4)
          - 4\,\*\BS(1,1,2,1,2)
          + 36\,\*\BS(1,1,2,2,1)
\right.
\\ && \nn \mbox{}
          + 40\,\*\BS(1,1,2,3)
          + 10\,\*\BS(1,1,3,2)
          - 22\,\*\BS(1,1,4,1)
          + 14\,\*\BS(1,2,1,3)
          - 44\,\*\BS(1,2,2,2)
          + 54\,\*\BS(1,2,3,1)
          + 14\,\*\BS(1,2,4)
\\ && \nn \mbox{}
          - 28\,\*\BS(1,3,1,2)
          + 13\,\*\BS(1,3,2,1)
          - 9\,\*\BS(1,4,1,1)
          - 14\,\*\BS(1,5,1)
          + 3\,\*\BS(2,1,1,2,1)
          + 14\,\*\BS(2,1,1,3)
          - 3\,\*\BS(2,1,2,1,1)
\\ && \nn \mbox{}
          - 23\,\*\BS(2,1,2,2)
          + 29\,\*\BS(2,1,3,1)
          + 2\,\*\BS(2,1,4)
          - \BS(2,2,1,2)
          - 38\,\*\BS(2,2,2,1)
          - 24\,\*\BS(2,2,3)
          + 19\,\*\BS(2,3,1,1)
\\ && \nn \mbox{}
          - 5\,\*\BS(2,3,2)
          + 27\,\*\BS(2,4,1)
          - 22\,\*\BS(3,1,1,2)
          - 8\,\*\BS(3,1,3)
          + 22\,\*\BS(3,2,1,1)
          + 13\,\*\BS(3,2,2)
          + 5\,\*\BS(3,3,1)
          - 18\,\*\BS(4,1,2)
\\ && \nn \left. \mbox{}
          + 22\,\*\BS(4,2,1)
          - 14\,\*\BS(5,1,1)
          \right\}
\\[1mm] && \nn \mbox{\hspn}
   + 16\,\* {\colourBlue{ \cf\*\ca\* ( \ca - \cf )^2 }} \* \left\{
            124\,\*\BS(1,1,1,3,1)
          - 70\,\*\BS(1,1,1,2,2)
          + 86\,\*\BS(1,1,1,4)
          - 4\,\*\BS(1,1,1,2,1,1)
          + 4\,\*\BS(1,1,2,1,1,1)
\right.
\\ && \nn \mbox{}
          + 2\,\*\BS(1,1,2,1,2)
          - 44\,\*\BS(1,1,2,2,1)
          - 78\,\*\BS(1,1,2,3)
          - 12\,\*\BS(1,1,3,1,1)
          - 28\,\*\BS(1,1,3,2)
          + 20\,\*\BS(1,1,4,1)
\\ && \nn \mbox{}
          - 8\,\*\BS(1,2,1,1,2)
          + 2\,\*\BS(1,2,1,2,1)
          - 21\,\*\BS(1,2,1,3)
          + 16\,\*\BS(1,2,2,1,1)
          + 79\,\*\BS(1,2,2,2)
          - 98\,\*\BS(1,2,3,1)
          - 44\,\*\BS(1,2,4)
\\ && \nn \mbox{}
          - 10\,\*\BS(1,3,1,1,1)
          + 54\,\*\BS(1,3,1,2)
          - 14\,\*\BS(1,3,2,1)
          - 2\,\*\BS(1,3,3)
          + 4\,\*\BS(1,4,2)
          + 42\,\*\BS(1,5,1)
          - 6\,\*\BS(2,1,1,1,2)
\\ && \nn \mbox{}
          - 39\,\*\BS(2,1,1,3)
          + 6\,\*\BS(2,1,2,1,1)
          + 41\,\*\BS(2,1,2,2)
          - 56\,\*\BS(2,1,3,1)
          - 10\,\*\BS(2,1,4)
          - 4\,\*\BS(2,2,1,2)
          + 54\,\*\BS(2,2,2,1)
\\ && \nn \mbox{}
          + 44\,\*\BS(2,2,3)
          + 4\,\*\BS(2,3,1,1)
          + 4\,\*\BS(2,3,2)
          - 38\,\*\BS(2,4,1)
          + 20\,\*\BS(3,1,1,2)
          - 4\,\*\BS(3,1,2,1)
          + 32\,\*\BS(3,1,3)
          - 24\,\*\BS(3,2,2)
\\ && \nn \left. \mbox{}
          - 12\,\*\BS(3,2,1,1)
          - 16\,\*\BS(3,3,1)
          - 4\,\*\BS(4,1,1,1)
          + 48\,\*\BS(4,1,2)
          - 32\,\*\BS(4,2,1)
          - 8\,\*\BS(4,3)
          - 8\,\*\BS(5,1,1)
          + 8\,\*\BS(5,2)
          \right\}
\\[1mm] && \nn \mbox{\hspn}
   + 16\,\* {\colourBlue{ \cf \* ( \ca - \cf )^3 }} \* \left\{
            60\,\*\BS(1,1,1,2,2)
          - 56\,\*\BS(1,1,1,3,1)
          - 92\,\*\BS(1,1,1,4)
          - 4\,\*\BS(1,1,2,1,2)
          + 8\,\*\BS(1,1,2,2,1)
\right.
\\ && \nn \mbox{}
          + 44\,\*\BS(1,1,2,3)
          - 8\,\*\BS(1,1,3,1,1)
          + 32\,\*\BS(1,1,3,2)
          + 16\,\*\BS(1,1,4,1)
          - 2\,\*\BS(1,2,1,2,1)
          + 2\,\*\BS(1,2,1,3)
          - 2\,\*\BS(1,2,2,1,1)
\\ && \nn \mbox{}
          - 54\,\*\BS(1,2,2,2)
          + 52\,\*\BS(1,2,3,1)
          + 48\,\*\BS(1,2,4)
          + 4\,\*\BS(1,3,1,1,1)
          - 20\,\*\BS(1,3,1,2)
          + 16\,\*\BS(1,3,2,1)
          + 4\,\*\BS(1,3,3)
\\ && \nn \mbox{}
          + 4\,\*\BS(1,4,1,1)
          - 8\,\*\BS(1,4,2)
          - 44\,\*\BS(1,5,1)
          - 4\,\*\BS(2,1,1,1,2)
          - 10\,\*\BS(2,1,1,2,1)
          + 38\,\*\BS(2,1,1,3)
          + 2\,\*\BS(2,1,2,1,1)
\\ && \nn \mbox{}
          - 30\,\*\BS(2,1,2,2)
          + 36\,\*\BS(2,1,3,1)
          + 12\,\*\BS(2,1,4)
          + 12\,\*\BS(2,2,1,1,1)
          + 12\,\*\BS(2,2,1,2)
          - 20\,\*\BS(2,2,2,1)
          - 24\,\*\BS(2,2,3)
\\ && \mbox{}
          - 36\,\*\BS(2,3,1,1)
          + 4\,\*\BS(2,3,2)
          + 8\,\*\BS(2,4,1)
          + 24\,\*\BS(3,1,1,2)
          + 16\,\*\BS(3,1,2,1)
          - 48\,\*\BS(3,1,3)
          - 32\,\*\BS(3,2,1,1)
\\ && \nn \left. \mbox{}
          + 20\,\*\BS(3,2,2)
          + 20\,\*\BS(3,3,1)
          - 8\,\*\BS(4,1,1,1)
          - 56\,\*\BS(4,1,2)
          + 8\,\*\BS(4,2,1)
          + 16\,\*\BS(4,3)
          + 56\,\*\BS(5,1,1)
          - 16\,\*\BS(5,2)
          \right\}
\:\: . \;\;
\eea
Here we have subtracted its large-$\nc$ limit from the quartic 
colour factor in SU($\nc$), 
$\dfRAnc$ $= 2\,\cf\dfRAnA = \cf ( 1/24\; \nct + 1/4\;\nc )$.
For the group invariants see table 1 of ref.~\cite{Ruijl:2017eht}.
Had we used $(\ca - 2\,\cf)$ for the rest of eq.~(\ref{eq:gns3W7}), 
then the $\cf \cat$ main bracket would have been identical to the 
large-$\nc$ result of ref.~\cite{Moch:2017uml}.

Instead, only the first two main brackets remain for $\cf = \ca$.
These contributions should be identical to the corresponding result for 
${\cal N} \!=\! 4$ maximally supersymmetric Yang-Mills (MSYM) theory,
first obtained to three and (for large $\nc$) four loops in 
refs.~\cite{Kotikov:2004er,Staudacher:2004tk} and 
refs.~\cite{Kotikov:2007cy,Bajnok:2008qj}. 
Indeed, the remaining two brackets of eq.~(\ref{eq:gns3W7}) agree 
with the r.h.s.~of
eqs.~(41) and (42) of ref.~\cite{Kniehl:2021ysp}. In the $\cf \nc$ case, 
this agreement can be viewed as confirmation of both the QCD and MSYM 
results.

Before we turn to the behaviour of the four-loop non-singlet splitting
in the large-$x$ and small-$x$ limits, we address one more
structural feature. This concerns the function 
\bea
\label{eq:fwrap}
    f(N) &\, = \,&
         5\:\! \* \z5
       - 2\, \* \S(-5)
       + 4\:\! \* \z3\,\*\S(-2) 
       - 4\, \* \S(-2,-3)
       + 8\, \* \S(-2,-2,1)
       + 4\, \* \S(3,-2)
       - 4\, \* \S(4,1)
       + 2\, \* \S(5)
\nn \\   &\, = \,&
         5\:\!\* \z5 
       - 2\:\!\* \z3\,\*\BS(2)
       - \BS(2,1,2)
       + \BS(3,1,1)
\:\: .
\eea
This function appeared first in the three-loop coefficient functions
for structure functions $F_2$ at even $N$ and $F_3$ at odd $N$
\cite{Vermaseren:2005qc,Moch:2008fj}. Due to the alternating sums,
it corresponds to different $x$-space functions in these two cases.
{The product $g(N) = N (N\!+\!1) f(N)$ behaves like $N^{\:\!0}$ for 
$N \ra \infty$ with Mellin inverses $ g^{\pm}(x) = (\z2 \pm \z3)\, \delta(1-x) 
+ {\cal O}\left( (1-x)^1 \right)$, where $g^{+}(x)$ corresponds to  
even $N$ and $g^{-}(x)$ to odd $N$.} 
Note: $ N (N\!+\!1) f $ is RR, hence no $(1-x)^0$ term.
For more details see refs.~\cite{Vermaseren:2005qc,Moch:2008fj}. 

At four loops, $S_1^{\,2} f(N)$ occurs as the `wrapping correction'
\cite{Bajnok:2008qj} in the $\nfz$ parts \cite{Gehrmann:2026qbl} 
of the non-singlet anomalous dimensions, in the form suggested by its 
$\z5$ and $\z3$ contributions 
\cite{Moch:2017uml,Vogt:2023unp,Kniehl:2025jfs},
\beq
  \label{eq:PnsPMwrap}
  \gamma_{\,\rm ns}^{\:(3)\pm}(N) \:=\: 
  -\,\frct{128}{3} \left\{ 3\,\cfs\,\cas - 2\,\cf\cat + 12\:\dfRAnc \right\}
   [S_1(N)]^2\, f(N) \;+\; \dots
\;\; .
\eeq
Its $w\!=\!7$ non-$\zeta$ part is not really visible in the `linear' 
representation of the sums used in eq.~(\ref{eq:gns3W7}). 
This is in contrast to the contribution of $f(N)$ to the third non-singlet 
anomalous dimension,
\beq
  \label{eq:PnsSf(N)}
  \gamma_{\,\rm ns}^{\:(3)\rm s}(N) \:=\: 
  -\,\frct{32}{3}\; \ca \nf\,\dabc2n \;
   f(N) \;+\; \mbox{ terms with denom's }
\:\: .
\eeq
This $k=0$ `wrapping' part with $f(N)$ is the new four-loop feature of 
$\gamma_{\,\rm ns}^{\;\rm s}$ mentioned below eq.~(\ref{eq:gamStr}).

\vspace*{2mm}
In the large-$N$ limit, the main non-singlet anomalous dimensions take
the form \cite{Korchemsky:1988si,Albino:2000cp,Dokshitzer:2005bf}
\beq
\label{eq:NtoInf}
  \gamma_{\,\rm ns}^{\:\pm (n\!-\!1)}(N) \:\: = \:\:
    A_{\rm q}^{(n)} \, \lnN 
  - B_{\rm q}^{(n)}
  + \, N^{-1}\, \Big\{ C_{\rm q}^{\,(n)} \lnN  + \Big[ \, 
    \frkt{1}{2}\: A_{\rm q}^{(n)}- \,\widetilde{D}_{\rm q}^{(n)} 
    \Big] \Big\}
  + {\cal O}\left( N^{-2} \right)
\eeq
with $\lnN = \ln N +\Ge$, where $\Ge \cong 0.5772156649$ is the Euler-%
Mascheroni constant. 
The four-loop (light-like) cusp anomalous dimensions $A_{\rm q/g}^{(4)}$ 
are known from refs.~\cite{Henn:2019swt,vonManteuffel:2020vjv}. 
As checked already in ref.~\cite{Gehrmann:2026qbl}, the $1/N$ coefficients in 
eq.~(\ref{eq:NtoInf}) are given by the fourth-order expansion coefficients of
\beq
\label{eq:CDofAB}
  C_{\rm q} \;=\; ( \:\! A_{\rm q} \:\! )^{2^{}}
\;\; , \quad
  D_{\rm q} \;=\; A_{\rm q} \cdot ( B_{\rm q}- \beta(\ars)/\ars )
\:\: ,
\eeq
where $\,\beta(\ars) = - \b0\,\ar(2) - \beta_1\,\ar(3) \,- \ldots\:$ 
with $\beta_0^{} = \frkt{11}{3}\:C_A - \frkt{2}{3}\:\nf$ etc.
 
The fact that $\gamma_{\,\rm u}^{}(N)$ in eq.~(\ref{eq:gu-exp}) can be 
expressed in terms of reciprocity-respecting denominators and binomial sums 
implies that it cannot include any $N^{-1}$ terms besides those generated
by $A^{(n)} S_1(N)$. Thus the coefficients $C^{\,(n)}$ and $D^{\,(n)}$ in
eq.~(\ref{eq:NtoInf}) must be `inherited' from lower orders, and the
agreement of ref.~\cite{Gehrmann:2026qbl} with eq.~(\ref{eq:CDofAB}) is
essential.

The `virtual' anomalous dimension $B_{\rm q}^{(4)}$, on the other hand, is 
not purely virtual but a genuine four-loop quantity. 
It was not completely known before ref.~\cite{Gehrmann:2026qbl}: 
the (large-$\nc$ related) coefficients of the $\cf \cat$ and $\dfRAnc$ were 
only known approximately with \cite{Moch:2023tdj}
\beq
\label{eq:b4FAappr} \left.
  B_{\,\rm q}^{\,(4)} \right|_{\dfRAnc}^{\,\rm approx.} 
  \;=\:\: \mbox{} - 998.02 \,\pm\, 0.02
\:\: ,  
\eeq
see eq.~(A.9) of ref.~\cite{Kniehl:2025ttz}. 
This result is now superseded by the exact value \cite{Gehrmann:2026qbl}
\bea \left.
\label{eq:b4FA}
  B_{\,\rm q}^{\,(4)} \right|_{\dfRAnc} &\!=\!&
    96
  - \frct{944}{3}\:\* \z2
  - \frct{1232}{3}\:\* \z3
  + \frct{32}{3}\:\* \z4
  - 896\:\* \z2\*\z3
  - 400\:\* \z5
\nn \\[-1mm] & & \mbox{}
  - 704\:\* \zts
  - \frct{1562}{9}\,\z6
  + 320\:\* \z2\*\z5 
  - 64\:\* \z3\*\z4
  + 2800\:\* \z7
\nn \\[1mm] & \cong & \mbox{}
  \mbox{} - 998.0155365
\:\: .
\eea
While the error in eq.~(\ref{eq:b4FAappr}) was small, it was not entirely 
irrelevant as mentioned in the next section.

The $x$-space analogue and extension (second line) of eq.~(\ref{eq:NtoInf}) 
reads, at four loops,
\bea
\label{eq:xto1}
  P_{\rm ns}^{\,(3)\pm}(x) &\! = \! &
        A_{\rm q}^{(4)} \, \frct{1}{(1-x)_+}
  \,+\, B_{\rm q}^{(4)} \, \delta \xm
  \,+\, C_{\rm q}^{\,(4)} \, \ln \xm
  \,-\, A_{\rm q}^{(4)} + \widetilde{D}_{\rm q}^{(4)}
\nn \\[0.2mm] & &  
  \,+\, \mbox{ $\xm^p \ln^{\,\ell} \xm$ terms} \quad
        \mbox{ with $\,\ell \leq 3\,$ and $\,p \geq 2\,$ for $\,\ell = 3\,$} 
\:\: .
\eea
The highest logarithms, so far generally $\ln^{\,n} \xm$ for 
$P_{\rm ns}^{\,\pm (n)}(x)$, can be collected to all $p$ as
\beq
  - \,\frct{128}{6}\:\cff \,\ln^{3\!} x \:\ln^{3\!} \xm \,\pqq(x)
  \quad \mbox{ with} \quad 
  \pqq(x) = 2\, (1 - x)^{-1} - 1 - x 
\;\; .
\eeq
The form of this result agrees with eq.~(3.26) of ref.~\cite{Moch:2009hr}.
Its prefactor provides hitherto undetermined coefficients, $\xi_{P_3^{}} 
= - 128$ for the present space-like case and, by virtue of 
ref.~\cite{Moch:2017uml}, $+128$ for its time-like counterpart. 
This continues a very regular pattern of these power-suppressed 
leading-logarithmic coefficients. 
These results can be used to extend the prediction of the 
$(1-x)^p \ln^{4} \xm $ terms to all $p$ for the four-loop non-singlet 
coefficient functions of the structure functions in inclusive DIS and 
of the fragmentation functions in semi-inclusive $e^+ e^-$ annihilation.
See~ref.~\cite{Moch:2009hr} for the corresponding $\ln^{\,\ell} \xm $
coefficient-function contributions with $\ell = $ 5, 6 and 7.

\vspace*{2mm}
In the small-$x$ limit, the results of ref.~\cite{Gehrmann:2026qbl} 
can be compared mainly to two types of predictions addressed in detail in 
ref.~\cite{Davies:2022ofz}. The first concerns the double logarithms,
$x^{\,k} \ln^{\,\ell\!} x$ with $\ell = 4,\,5,\,6$ at even $k$ for 
$P_{\rm ns}^{\,(3)+}$ and at odd $k$ for $P_{\rm ns}^{\,(3)-}$. 
These have been constructed from the structure of the unfactorized
structure functions in DIS in terms of log-generating phase-space and 
virtual-correction contributions with $x^{\,\ell \ep}$ in $D = 4 - 2\:\ep$ 
dimensions \cite{Vogt:2012gb}, analogous its fragmentation-function 
counterparts \cite{Vogt:2011jv} and to a large-$x$ approach 
\cite{Almasy:2010wn} complementary to that in ref.~\cite{Moch:2009hr}.

The second types of predictions concern all $x^{\,0} \ln^{\,\ell\!} x$ 
terms for $\ell > 0$ of $P_{\rm ns}^{\,(3)+}$. These can be generated 
by a surprisingly simple generalization \cite{Velizhanin:2014dia} of a 
leading-log relation \cite{Kirschner:1983di,Blumlein:1995jp},
\beq
\label{eq:gnsto0V}
  \gamma_{\rm ns}^{}(N,\ars) \cdot \left( \,\gamma_{\rm ns}^{}(N,\ars)
  + N - \beta(\ars) / \ars \right) \;=\; O(1)
\;\; ,
\eeq
that seems to hold except for terms with $\z2^{}$ beyond the large-$\nc$ limit.
See ref.~\cite{Manashov:2025kgf} for research on the origin of this and 
similar relations.

All these predictions agree with the exact splitting functions 
$P_{\rm ns}^{\,(3)\pm}(x)$ with one exception that was already noted in 
ref.~\cite{Gehrmann:2026qbl}: the $\z2$ coefficients of the lowest 
double-logarithms $x^{\,k} \ln^{\,4} x$ disagree for all $\nfz$ terms;
this disagreement vanishes in the large-$\nc$ limit. 
The $\dfRAnc$ result is especially striking, 
\bea
  P_{\rm ns}^{\,(3)+}(x) \big|_{\dfRAnc} &\! = \! & 
    -\,  48\,\* \z2\,\* \ln^{\,4} x
  \,-\, 256\,\* \z2\,\* \ln^{\,3} x
  \,-  \left(  2112\,\* \z2 -  768\,\* \zz(2,2) \right) \* \ln^{\,2} x
\nn \\ & & \mbox{}
  - \left( 9984\,\* \z2 + 2304\,\* \zz(2,2) + 6144\,\* \z2\*\z3 \right) \ln x
  \:+\: \dots
\:\: .
\eea
It does not include any $x^{\,0}$ terms enhanced by $\ln^{\,\ell\!} x$ 
without powers of $\pi$, a pattern that did not occur anywhere at lower orders.
The difference between the new result of ref.~\cite{Gehrmann:2026qbl} and the 
double logarithmic predictions for the $\z2 x^{\,k} \ln^{\,4} x$ can be cast in a simple form for all values of $k$,
\bea
    \left[ P_{\rm ns, [1] }^{\,(3)\pm} - P_{\rm ns, predicted}^{\,(3)\pm}
    \right]_{\,\z2 \ln^{\,4} x} &\!\!=\!\!&
  - \left( \dfRAnc - \frkt{1}{24}\:\* \cf \* \cat \right) 
    \left\{ 256\,(1-x)^{-1} - 160\, (1+x) \right\}
\nn \\[-1mm] & & \mbox{}
  + \;\cf \cas (\ca \!-\! 2\,\cf)  \;
      \left\{ 36 \,(1-x)^{-1} - 22\, (1+x) \right\}
\nn \\[1mm] & & \mbox{}
  + \,\cf \ca (\ca \!-\! 2\,\cf)^2 
      \left\{ 44 \,(1-x)^{-1} - 24\, (1+x) \right\}
\nn \\[1mm] & & \mbox{}
  + \;\;\cf (\ca \!-\! 2\,\cf)^3 \;\;
      \left\{  8 \,(1-x)^{-1} -  6\, (1+x) \right\}
\;\; .
\eea
This applies to even (odd) powers of $x$ for $P_{\rm ns}^{\,+}$
($\,P_{\rm ns}^{\,-\,}$) after expanding the curly brackets about $x=0$.

These results suggest that an entirely new structure occurs at the fourth 
order, which generates large-$\nc$ suppressed small-$x$ logarithms even 
beyond the level of a single-log higher-order enhancement (here $\ln^{\,3} x$)
but only with powers of $\z2$ which may be generated via $(i\:\!\pi)^2$.
This pattern is very similar to that of the `super-leading' logarithms
found and studied in refs.~\cite{Forshaw:2006fk,Catani:2011st,Forshaw:2012bi},
for recent developments we refer the reader to refs.~%
\cite{Becher:2023mtx,Becher:2024kmk,Becher:2025igg} and references therein.
It invites further research.

We remind the reader of eqs.~(21) and (22) of \cite{Moch:2009mu},
where similarly large-$\nc$ suppressed powers of $\pi^2$ appeared in the
threshold resummation of the DIS structure function $F_L$.

So far, we have addressed the $x^{2m} \ln^{\,\ell\!} x$ terms of 
$P_{\rm ns}^{\,+}(x)$ and the $x^{2m+1} \ln^{\,\ell\!} x$ terms of 
$P_{\rm ns}^{\,-}(x)$. 
As~far as we know, there are no resummation predictions for the remaining 
powers of $x$, with the exception of the leading $x^0$ logarithms 
$\ln^{\,2n\!} x$ of $P_{\rm ns}^{\,(n)-}(x)$ with have been obtained
long ago from the results of ref.~\cite{Kirschner:1983di} in 
ref.~\cite{Blumlein:1995jp}. There only the third-order coefficient was given 
explicitly, the higher-order results were presented only numerically in 
table~1 of ref.~\cite{Blumlein:1996aw}. The result of 
ref.~\cite{Gehrmann:2026qbl},
\bea
  \hspn \left. P_{\rm ns}^{\,(3)-} \right|_{\,\ln^{\,6} x} &\!\!=\!&
  \frct{1}{9}\:\cff + \frct{28}{45}\: \cft (\ca \!-\! 2\,\cf)
  + \frct{32}{45}\: \cfs (\ca \!-\! 2\,\cf)^2
  + \frct{4}{15}\: \cf (\ca \!-\! 2\,\cf)^3
\nn \\[1mm]  &\!\cong\!& 0.99643347 \quad \mbox{ for  QCD }
\;\; ,
\eea
agrees with that prediction. 
As it has to, it reduces to the corresponding simpler coefficient for 
$P_{\rm ns}^{\,(3)-}(x)$ in the large-$\nc$ limit 
$(\ca \!-\! 2\,\cf) = 0$.
 
%
\section{Implications for other four-loop quantities}
\label{sec:inpls}
\setcounter{equation}{0}
%

There is a well-known connection between the coefficients 
$B^{\,(n)}$ in eq.~(\ref{eq:NtoInf}), the $1/(1-x)_+$ soft-gluon
contributions to $n^{\,\rm th}$-order coefficients functions, e.g., 
for inclusive DIS, and the $\ep^{-1}$ poles of the $n$-loop quark or 
gluon form factor. At three loops, we have employed the results of 
refs.~\cite{Moch:2004pa,Vogt:2004mw,Vermaseren:2005qc} to derive the
$1/\ep$ terms of the form factors \cite{Moch:2005id,Moch:2005tm}, 
and then used these to obtain new results for Drell-Yan lepton-pair
production and for the total cross section for Higgs-boson production 
in proton-proton collisions \cite{Moch:2005ky}. 
See also refs.~\cite{Laenen:2005uz,Idilbi:2005ni}.

A corresponding four-loop effort, partially based on numerical 
information of rather limited accuracy, was undertaken in 
refs.~\cite{Das:2019btv,Das:2020adl}. Taking into account the $\ep^{-1}$
form-factor results of refs.~\cite{Agarwal:2021zft,Lee:2021uqq}
and the $\nfo$ contributions to the non-singlet splitting function 
contributions ref.~\cite{Kniehl:2025ttz}, the new $\nfz$ results 
\cite{Gehrmann:2026qbl} remove the last coefficient (\ref{eq:b4FAappr})
that was known only approximately. 

The four-loop virtual gluon anomalous dimension $B_{\rm q}^{\,(4)}$ in 
eq.~(13) of ref.~\cite{Gehrmann:2026qbl} is related to its gluon
counterpart $B_{\rm g}^{\,(4)}$ via eikonal anomalous dimensions 
$f_{\rm q/g}^{(\,(4)}$ corresponding to colour-charged Wilson lines 
\cite{Dixon:2008gr,Falcioni:2019nxk}.
These anomalous dimensions exhibit the same maximal non-Abelian color 
structure as the cusp anomalous dimensions. 
Therefore $f_{\rm q}^{\,(4)}$ and $f_{\rm g}^{\,(4)}$ are related by 
(generalized \cite{Moch:2018wjh}) Casimir scaling. 
Inserting eq.~(\ref{eq:b4FA}) into the previous approximate result, 
one obtains
\bea
\label{eq:fI4}
  {\lefteqn{
  f_{\,\rm q/g}^{\,(4)} \:\: = }}
\nn \\ && \mbox{\hspn} \phantom{+}
  \colourBlue{\cI\,\*\cat} \* \left( 
            \frct{9364079}{6561}
          - \frct{1183819}{729}\,\*\z2
          - \frct{837520}{243}\,\*\z3
          + \frct{115783}{27}\,\*\z4
          + \frct{11896}{9}\,\*\z3\*\z2
          + \frct{106034}{27}\,\*\z5
\right. \nn \\ && \left. \mbox{}
          - \frct{4906}{9}\,\*\zts
          - \frct{131197}{54}\,\*\z6
          - 710\,\*\z4\*\z3
          - 432\,\*\z5\*\z2
          - \frct{9671}{6}\,\*\z7
  \right)
  + \colourBlue{\dfIAnI} \* \Big( 
          - 96\,\*\z2
\nn \\ && \left. \mbox{}
          - \frct{416}{9}\,\*\z3
          + 16\,\*\z4
          + \frct{5360}{9}\,\*\z5
          + \frct{880}{3}\,\*\zts
          + \frct{2200}{3}\,\*\z6
          - 240\,\*\z4\*\z3
          + 384\,\*\z5\*\z2
          - 2116\,\*\z7
  \right)
\nn \\[1mm] && \mbox{\hspn}
  + \colourBlue{\nf\,\*\cI\,\*\cas} \* \left( 
          - \frct{394109}{1944}
          + \frct{294539}{729}\,\*\z2
          - \frct{31340}{243}\,\*\z3
          - \frct{11050}{9}\,\*\z4
          - \frct{104}{9}\,\*\z3\*\z2
          - \frct{692}{27}\,\*\z5
          + \frct{4420}{9}\,\*\zts
\right. \nn \\ && \left. \mbox{}
          + \frct{16895}{27}\,\*\z6
   \right)
   + \colourBlue{\nf\,\*\dfIFnI} \* \left( 
            256\,\*\z2
          + \frct{640}{9}\,\*\z3
          - 32\,\*\z4
          - \frct{1600}{9}\,\*\z5
          - \frct{320}{3}\,\*\zts
          - \frct{800}{3}\,\*\z6
  \right) 
\nn \\[1mm] && \mbox{\hspn}
  + \colourBlue{\nf\,\*\cI\,\*\cf\*\ca} \* \left( 
          - \frct{813475}{972}
          + \frct{2819}{9}\,\*\z2
          + \frct{68882}{81}\,\*\z3
          - \frct{988}{9}\,\*\z4
          - 160\,\*\z3\*\z2
          + \frct{1448}{9}\,\*\z5
          - 312\,\*\zts
\right. \nn \\ &&  \mbox{}
          + 16\,\*\z6
  \Big) 
  + \colourBlue{\nf\,\*\cI\,\*\cfs} \* \left( 
            \frct{21037}{108}
          - 2\,\*\z2
          + \frct{4424}{9}\,\*\z3
          + 74\,\*\z4
          - \frct{1600}{3}\,\*\z5
          - 80\,\*\zts
          - 200\,\*\z6
  \right) 
\nn \\[1mm] && \mbox{\hspn}
  + \colourBlue{\nfs\,\*\cI\*\ca} \* \left( 
            \frct{27875}{17496}
          - \frct{15481}{729}\,\*\z2
          + \frct{32152}{243}\,\*\z3
          + \frct{388}{9}\,\*\z4
          - \frct{224}{9}\,\*\z3\*\z2
          - 112\,\*\z5
  \right) 
\nn \\[1mm] && \mbox{\hspn}
  + \colourBlue{\nfs\,\*\cI\*\cf} \* \left( 
            \frct{16733}{486}
          - \frct{172}{9}\,\*\z2
          - \frct{4568}{81}\,\*\z3
          + \frct{64}{9}\,\*\z4
          + \frct{32}{3}\,\*\z3\*\z2
          + \frct{304}{9}\,\*\z5
  \right) 
\nn \\[1mm] && \mbox{\hspn}
  + \colourBlue{\nft\,\*\cI} \* \left( 
          - \frct{16160}{6561}
          - \frct{16}{81}\,\*\z2
          - \frct{400}{243}\,\*\z3
          + \frct{128}{27}\,\*\z4
  \right)
\eea
with $\cI= \cf$, $d_I^{\,abcd} = d_F^{\,abcd}$ and $\nI = \nF$ 
for the quark case, and $\cI= \ca$, 
$d_I^{\,abcd} = d_A^{\,abcd}$ and $\nI = \nA$
for the gluon case. Its lower-order counterparts can be found, e.g.,
in eq.~(3.22) of ref.~\cite{Das:2019btv}. 

\pagebreak

Following the procedure applied in ref.~\cite{Das:2020adl}, the four-loop 
gluon virtual anomalous dimension is found to be \cite{avLL2026}
\bea
\label{eq:Bg4}
  {\lefteqn{
  B_{\rm g}^{\,(4)}  =\! }}
\nn \\ && \mbox{\hspn} \phantom{+}
  \colourBlue{\caf} \* \left(
            \frct{48443}{486}
          + \frct{2452}{27}\,\*\z2
          + \frct{48550}{27}\,\*\z3
          + \frct{8941}{54}\,\*\z4
          - \frct{3566}{9}\,\*\z3\*\z2
          - \frct{14467}{9}\,\*\z5
          + \frct{770}{3}\,\*\zts
\right. \nn \\ &&  \left. \mbox{}
          - \frct{37477}{108}\,\*\z6
          + \frct{512}{3}\,\*\z4\*\z3
          + \frct{200}{3}\,\*\z5\*\z2
          + \frct{1750}{3}\,\*\z7  
  \right)
  + \colourBlue{\dfAAna} \* \left(
            \frct{64}{9}
          + 80\,\*\z2
          - 672\,\*\z3
\right. \nn \\ &&  \left. \mbox{}
          - \frct{476}{3}\,\*\z4
          - 1168\,\*\z3\*\z2
          - \frct{440}{3}\,\*\z5
          - 704\,\*\zts
          - \frct{814}{9}\,\*\z6
          - 64\,\*\z4\*\z3
          + 320\,\*\z5\*\z2
          + 2800\,\*\z7
  \right)
\nn \\ && \mbox{\hspn}
  + \colourBlue{\nf\,\*\cat} \* \left(
          - 7
          - \frct{2579}{27}\,\*\z2
          - \frct{12749}{27}\,\*\z3
          - \frct{1601}{54}\,\*\z4
          + \frct{718}{9}\,\*\z3\*\z2
          + \frct{931}{3}\,\*\z5
          - \frct{836}{3}\,\*\zts
          + \frct{1727}{54}\,\*\z6
  \right)
\nn \\ && \mbox{\hspn}
  + \colourBlue{\nf\,\*\dfFAna} \* \left(
            \frct{224}{9}
          - 160\,\*\z2
          + \frct{320}{3}\,\*\z3
          + \frct{664}{3}\,\*\z4
          + 608\,\*\z3\*\z2
          - \frct{4880}{3}\,\*\z5
          + 256\,\*\zts
          + \frct{296}{9}\,\*\z6
  \right)
\nn \\ && \mbox{\hspn}
  + \colourBlue{\nf\,\*\cas \* \cf} \* \left(
          - \frct{22627}{486}
          + \frct{34}{3}\,\*\z2
          - \frct{2182}{9}\,\*\z3
          + \frct{65}{3}\,\*\z4
          - 8\,\*\z3\*\z2
          - 80\,\*\z5
          + 232\,\*\zts
          + \frct{280}{9}\,\*\z6
  \right)
\nn \\[1mm] && \mbox{\hspn}
  + \colourBlue{\nf\,\*\ca \* \cfs} \* \left(
          - \frct{1859}{27}
          + \frct{176}{9}\,\*\z3
  \right)
  + 23\,\* \colourBlue{\nf\,\*\cft}
  + \colourBlue{\nfs\,\*\dfFFna} \* \left(
          - \frct{704}{9}
          + \frct{512}{3}\,\*\z3
  \right)
\nn \\[1mm] && \mbox{\hspn}
  + \colourBlue{\nfs\,\*\cas} \* \left(
            \frct{1352}{81}
          + \frct{37}{27}\,\*\z2
          + \frct{289}{27}\,\*\z3
          + \frct{200}{27}\,\*\z4
          - \frct{32}{9}\,\*\z3\*\z2
          - \frct{8}{9}\,\*\z5
  \right)
  + \colourBlue{\nfs\,\*\ca \* \cf} \* \left(
            \frct{3910}{243}
          + \frct{160}{9}\,\*\z3
  \right)
\nn \\[1mm] && \mbox{\hspn}
  + \colourBlue{\nfs\,\*\cfs} \* \left(
            \frct{338}{27}
          - \frct{176}{9}\,\*\z3
  \right)
  + \frct{5}{243}\,\* \colourBlue{\nft\,\*\ca}
  + \frct{154}{243}\,\* \colourBlue{\nft\,\*\cf}
\:\: .
\eea
The QED terms without $\ca$ and $d_{A}^{\,abcd}$ have no counterparts 
in $B_{\rm q}^{\,(4)}$. As at lower orders they are identical to the 
corresponding terms in the $\beta$-function  
\cite{Gorishnii:1990kd,vanRitbergen:1997va,Czakon:2004bu}. 
The coefficients of $\nft\,\ca$, $\nft\,\cf$ and $\nfs\,\cfs$ have also been 
determined as part of all-$N$ expressions for $\gamma_{\rm gg}^{\:(3)}$ in 
refs.~\cite{Bennett:1997ch,Davies:2016jie,Kniehl:2026eij}.

Eq.(\ref{eq:Bg4}) leads to the numerical QCD result
\beq
  B_{\rm g}^{\,(4)} \:\cong\:\:
    68587.703
  - 18143.983\,\*\nf
  + 423.81135\,\*\nfs 
  + 0.90672154\,\*\nft
\:\: .
\eeq
The $\nfz$ part supersedes the approximation $68587.64 \pm 0.2$ employed 
in ref.~\cite{Falcioni:2024qpd} where this uncertainty, despite being tiny, 
was relevant for the width of the error band for the gluon-gluon splitting 
function $P_{\rm gg}^{\,(3)}(x)$ at small $x$.

The 4-loop results for the virtual and eikonal anomalous dimensions 
determine the {$n=4, \ell=0$} plus-distribution coefficients in the soft-gluon
(threshold) expansion of the coefficient functions,
\beq
  c_{\!P}^{(n)}(x) \;=\; 
    \sum_{\ell\,=\,0}^{2\,n-1} \ar(n)\, c^{(n,\ell)}_{\!P}
       \Bigg[ \frac{\ln^{\,\ell\!} \xm}{\xm} \Bigg]_+
  + \: c^{(n,\delta)}_{\!P} \, \delta \xm
\eeq
with $c^{(0,\delta)}_{P} = 1$,
for the dominant $x\!\ra 1$ contributions to the structure functions $F_2^{}$ 
and $F_3^{}$ in inclusive DIS ($\,P = 2/3,{\rm q\,}$), the corresponding 
fragmentation functions in semi-inclusive $e^+e^-$ annihilation (SIA), and 
for lepton-pair production ($\,P = {\rm DY, q\,}$) and inclusive Higgs-boson 
production ($\,P = {\rm H,gg\,}$) at hadron colliders. 
Taking into account eq.~(32) of~\cite{Moch:2009my}, these coefficient
have been provided in eq.~(A.12) -- (A.14) of ref.~\cite{Kniehl:2025ttz}
without the exact expression for the coefficient (\ref{eq:b4FA}).

These results can be used to fully specify the coefficients 
$B^{\,(4)}_{\rm q,DIS}$ and $D^{\,(4)}_{\rm DY/H}$ 
of the next-to-next-to-next-to-next-to-leading logarithmic (N$^{\ell=4}$LL) 
terms in the soft-gluon exponentiation 
\cite{Sterman:1986aj,Catani:1989ne,Magnea:1990qg,Catani:1990rp}
\beq
\label{eq:cNres}
  c_{P}(N,\ars) \; =\;
  g_{P,0}^{}(\ars) \, \exp\, [G_{P}(N,\ars)] \: + \:
  {\cal O}(N^{-1}\ln^n N) 
\eeq
with
\beq
\label{eq:GN}
  G_P \;=\; \ln \widetilde{N} g_{P,1}^{}(\lambda) 
           + g_{P,2}^{}(\lambda)
           + \sum_{\ell=2} \ar(\ell-1) g_{P,\ell+1}^{}(\lambda) 
\;\; , \quad \lambda \,=\, \b0 \,\ars \ln N 
\:\: .
\eeq
The functions $g_{P,\ell+1}^{}$ at $\,2 \leq \ell \leq 4$,
{which include the coefficients $A_{\rm q/g}^{(\ell+1)}$ and 
$B_{\rm q,DIS}^{\,(\ell)}$ or $D^{\,(\ell)}_{\rm DY/H}$, see, e.g., 
eqs.~(2.2) - (2.7) of ref.~\cite{Moch:2005ba},} 
have been determined in 
refs.~\cite{Vogt:2000ci,Catani:2003zt,Moch:2005ba,Das:2019btv}.
The SIA expression (\ref{eq:cNres}) differs from the DIS case only in the
prefactor $g_0^{}$, see ref.~\cite{Moch:2009my}. The N$^4$LL coefficients read
\bea
\label{eq:B4dis}
  {\lefteqn{
  B_{\rm q,DIS}^{\,(4)} \:\: = }}
\nn \\ && \mbox{\hspn} \phantom{+}
  \colourBlue{\cf\*\cat} \* \left(
            \frct{1778641645}{157464}
          + \frct{50250284}{2187}\,\*\z2
          - \frct{17964269}{1458}\,\*\z3
          - \frct{3098371}{324}\,\*\z4
          - \frct{71684}{9}\,\*\z3\*\z2
\right. \nn \\[1mm] &&  \left. \mbox{\hspn}
          - \frct{94642}{27}\,\*\z5
          + \frct{36718}{27}\,\*\zts
          + \frct{1605583}{324}\,\*\z6
          + \frct{2098}{3}\,\*\z4\*\z3
          - \frct{176}{3}\,\*\z5\*\z2
          + \frct{9197}{2}\,\*\z7
  \right)
\nn \\ &&  \mbox{\hspn}
  + \colourBlue{\dfRAnr} \* \left(
          - 96
          + \frct{1232}{3}\,\*\z2
          + \frct{4112}{9}\,\*\z3
          - \frct{80}{3}\,\*\z4
          + 896\,\*\z3\*\z2
          - \frct{1760}{9}\,\*\z5
          + \frct{1232}{3}\,\*\zts
\right. \nn \\ &&  \left. \mbox{\hspn}
          - \frct{5038}{9}\,\*\z6
          + 304\,\*\z4\*\z3
          - 704\,\*\z5\*\z2
          - 684\,\*\z7
  \right)
  + \colourBlue{\cfs\*\cas} \* \left(
          - \frct{190583}{108}
          - \frct{1961281}{162}\,\*\z2
\right. \nn \\[1mm] &&  \left. \mbox{\hspn}
          + \frct{960862}{81}\,\*\z3
          - \frct{1815931}{162}\,\*\z4
          + \frct{279844}{27}\,\*\z3\*\z2
          + \frct{142822}{27}\,\*\z5
          + \frct{362}{9}\,\*\zts
          + \frct{328445}{54}\,\*\z6
\right. \nn \\[1mm] &&  \left. \mbox{\hspn}
          + 32\,\*\z4\*\z3
          + 2104\,\*\z5\*\z2
          - 8610\,\*\z7
  \right)
  + \colourBlue{\cft\*\ca} \* \left(
            \frct{34679}{12}
          + \frct{2567}{18}\,\*\z2
          + \frct{50755}{3}\,\*\z3
          + \frct{21164}{3}\,\*\z4
\right. \nn \\[1mm] &&  \left. \mbox{\hspn}
          - 5028\,\*\z3\*\z2
          - \frct{50048}{3}\,\*\z5
          - \frct{7900}{3}\,\*\zts
          - \frct{361663}{54}\,\*\z6
          - 128\,\*\z4\*\z3
          - 2064\,\*\z5\*\z2
          + 10920\,\*\z7
  \right)
\nn \\ && \mbox{\hspn}
  + \colourBlue{\cff} \* \left(
          - \frct{4873}{24}
          + 450\,\*\z2
          - 2004\,\*\z3
          + 342\,\*\z4
          + 120\,\*\z3\*\z2
          + 2520\,\*\z5
          + 1152\,\*\zts
          + 2111\,\*\z6
\right. \nn \\ &&  \left. \mbox{\hspn}
          - 64\,\*\z4\*\z3
          + 384\,\*\z5\*\z2
          - 5880\,\*\z7
  \right)
  + \colourBlue{\nf\,\*\cf\*\cas} \* \left(
          - \frct{100892009}{17496}
          - \frct{23133961}{2187}\,\*\z2
          + \frct{4355635}{729}\,\*\z3
\right. \nn \\ &&  \left. \mbox{\hspn}
          + \frct{140921}{54}\,\*\z4
          + \frct{14200}{9}\,\*\z3\*\z2
          - \frct{11714}{27}\,\*\z5
          - \frct{23740}{27}\,\*\zts
          - \frct{86146}{81}\,\*\z6
  \right)
  + \colourBlue{\nf\,\*\dfRRnr} \* \left(
            192
\right. \nn \\ &&  \left. \mbox{\hspn}
          - \frct{2656}{3}\,\*\z2
          + \frct{2336}{9}\,\*\z3
          + \frct{448}{3}\,\*\z4
          - 64\,\*\z3\*\z2
          + \frct{11680}{9}\,\*\z5
          - \frct{448}{3}\,\*\zts
          + \frct{608}{9}\,\*\z6
  \right)
\nn \\[1mm] &&  \mbox{\hspn}
  + \colourBlue{\nf\,\*\cfs\*\ca} \* \left(
          - \frct{167965}{1458}
          + \frct{501947}{162}\,\*\z2
          - \frct{2060548}{243}\,\*\z3
          + \frct{294283}{81}\,\*\z4
          - \frct{51584}{27}\,\*\z3\*\z2
          + \frct{2168}{3}\,\*\z5
\right. \nn \\[1mm] &&  \left. \mbox{\hspn}
          + \frct{10312}{9}\,\*\zts
          - \frct{26275}{27}\,\*\z6
  \right)
  + \colourBlue{\nf\,\*\cft} \* \left(
          - \frct{71275}{108}
          - \frct{4267}{9}\,\*\z2
          - \frct{24218}{9}\,\*\z3
          - \frct{4148}{3}\,\*\z4
          + 1120\,\*\z3\*\z2
\;\;\;\right. \nn \\[1mm] &&  \left. \mbox{\hspn}
          + 2832\,\*\z5
          - \frct{752}{3}\,\*\zts
          + \frct{35954}{27}\,\*\z6
  \right)
  + \colourBlue{\nfs\,\*\cf\*\ca} \* \left(
            \frct{49221719}{52488}
          + \frct{3375704}{2187}\,\*\z2
          - \frct{453428}{729}\,\*\z3
\right. \nn \\[1mm] &&  \left. \mbox{\hspn}
          - \frct{728}{9}\,\*\z4
          - \frct{344}{9}\,\*\z3\*\z2
          + \frct{1136}{9}\,\*\z5
  \right)
  + \colourBlue{\nfs\,\*\cfs} \* \left(
            \frct{159551}{729}
          - \frct{14705}{81}\,\*\z2
          + \frct{237748}{243}\,\*\z3
          - \frct{24260}{81}\,\*\z4
\right. \nn \\[1mm] &&  \left. \mbox{\hspn}
          - \frct{560}{27}\,\*\z3\*\z2
          - \frct{7552}{27}\,\*\z5
  \right)
  + \colourBlue{\nft\,\*\cf} \* \left(
          - \frct{881495}{19683}
          - \frct{17176}{243}\,\*\z2
          - \frct{5104}{729}\,\*\z3
          - \frct{956}{81}\,\*\z4
  \right)
\eea
and
\bea
\label{eq:D4HDY}
  {\lefteqn{ 
  D_{\rm DY/H}^{(4)} \:\: = }}
\nn \\ && \mbox{\hspn} \phantom{+}
  \colourBlue{\cI\,\*\cat} \* \left( 
          - \frct{28325071}{2187}
          + \frct{5759726}{243}\,\*\z2
          + \frct{289160}{9}\,\*\z3
          - \frct{150620}{9}\,\*\z4
          - \frct{119624}{9}\,\*\z3\*\z2
          - \frct{148180}{27}\,\*\z5
\right. \nn \\ && \left. \mbox{\hspn}
          - \frct{13772}{9}\,\*\zts
          + \frct{14113}{3}\,\*\z6
          + 1420\,\*\z4\*\z3
          + 864\,\*\z5\*\z2
          + \frct{9671}{3}\,\*\z7
  \right)
  + \colourBlue{\dfIAnI} \* \Big( 
            192\,\*\z2
\nn \\ && \left. \mbox{\hspn}
          + \frct{832}{9}\,\*\z3
          - 32\,\*\z4
          - \frct{10720}{9}\,\*\z5
          - \frct{1760}{3}\,\*\zts
          - \frct{4400}{3}\,\*\z6
          + 480\,\*\z4\*\z3
          - 768\,\*\z5\*\z2
          + 4232\,\*\z7
  \right)
\nn \\&& \mbox{\hspn}
  + \colourBlue{\nf\,\*\cI\,\*\cas} \* \left( 
            \frct{11551831}{2916}
          - \frct{2400868}{243}\,\*\z2
          - \frct{829304}{81}\,\*\z3
          + \frct{54448}{9}\,\*\z4
          + \frct{23440}{9}\,\*\z3\*\z2
          - \frct{7064}{27}\,\*\z5
\right. \nn \\ && \left. \mbox{\hspn}
          - \frct{4552}{9}\,\*\zts
          - \frct{3670}{3}\,\*\z6
   \right)
   + \colourBlue{\nf\,\*\dfIFnI} \* \left( 
          - 512\,\*\z2
          - \frct{1280}{9}\,\*\z3
          + 64\,\*\z4
          + \frct{3200}{9}\,\*\z5
          + \frct{640}{3}\,\*\zts
\right. \nn \\ && \left. \mbox{\hspn}
          + \frct{1600}{3}\,\*\z6
  \right) 
  + \colourBlue{\nf\,\*\cI\,\*\cf\*\ca} \* \left( 
            \frct{1870013}{486}
          - \frct{23930}{9}\,\*\z2
          - \frct{88396}{27}\,\*\z3
          - 72\,\*\z4
          + \frct{4832}{3}\,\*\z3\*\z2
          - \frct{2608}{3}\,\*\z5
\right. \nn \\ && \mbox{\hspn}
          + 624\,\*\zts
          - 32\,\*\z6
  \Big) 
  + \colourBlue{\nf\,\*\cI\,\*\cfs} \* \left( 
          - \frct{21037}{54}
          + 16\,\*\z2
          - \frct{8848}{9}\,\*\z3
          - 148\,\*\z4
          + \frct{3200}{3}\,\*\z5
          + 160\,\*\zts
\right. \nn \\ && \mbox{\hspn}
          + 400\,\*\z6
  \Big) 
  + \colourBlue{\nfs\*\cI\*\ca} \* \left( 
          - \frct{898033}{2916}
          + \frct{293528}{243}\,\*\z2
          + \frct{87280}{81}\,\*\z3
          - \frct{1744}{3}\,\*\z4
          - \frct{608}{9}\,\*\z3\*\z2
          + \frct{608}{3}\,\*\z5
  \right) 
\nn \\[1mm] && \mbox{\hspn}
  + \colourBlue{\nfs\*\cI\*\cf} \* \left( 
          - \frct{110059}{243}
          + 384\,\*\z2
          + \frct{10768}{27}\,\*\z3
          + \frct{160}{3}\,\*\z4
          - 256\,\*\z3\*\z2
          + 32\,\*\z5
  \right) 
\nn \\[1mm] && \mbox{\hspn}
  + \colourBlue{\nft\*\cI} \* \left( 
            \frct{10432}{2187}
          - \frct{3200}{81}\,\*\z2
          - \frct{3680}{81}\,\*\z3
          + \frct{112}{9}\,\*\z4
  \right)
\eea
with $\cI= \cf$, $d_I^{\,abcd} = d_F^{\,abcd}$ and $\nI = \nF$ 
for the Drell-Yan (quark) case, and $\cI= \ca$, $d_I^{\,abcd} 
= d_A^{\,abcd}$ and $\nI = \nA$ for the Higgs-production (gluon) 
case. 
The lower-order coefficients corresponding to eq.~(\ref{eq:B4dis}) 
and eq.~(\ref{eq:D4HDY}) can be found in eqs.~(4.10) -- (4.12) of 
ref.~\cite{Moch:2005ba} and eqs.~(33) -- (35) of ref.~\cite{Moch:2005ky},
respectively. They are included in an ancillary file 
for the convenience of the reader.

The expansion coefficients of $B_{\rm DIS}$ and $D_{\rm DY/H}$ have 
properties that are not obvious from the explicit four-loop expressions 
in eqs.~(\ref{eq:B4dis}) and (\ref{eq:D4HDY}) and the lower-order results. 
They are related to the Wilson-line \cite{Dixon:2008gr,Falcioni:2019nxk}
anomalous dimension $f_{\rm q/g}$ and the virtual anomalous dimensions 
$B_{\rm q/g}$~by
\bea
\label{Bdis-fB}
  B_{\rm p,DIS}^{(n)}\, &\! = \!& \mbox{} \;
  - f_{\rm q/g}^{(n)} \:-\: B_{\rm q/g}^{(n)}
  \: + \: \mbox{ $\beta$-function terms }
\\[0.5mm]
\label{DHDY-f}
  D_{\rm DY/H}^{(n)}    &\! = \!&  \mbox{}
  - 2 f_{\rm q/g}^{(n)} 
  \; + \: \mbox{ $\beta$-function terms } \; ,
\eea
where $\rm p=q$ refers to standard gauge-boson exchange DIS, and $\rm p=g$ 
to DIS via the exchange of a Higgs boson in the heavy-top limit.
The latter process was required in 
refs.~\cite{Vogt:2004mw,Moch:2021qrk,Moch:2023tdj} for the determination
of the singlet splitting functions $P_{\rm qg}$ and $P_{\rm gg}$, its
coefficients functions were studied and/or employed in 
refs.~\cite{Moch:2005ba,Moch:2005tm,Soar:2009yh,deFlorian:2014vta}.
The $\beta$-function terms in eqs.~(\ref{Bdis-fB}) and (\ref{DHDY-f})
consist of lower-order quantities multiplied by (powers of) coefficients
of the $\beta$-function. Their explicit form for $B_{\rm q,DIS}^{\,(n)}$
to four loops can be found in eqs.~(3.35) -- (3.38) of 
ref.~\cite{Das:2019btv}. 
See also refs.~\cite{Ravindran:2005vv,Goyal:2025jgt}.

The N$^3$LO timelike non-singlet splitting functions, obtained by
combining results of refs.~\cite{Gehrmann:2026qbl,Moch:2017uml}, 
provide new input to the resummation of energy-energy correlations 
\cite{Dixon:2019uzg}, see appendix A.

%
\section{Constraints on the five-loop splitting functions}
\label{sec:5loop}
\setcounter{equation}{0}
%

The lowest non-vanishing values, $N=2$ and $N=3$, have been computed
for the five-loop non-singlet anomalous dimensions 
$\gamma_{\,\rm ns}^{\:(4)\pm,\rm s}$ in ref.~\cite{Herzog:2018kwj} by 
using the implementation of ref.~\cite{Herzog:2017bjx} of the R$^\ast\!$ 
operation \cite{RSTAR1984,RSTAR1985,RSTAR1991} to obtain the $1/\ep$ pole 
terms of their Feynman diagrams via {\sc Forcer} \cite{Ruijl:2017cxj}.
These results were used, together with the large-$\nc$ small-$x$
constraints resulting from eq.~(\ref{eq:gnsto0V}), see eq.~(3.4) and
(3.11) -- (3.15) in ref.~\cite{Davies:2022ofz}, and the corresponding
fifth-order coefficients of eq.~(\ref{eq:CDofAB}) to produce a first
reliable estimate of the five-loop cusp anomalous dimension 
$A_{\rm q}^{(5)}$.

The four-loop results of ref.~\cite{Gehrmann:2026qbl} provide additional 
constraints also for the splitting functions at the next (five-loop) order. 
One may assume that progress at this order will be limited for quite some
time to low-$N$ values for both the non-singlet and singlet anomalous 
dimensions. 
Hence, for now, the most important constraints are those that can assist 
constructing numerical $x$-space approximations from a rather small 
number of $N$-values. 
This applies to the coefficients $C^{(5)}$ and $D^{(5)}$ in 
eqs.~(\ref{eq:NtoInf}) and (\ref{eq:xto1}) for which eq.~(\ref{eq:CDofAB})
yields
\bea
\label{eq:Cp5}
  C_{\rm p}^{\,(5)} &\!\! = \!& 
         2\, A_{\rm p}^{(1)} A_{\rm p}^{(4)}
     \;+\; 2\, A_{\rm p}^{(2)} A_{\rm p}^{(3)}
\:\: ,\\
\label{eq:Dp5}
  D_{\rm p}^{\,(5)} &\!\! = \!&
    \sum_{k=1}^{4} A_{\rm p}^{\,(k)} 
     \left( B_{\rm p}^{(5-k)} - \beta_{\:\!4-k} \right)
\:\: .
\eea
%
With the $\nfz$ parts of $B_{\rm q}^{(4)}$ in eq.~(13) of 
ref.~\cite{Gehrmann:2026qbl} and of $B_{\rm g}^{(4)}$ in 
eq.~(\ref{eq:Bg4}) above, also $D_{\rm p}^{\,(5)}$ is now 
known analytically for both $\rm p = \rm q$ and $\rm p = \rm g$.
The explicit results needed for future applications~are
\bea
\label{eq:Cq5}
  {\lefteqn{
  C_{\rm q}^{\,(5)} \:\: = }}
\nn \\ && \mbox{\hspn} \phantom{+}
   \colourBlue{\cfs \* \cat} \* \left(
            \frct{1462144}{81}
          - \frct{497824}{27}\,\*\z2
          + \frct{214720}{27}\,\*\z3
          + \frct{73312}{3}\,\*\z4
          - 1408\,\*\z3\*\z2
          - \frct{28864}{9}\,\*\z5
          - 128\,\*\zts
\right. \nn \\ && \left. \mbox{}
          - \frct{27424}{3}\,\*\z6
  \right)
 + \colourBlue{\cf\,\*\dfRAnr} \* \left(
          - 1024\,\*\z2
          + \frct{1024}{3}\,\*\z3
          + \frct{28160}{3}\,\*\z5
          - 3072\,\*\zts
          - 7936\,\*\z6
  \right)
\nn \\ && \mbox{\hspn}
 + \colourBlue{\nf\,\*\cfs \* \cas} \* \left(
          - \frct{1380184}{243}
          + \frct{124736}{27}\,\*\z2
          - \frct{83968}{9}\,\*\z3
          - \frct{5888}{3}\,\*\z4
          + 1792\,\*\z3\*\z2
          + \frct{16768}{9}\,\*\z5
  \right)
\nn \\ && \mbox{\hspn} 
 + \colourBlue{\nf\,\*\cft \* \ca} \* \left(
          - \frct{449408}{81}
          + 1760\,\*\z2
          + \frct{15616}{3}\,\*\z3
          - 1408\,\*\z4
          - 1536\,\*\z3\*\z2
          + 1280\,\*\z5
  \right)
\nn \\ && \mbox{\hspn} 
 + \colourBlue{\nf\,\* \cf \* \dfRRnr} \* \left(
            2048\,\*\z2
          - \frct{2048}{3}\,\*\z3
          - \frct{10240}{3}\,\*\z5
  \right)
 + \colourBlue{\nf\,\* \cff} \* \left(
            \frct{4576}{9}
          + \frct{4736}{3}\,\*\z3
          - 2560\,\*\z5
  \right)
\nn \\ && \mbox{\hspn} 
 + \colourBlue{\nfs\,\*\cfs \* \ca} \* \left(
            \frct{80456}{243}
          - \frct{5632}{27}\,\*\z2
          + \frct{8960}{9}\,\*\z3
          - \frct{896}{3}\,\*\z4
  \right)
 + \colourBlue{\nfs\,\* \cft} \* \left(
            \frct{45536}{81}
          - \frct{2560}{3}\,\*\z3
          + 256\,\*\z4
  \right)
\nn \\[1mm] && \mbox{\hspn} 
 + \colourBlue{\nft\,\* \cfs} \* \left(
            \frct{512}{243}
          + \frct{512}{27}\,\*\z3
  \right)
\;\; , \\[2mm]
\label{eq:Dq5}
  {\lefteqn{
  D_{\rm q}^{\,(5)} \:\: = }}
\nn \\ && \mbox{\hspn} \phantom{+}
   \colourBlue{\cf\* \caf} \* \left(
          - \frct{2061022}{243}
          + \frct{1403284}{243}\,\*\z2
          - \frct{255728}{81}\,\*\z3
          - 7612\,\*\z4
          + \frct{3872}{9}\,\*\z3\*\z2
          + \frct{39688}{27}\,\*\z5
\right. \nn \\[1mm] && \left. \mbox{\hspn} 
          + \frct{176}{3}\,\*\zts
          + \frct{27544}{9}\,\*\z6
  \right)
 + \colourBlue{\cfs\* \cat} \* \left(
          - \frct{15193}{162}
          + \frct{2496230}{243}\,\*\z2
          - \frct{988604}{81}\,\*\z3
          - \frct{103828}{27}\,\*\z4
\right. \nn \\ && \left. \mbox{\hspn} 
          + \frct{32992}{9}\,\*\z3\*\z2
          + \frct{15328}{3}\,\*\z5
          + \frct{5488}{3}\,\*\zts
          + \frct{68599}{27}\,\*\z6
          - \frct{3040}{3}\,\*\z4\*\z3
          + \frct{4928}{3}\,\*\z5\*\z2
          - \frct{35840}{3}\,\*\z7
  \right)
\nn \\[1mm] && \mbox{\hspn} 
 + \colourBlue{\cft\* \cas} \* \left(
            \frct{13987}{3}
          - \frct{40318}{3}\,\*\z2
          + \frct{94652}{3}\,\*\z3
          - \frct{67316}{9}\,\*\z4
          - \frct{12160}{3}\,\*\z3\*\z2
          + \frct{53576}{9}\,\*\z5
          - \frct{26296}{3}\,\*\zts
\right. \nn \\ && \mbox{\hspn} 
          - \frct{31610}{3}\,\*\z6
          + 1664\,\*\z4\,\*\z3
          - 9376\,\*\z5\,\*\z2
          + 34440\,\*\z7
  \Big)
 + \colourBlue{\cff\* \ca} \* \left(
          - \frct{14879}{9}
          + 5088\,\*\z2
          - \frct{99136}{9}\,\*\z3
\right. \nn \\ && \left. \mbox{\hspn} 
          + 12596\,\*\z4
          - \frct{37328}{9}\,\*\z3\*\z2
          - \frct{33152}{3}\,\*\z5
          + 12880\,\*\zts
          + \frct{140450}{9}\,\*\z6
          + 1152\,\*\z4\,\*\z3
          + 10176\,\*\z5\,\*\z2
\right. \nn \\ && \left. \mbox{\hspn} 
          - 43680\,\*\z7
  \right)
 + \colourBlue{\cfi} \* \left(
            \frct{4873}{6}
          - 1800\,\*\z2
          + 8016\,\*\z3
          - 1368\,\*\z4
          - 480\,\*\z3\*\z2
          - 10080\,\*\z5
          - 4608\,\*\zts
\right. \nn \\ && \mbox{\hspn} 
          - 8444\,\*\z6
          + 256\,\*\z4\*\z3
          - 1536\,\*\z5\*\z2
          + 23520\,\*\z7
  \Big)
 + \colourBlue{\cf\* \dfRAnr} \* \left(
            384
          - \frct{4928}{3}\,\*\z2
          - \frct{4544}{3}\,\*\z3
\right. \nn \\ && \left. \mbox{\hspn} 
          + \frct{128}{3}\,\*\z4
          - 3584\,\*\z3\,\*\z2
          + 1920\,\*\z5
          - 3968\,\*\zts
          - \frct{33032}{9}\,\*\z6
          - 256\,\*\z4\,\*\z3
          + 1280\,\*\z5\,\*\z2
          + 11200\,\*\z7
  \right)
\nn \\ && \mbox{\hspn}
 + \colourBlue{\cf\* \dfAAna} \* \left(
            \frct{320}{9}
          - \frct{2816}{3}\,\*\z3
  \right)
 + \colourBlue{\ca\* \dfRAnr} \* \left(
            \frct{1408}{3}\,\*\z2
          - \frct{1408}{9}\,\*\z3
          - \frct{38720}{9}\,\*\z5
\right. \nn \\ && \left. \mbox{\hspn} 
          + 1408\,\*\zts
          + \frct{10912}{3}\,\*\z6
  \right)
 + \colourBlue{\nf\,\*\cf\* \cat} \* \left(
            \frct{1133243}{243}
          - \frct{22100}{9}\,\*\z2
          + \frct{330880}{81}\,\*\z3
          + \frct{15400}{9}\,\*\z4
\right. \nn \\[1mm] && \left. \mbox{\hspn} 
          - \frct{5632}{9}\,\*\z3\*\z2
          - \frct{30272}{27}\,\*\z5
          - \frct{32}{3}\,\*\zts
          - \frct{5008}{9}\,\*\z6
  \right)
 + \colourBlue{\nf\,\*\cfs\* \cas} \* \left(
            \frct{746242}{243}
          - \frct{47852}{9}\,\*\z2
          - \frct{30728}{27}\,\*\z3
\right. \nn \\[1mm] && \left. \mbox{\hspn} 
          + \frct{33136}{9}\,\*\z4
          - \frct{7856}{9}\,\*\z3\*\z2
          + \frct{3928}{9}\,\*\z5
          + \frct{3008}{3}\,\*\zts
          - \frct{27496}{27}\,\*\z6
  \right)
 + \colourBlue{\nf\,\*\cft\* \ca} \* \left(
          - \frct{75337}{27}
          + \frct{6496}{9}\,\*\z2
\right. \nn \\[1mm] && \left. \mbox{\hspn} 
          - \frct{14144}{3}\,\*\z3
          + \frct{43336}{9}\,\*\z4
          + \frct{21952}{9}\,\*\z3\*\z2
          - \frct{19648}{9}\,\*\z5
          - \frct{8768}{3}\,\*\zts
          + \frct{2588}{3}\,\*\z6
  \right)
 + \colourBlue{\nf\,\*\cff} \* \left(
           \frct{965}{9}
\right. \nn \\[1mm] && \left. \mbox{\hspn} 
          + 1008\,\*\z2
          - \frct{15968}{9}\,\*\z3
          - 1456\,\*\z4
          - \frct{5248}{9}\,\*\z3\*\z2
          + \frct{11264}{3}\,\*\z5
          + 1664\,\*\zts
          - \frct{25736}{9}\,\*\z6
  \right)
\nn \\ && \mbox{\hspn}
 + \colourBlue{\nf\,\*\cf\* \dfRRnr} \* \left(
          - 768
          + \frct{9856}{3}\,\*\z2
          - \frct{4736}{3}\,\*\z3
          - \frct{1408}{3}\,\*\z4
          + 256\,\*\z3\,\*\z2
          - 5760\,\*\z5
          + 1024\,\*\zts
\right. \nn \\ && \left. \mbox{\hspn} 
          + \frct{7168}{9}\,\*\z6
  \right)
 + \colourBlue{\nf\,\*\cf\* \dfFAna} \* \left(
          - \frct{2048}{9}
          + \frct{6656}{3}\,\*\z3
  \right)
 + \colourBlue{\nf\,\*\ca\* \dfRRnr} \* \left(
          - \frct{2816}{3}\,\*\z2
\right. \nn \\ && \left. \mbox{\hspn} 
          + \frct{2816}{9}\,\*\z3
          + \frct{14080}{9}\,\*\z5
  \right)
 + \colourBlue{\nf\,\* \dfRAnr} \* \left(
          - \frct{256}{3}\,\*\z2
          + \frct{256}{9}\,\*\z3
          + \frct{7040}{9}\,\*\z5
          - 256\,\*\zts
          - \frct{1984}{3}\,\*\z6
  \right)
\nn \\ && \mbox{\hspn}
 + \colourBlue{\nfs\,\* \cf\* \cas} \* \left(
          - \frct{144551}{243}
          + \frct{21524}{81}\,\*\z2
          - 1024\,\*\z3
          + \frct{880}{9}\,\*\z4
          + \frct{896}{9}\,\*\z3\,\*\z2
          + \frct{4192}{27}\,\*\z5
  \right)
\nn \\[1mm] && \mbox{\hspn}
 + \colourBlue{\nfs\,\* \cfs\* \ca } \* \left(
          - \frct{205823}{243}
          + \frct{54136}{81}\,\*\z2
          + \frct{4432}{9}\,\*\z3
          - \frct{5120}{9}\,\*\z4
          + \frct{1216}{9}\,\*\z3\,\*\z2
          + \frct{608}{9}\,\*\z5
  \right)
\nn \\[1mm] && \mbox{\hspn}
 + \colourBlue{\nfs\,\* \cft} \* \left(
            \frct{2638}{27}
          + \frct{2336}{9}\,\*\z2
          + \frct{736}{3}\,\*\z3
          - \frct{3488}{9}\,\*\z4
          - \frct{1408}{9}\,\*\z3\,\*\z2
          - \frct{448}{9}\,\*\z5
  \right)
\nn \\ && \mbox{\hspn}
 + \colourBlue{\nfs\,\* \dfRRnr} \* \left(
            \frct{512}{3}\,\*\z2
          - \frct{512}{9}\,\*\z3
          - \frct{2560}{9}\,\*\z5
  \right)
 + \colourBlue{\nfs\,\* \cf\* \dfFFna} \* \left(
            \frct{2816}{9}
          - \frct{2048}{3}\,\*\z3
  \right)
\nn \\ && \mbox{\hspn}
 + \colourBlue{\nft\,\* \cf\* \ca} \* \left(
            \frct{3086}{243}
          - \frct{1216}{243}\,\*\z2
          + \frct{3776}{81}\,\*\z3
          - \frct{224}{9}\,\*\z4
  \right)
 + \colourBlue{\nft\,\* \cfs} \* \left(
            \frct{5428}{243}
          - \frct{2432}{243}\,\*\z2
          - \frct{1408}{81}\,\*\z3
\right. \nn \\[1mm] && \left. \mbox{\hspn} 
          + \frct{448}{27}\,\*\z4
  \right)
 + \colourBlue{\nff\,\* \cf } \* \left(
          - \frct{64}{243}
          + \frct{128}{81}\,\*\z3
  \right)
\eea

\pagebreak
 
\noindent
and
\bea
\label{eq:Cg5}
  {\lefteqn{
  C_{\rm g}^{\,(5)} \:\: = }}
\nn \\ && \mbox{\hspn} \phantom{+}
   \colourBlue{\cai} \* \left(
            \frct{1462144}{81}
          - \frct{497824}{27}\,\*\z2
          + \frct{214720}{27}\,\*\z3
          + \frct{73312}{3}\,\*\z4
          - 1408\,\*\z3\*\z2
          - \frct{28864}{9}\,\*\z5
          - 128\,\*\zts
\right. \nn \\ && \left. \mbox{} 
          - \frct{27424}{3}\,\*\z6
  \right)
 + \colourBlue{\ca \*\dfAAna} \* \left(
          - 1024\,\*\z2
          + \frct{1024}{3}\,\*\z3
          + \frct{28160}{3}\,\*\z5
          - 3072\,\*\zts
          - 7936\,\*\z6
  \right)
\nn \\ && \mbox{\hspn} 
 + \colourBlue{\nf\,\*\caf} \* \left(
          - \frct{1380184}{243}
          + \frct{124736}{27}\,\*\z2
          - \frct{83968}{9}\,\*\z3
          - \frct{5888}{3}\,\*\z4
          + 1792\,\*\z3\*\z2
          + \frct{16768}{9}\,\*\z5
  \right)
\nn \\ && \mbox{\hspn}
 + \colourBlue{\nf\,\*\ca\*\dfFAna} \* \left(
            2048\,\*\z2
          - \frct{2048}{3}\,\*\z3
          - \frct{10240}{3}\,\*\z5
  \right)
 + \colourBlue{\nf\,\*\cf\*\cat} \* \left(
          - \frct{449408}{81}
          + 1760\,\*\z2
\right. \nn \\ && \left. \mbox{} 
          + \frct{15616}{3}\,\*\z3
          - 1408\,\*\z4
          - 1536\,\*\z3\*\z2
          + 1280\,\*\z5
  \right)
 + \colourBlue{\nf\,\*\cfs\*\cas} \* \left(
            \frct{4576}{9}
          + \frct{4736}{3}\,\*\z3
          - 2560\,\*\z5
  \right)
\nn \\[1mm] && \mbox{\hspn}
 + \colourBlue{\nfs\,\*\cat } \* \left(
            \frct{80456}{243}
          - \frct{5632}{27}\,\*\z2
          + \frct{8960}{9}\,\*\z3
          - \frct{896}{3}\,\*\z4
  \right)
 + \colourBlue{\nfs\,\*\cf \*\cas } \* \left(
            \frct{45536}{81}
          - \frct{2560}{3}\,\*\z3
          + 256\,\*\z4
  \right)
\nn \\[1mm] && \mbox{\hspn}
 + \colourBlue{\nft\,\*\cas } \* \left(
            \frct{512}{243}
          + \frct{512}{27}\,\*\z3
  \right)
\;\; ,
\\[2mm]
\label{eq:Dg5}
  {\lefteqn{
  D_{\rm g}^{\,(5)} \:\: = }}
\nn \\ && \mbox{\hspn} \phantom{+}
   \colourBlue{\cai} \* \left(
          - \frct{327896}{243}
          + \frct{1888}{3}\,\*\z2
          + \frct{130256}{9}\,\*\z3
          + \frct{9866}{9}\,\*\z4
          - 4920\,\*\z3\*\z2
          - 8812\,\*\z5
          + \frct{4136}{3}\,\*\zts
\right. \nn \\ && \left. \mbox{\hspn}
          - \frct{44407}{27}\,\*\z6
          + \frct{6176}{3}\,\*\z4\*\z3
          + \frct{2720}{3}\,\*\z5\*\z2
          + \frct{7000}{3}\,\*\z7
   \right)
 + \colourBlue{\ca \*\dfAAna} \* \left(
            64
          + 320\,\*\z2
          - \frct{10880}{3}\,\*\z3
\right. \nn \\ && \left. \mbox{\hspn}
          - \frct{1904}{3}\,\*\z4
          - 4672\,\*\z3\*\z2
          - \frct{1760}{3}\,\*\z5
          - 2816\,\*\zts
          - \frct{3256}{9}\,\*\z6
          - 256\,\*\z4\*\z3
          + 1280\,\*\z5\*\z2
          + 11200\,\*\z7
   \right)
\nn \\ && \mbox{\hspn}
 + \colourBlue{\nf\,\*\caf} \* \left(
            \frct{13702}{9}
          - \frct{2012}{3}\,\*\z2
          - \frct{11684}{3}\,\*\z3
          - \frct{562}{3}\,\*\z4
          + \frct{7352}{9}\,\*\z3\*\z2
          + \frct{14372}{9}\,\*\z5
          - \frct{4688}{3}\,\*\zts
          + \frct{4714}{27}\,\*\z6
   \right)
\nn \\ && \mbox{\hspn}
 + \colourBlue{\nf\,\*\cf\*\cat} \* \left(
          - \frct{2516}{9}
          + \frct{184}{3}\,\*\z2
          - \frct{3856}{3}\,\*\z3
          + \frct{260}{3}\,\*\z4
          - 32\,\*\z3\*\z2
          - 320\,\*\z5
          + 1312\,\*\zts
          + \frct{1120}{9}\,\*\z6
   \right)
\nn \\ && \mbox{\hspn}
 + \colourBlue{\nf\,\*\ca\*\dfFAna} \* \left(
          - 128
          - 640\,\*\z2
          + \frct{7936}{3}\,\*\z3
          + \frct{2656}{3}\,\*\z4
          + 2432\,\*\z3\*\z2
          - \frct{19520}{3}\,\*\z5
          + 1024\,\*\zts
\right. \nn \\ && \left. \mbox{\hspn}
          + \frct{1184}{9}\,\*\z6
   \right)
 + 36\,\* \colourBlue{\nf\,\*\cfs\*\cas} 
 + \colourBlue{\nfs\,\*\cat} \* \left( 
          - \frct{8578}{81}
          + 28\,\*\z2
          + \frct{940}{9}\,\*\z3
          + \frct{400}{9}\,\*\z4
          - \frct{128}{9}\,\*\z3\*\z2
          - \frct{32}{9}\,\*\z5
    \right)
\nn \\[1mm] && \mbox{\hspn}
 + \colourBlue{\nfs\,\*\cf\*\cas} \* \left( 
          - \frct{196}{9}
          + \frct{128}{3}\,\*\z3
    \right)
 + \colourBlue{\nft\,\*\cas} \* \left( 
          - \frct{448}{243}
    \right)
\:\: .
\eea

Obviously information about the five-loop cusp anomalous dimensions
$A_{\rm p}^{(5)}$ and the five-loop `virtual' anomalous dimensions
$B_{\rm p}^{(5)}$ would be most valuable. Beyond the leading 
large-$\nf$ terms of refs.~\cite{Gracey:1994nn,Bennett:1997ch}, so far
only the QED terms, i.e., the coefficients of $n_{\!f}^{\,k}\,C_F^{\,5-k}$
for $1 \leq k \leq 4$ and $\{\nfs \cf,\, \nft\} \dfRRnA$ are known
for $B_{\rm g}^{(5)}$. These colour factors cannot occur in the maximally 
non-abelian quantity $f_g^{\,(5)}$. Thus their contributions to 
$B_{\rm g}^{(5)}$ are, as at lower orders, identical to the
corresponding terms in the five-loop $\beta$-function of QED
\cite{Baikov:2012zm,Herzog:2017ohr,Luthe:2017ttg}, see also
ref.~\cite{betaQED1}. For a more detailed discussion the reader is 
referred to ref.~\cite{Das:2020adl}.

%
\section{Summary}
\label{sec:summ}
%

We have investigated the recently completed \cite{Gehrmann:2026qbl}
analytical fourth-order result for the splitting functions for non-singlet 
quark distributions in perturbative QCD and its generalization to other 
gauge groups.
These results have been subjected to additional numerical and, 
more importantly, structural checks.
In particular, we have transformed the \MSb\ results to `universal' 
twist-2 anomalous dimensions that must turn out to be reciprocity 
respecting (they are), 
and compared the highest weight terms in the supersymmetric limit to 
results obtained earlier for ${\cal N} = 4$ supersymmetric 
Yang-Mills theory (they agree, as they have to).
Given the manner in which the result of ref.~\cite{Gehrmann:2026qbl}
were obtained, these are highly non-trivial checks strongly suggesting 
that the results are correct.

The new non-fermionic contributions deviate from one prediction generated by 
extrapolating the structure of the lower-order contribution to four loops: 
the function $P_{\rm ns}^{\,(3)+\!}(x)$ includes additional double-logarithmic 
$\ln^{\,4\!} x$ terms proportional to $\pi^2$ for all $\nfz$ colour factors 
which cancel in the limit of a large number of colours $\nc$. 
The case of the quartic group invariants is particularly striking: 
no $\ln x$ terms with $\nf d_{F}^{\,abcd}d_{F}^{\,abcd}$, yet terms up to
$\ln^{\,4\!} x$, all with $\pi^2$, in the $d_{A}^{\,abcd}d_{F}^{\,abcd}$
contribution.
This suggests an entirely new structure at four loops, which shares some
main characteristics ($\pi^2$~coefficients, large-$\nc$ suppression) 
with the `super-leading' logarithms studied in 
refs.~\cite{Forshaw:2006fk,Catani:2011st,Forshaw:2012bi,Becher:2023mtx,
Becher:2024kmk,Becher:2025igg}.

By a known connection via a maximally non-abelian Wilson-line anomalous 
dimension $f_{\rm q/g}$, 
the quark result of 
ref.~\cite{Gehrmann:2026qbl} for the four-loop `virtual' anomalous dimension 
(i.e., the coefficient of $\delta \xm$ in $P_{\rm ns}^{\,(3)\pm\,}$) 
can be transferred to the gluon case.
We~have presented the resulting expression, already shown in 
ref.~\cite{avLL2026}, together with $f_{\rm q/g}^{\,(4)}$, and used this
information to finalize the coefficients $B_{\rm q,DIS}^{(4)}$ and
$D_{\rm DY/H}^{(4)}$ for the next-to-next-to-next-to-next-to-leading 
logarithmic contributions to the soft-gluon exponentiation for important
benchmark processes.
Finally we have briefly addressed some first constraints on the large-$x$
behaviour of the five-loop splitting functions.

%
\subsection*{Note added}
%

While this write-up was being finalized, we were informed by E. Gardi
that he and others have determined the anomalous dimensions 
$f_{\rm q/g}^{\,(4)}$ and $B_{\rm g}^{\,(4)\!}$. 
The latter result, which requires the former, had been presented already 
in ref.~\cite{avLL2026}, as eqs.~(\ref{eq:fI4}) and (\ref{eq:Bg4}) 
represent a simple update of our earlier results included in 
ref.~\cite{Kniehl:2025ttz} with eq.~(\ref{eq:b4FAappr}) instead of the 
exact coefficient (\ref{eq:b4FA}) of ref.~\cite{Gehrmann:2026qbl}. 
The results communicated to us agree with our findings.
 
%
\subsection*{Acknowledgements}
%
 
We compliment the authors of ref.~\cite{Gehrmann:2026qbl} on their
achievement.
This work has been supported by
the Deutsche Forschungsgemeinschaft (DFG) through Research Unit
FOR 2926, project number 409651613, and DFG grant MO~1801/4-2;
the European Research Council (ERC) Advanced Grant 101095857;
the Consolidated Grant ST/X000699/1 of the UK Science and Technology
Facilities Council (STFC),
and the Alexander-von-Humboldt Foundation through a Research Award.

%
%

{\small
\addtolength{\baselineskip}{-0.3mm}

\providecommand{\href}[2]{#2}\begingroup\raggedright\endgroup

}

\newpage

\appendix
%
\section{Some fixed-\emph{N} values of the anomalous dimensions}
\label{sec:appA}
\renewcommand{\theequation}{\ref{sec:appA}.\arabic{equation}}
\setcounter{equation}{0}
%

After ref.~\cite{Moch:2017uml}, we slowly extended the {\sc Forcer}
computations of $\,\gamma_{\,\rm ns}^{\,(3)\pm}$ to $N \leq 42$, 32 and 24,
respectively, for the $\nf\,\cft$, $\nf\,\cfs\,\ca$ and $\nf\, \dfRRnc$ 
contributions -- the hard $\nf\,\cf\,\cas$ part was not needed, as its 
coefficients are fixed by these and the known large-$\nc$ all-$N$ result.
These results led to the determination of the all-$N$ expressions for
$\,\gamma_{\,\rm ns}^{\,(3)\pm}(N)|_\nf$ in ref.~\cite{Kniehl:2025ttz} 
where, just as in refs.~\cite{Davies:2016jie,Moch:2017uml}, 
programs based on the LLL algorithm \cite{Lenstra1982,axbAlg,Calc,fplll} 
were employed to solve the Diophantine equations for its coefficients.

For the computationally hardest $\nfz$ terms, we reached $N \leq 22$ for 
$\gamma_{\,\rm ns}^{\,(3)+}$ at even $N$ and $\gamma_{\,\rm ns}^{\,(3)-}$  
at odd $N$.
These results are still far from sufficient for an all-$N$ determination. 
They were used, though, together with the exact results for the four-loop 
cusp anomalous dimension \cite{Henn:2019swt,vonManteuffel:2020vjv}, 
to construct improved approximations of $P_{\,\rm ns}^{\,(3)\pm}(x)$
including the numerical coefficient (\ref{eq:b4FAappr}) entering 
$B_{\rm q}^{(4)}$ and the other large-$x$ coefficient addressed in 
section~4 above, see ref.~\cite{Kniehl:2025ttz}.

Since the all $\nf$ parts of $\gamma_{\,\rm ns}^{\,(3)\pm}(N)$ have been 
obtained at least twice by now \cite{Gracey:1994nn,Davies:2016jie,
Basdew-Sharma:2022vya,Gehrmann:2023iah,Kniehl:2025ttz,Gehrmann:2026qbl},
the corresponding fixed-$N$ results are not relevant any more. 
The $\nfz$ values above $N=16$, on the other hand, provide a (successful) 
check of the new results of ref.~\cite{Gehrmann:2026qbl}. 
For brevity, we show only the results for $N=21$ and $N=22$ here, the
values for lower $N$ can be found in an ancillary file:
 
\vspace*{-5mm}
{\small
\bea
 {\lefteqn{ \gamma_{\,\rm ns}^{\:(3)-}(N\!=\!21)\Big|_{\nfz} \;=\; \quad \quad
}}
\nn
\\[0.5mm]  && \nn \mbox{\hspn} \phantom{+}
 \colourBlue{ \cff }\,\*\biggl(
 \frct{2505258783993812390986962438015709611064516350159320207603}{9648336668885750946751819767831485412524410159104000000} 
\nn \\[0.5mm] && 
 + \frct{5213094630823898560101782369209}{17381709372562962123472776000}\,\*\z3 
 - \frct{4376729092604387}{5645039165766}\,\*\z5
 \biggr)
\nn \\[0.5mm] && \mbox{\hspn}
 + \colourBlue{ \cft\,\*\ca }\,\*\biggl(
 - \frct{18772951650381978208536956546455408593112651820538277}{67581473662396865827660785956260492081618944000000} 
\nn \\ && 
 - \frct{40665760093759227471778309926523}{764795212392770333432802144000}\,\*\z3 
 - \frct{713396958068741}{3421235858040}\,\*\z4 
 + \frct{6613267311002551}{5645039165766}\,\*\z5
 \biggr)
\nn \\[0.5mm] && \mbox{\hspn}
 + \colourBlue{ \cfs\,\*\cas }\,\*\biggl(
 \frct{35776759924241225622497449467269936827804330007}{132659008538627144306535710379115185045504000} 
\nn \\[0.5mm] && \mbox{}
 + \frct{365179722900213960791996810543671}{191198803098192583358200536000}\,\*\z3
 + \frct{713396958068741}{2280823905360}\,\*\z4 
 - \frct{86640697944607297}{22580156663064}\,\*\z5
 \!\biggr)
\nn \\[0.5mm] && \mbox{\hspn}
 + \colourBlue{ \cf\,\*\cat }\,\*\biggl(
 \frct{2021038985620509544696878885462979075473277187}{1916487154581052107169983768242614272000000}
\nn \\[0.5mm] &&  \mbox{}
 - \frct{366695307313896383540104897151}{364274928503343812066112000}\,\*\z3
 - \frct{713396958068741}{6842471716080}\,\*\z4 
 + \frct{48537547888064141}{22580156663064}\,\*\z5
 \biggr)
\nn \\[0.5mm] && \mbox{\hspn}
 + \colourBlue{ \dfRAnr }\,\*\biggl(
 \frct{26550211375226203767990908816719908871}{8096520270886720904506231756800000}
\nn \\[0.5mm] && \mbox{} 
 + \frct{10054178201586838481463752147}{1011874801398177255739200}\,\*\z3
 - \frct{4859499530491750}{313613286987}\,\*\z5
 \biggr)
\eea
}
and
%
{\small
\bea
 {\lefteqn{ \gamma_{\,\rm ns}^{\:(3)+}(N\!=\!22)\Big|_{\nfz} \;=\; \quad \quad
}}
\nn
\\[0.5mm]  && \nn \mbox{\hspn} \phantom{+}
 \colourBlue{ \cff }\*\biggl( 
 \frct{7662059824912275864925540133686907309060980486708043736229228796933}{32850902211445417959204937939070273540372404218382617919488000000} 
\nn \\[0.5mm] &&  \mbox{}
 + \frct{1107142926156179968736546192950929793}{4864114932527391883594745108616000}\*\z3 
 - \frct{284196114428789909}{426603674098602}\*\z5
 \biggr)
\nn \\[0.5mm] && \mbox{\hspn}
 + \colourBlue{ \cft\*\ca }\*\biggl( 
 - \frct{997180256013840021492229204001182803738845304226327000538147}{4360928719749515393881929463360346923973298868789248000000} 
\nn \\[0.5mm] && 
 + \frct{20053102452839311918681857864156535733}{214021057031205242878168784779104000}\*\z3 
 - \frct{377403375512541269}{1809833768903160}\*\z4 
\nn \\[0.5mm] &&  \mbox{}
 + \frct{2846535475465434919}{2986225718690214}\*\z5
 \biggr)
\nn \\[0.5mm] && \mbox{\hspn}
 + \colourBlue{ \cfs\*\cas }\*\biggl( 
 \frct{85660847909123669084908027840815730017725468952028000999}{320189580372949893700260365155742012798000381952000000} 
\nn \\[0.5mm] &&  \mbox{}
 + \frct{101240241200046949006967654041206922123}{53505264257801310719542196194776000}\*\z3
 + \frct{377403375512541269}{1206555845935440}\*\z4 
\nn \\[0.5mm] &&  \mbox{}
 - \frct{45588585147633555793}{11944902874760856}\*\z5
 \biggr)
\nn \\[0.5mm] && \mbox{\hspn}
 + \colourBlue{ \cf\*\cat }\*\biggl(
 \frct{39030722141141809021500491049354573210109886369077}{36912140044643625288596932270078073475072000000}
\nn \\[0.5mm] && \mbox{}
 - \frct{169147821771065060277955472183051}{164153076114821635607717952000}\*\z3
 - \frct{377403375512541269}{3619667537806320}\*\z4
\nn \\ && \mbox{}
 + \frct{1142307872682791083}{519343603250472}\*\z5
 \biggr)
\nn \\ && \mbox{\hspn}
 + \colourBlue{ \dfRAnr }\*\biggl(
 \frct{2895186169590963427098633028347724589083}{856611844659815071696759319869440000}
\nn \\ && \mbox{}
 + \frct{16434770960792289512175674809817}{1605845309818907304858110400}\*\z3
 - \frct{5005309822959590}{313613286987}\*\z5
 \biggr)
\, , 
\eea
}
The odd-$N$ values of $\gamma_{\,\rm ns}^{\:(3){\rm s}}$ were extended in 
the past years to $N \leq 29$ with
{\small
\bea
 {\lefteqn{ \gamma_{\,\rm ns}^{\:(3)\rm s}(N\!=\!29) \;=\; \quad \quad
}}
\nn
\\[0.5mm]  && \nn \mbox{\hspn} \phantom{+}
 \colourBlue{ \nf\,\*\ca\*\dabctnr }\*\biggl(
 - \frct{14393229889420108654575655735570293788537885459112125303}{1499797657540005498879928020553842464326648200000000000}
\nn \\[0.5mm] && 
 - \frct{255522657563225731056084043}{6674872384126277227771875}\*\z3
 + \frct{222208}{4205}\*\z5
 \biggr) \mbox{}
\nn \\[0.5mm] && \mbox{\hspn}
 + \colourBlue{ \nf\,\*\cf\*\dabctnr }\*\biggl(
 - \frct{355828132917944211970258488791897378180836360251643765086113}{16829057663441475243968258993548977831093615357093750000000}
\nn \\[0.5mm] && \mbox{}
 + \frct{1441702324191972908540284}{76371537575815528921875}\*\z3
 \biggr)
\nn \\[0.5mm] && \mbox{\hspn}
 + \colourBlue{ \nfs\,\*\dabctnr }\*\biggl(
 \frct{206017438842923017721694466180412309013478383539}{749149679090911837602361648628292939224100000000}
 \biggr)
\: . 
\eea
}

The above expressions show a general pattern concerning the $\zeta$-function
in these anomalous dimensions that is seen, to an even larger extent, in
the expansion coefficients of the $\beta$-function, see refs.~%
\cite{Gorishnii:1990kd,vanRitbergen:1997va,Czakon:2004bu,Baikov:2012zm,
Herzog:2017ohr,Luthe:2017ttg}: the values $\zeta_{\,m}^{}$ occur to a lower
maximal $m$ than one may expect, in the present case from the general form of 
the $x$-space expressions in terms of HPLs. 

Indeed, if one employs the exact $x$-space results to compute the moments 
(\ref{eq:PvsGam}) at odd $N$ for $P_{\,\rm ns}^{\,(n)+}$ or at even $N$ for
$P_{\,\rm ns}^{\,(n)-,\rm s}$, then terms up to $\zeta_{2n+1}^{}$ do occur.
This was demonstrated at three loops for $N=1$ in 
ref.~\cite{Broadhurst:2004jx}. The same effect occurs in the third-order
coefficient functions as discussed in ref.~\cite{Moch:2007rq}.
These additional Riemann-$\zeta$ values do vanish, of course, in the 
large-$\nc$ limit where $P_{\,\rm ns}^{\,(n)+}$ and 
$P_{\,\rm ns}^{\,(n)-}$ coincide.
 
Here we illustrate this situation for $N=2$ at four loops:
%
\bea
 {\lefteqn{ 
     \, P_{\,\rm ns}^{\,(3)-}(N\!=\!2) 
  \,-\, P_{\,\rm ns}^{\,(3)+}(N\!=\!2) \;=\; \quad \quad
}}
\nn
\\[1.5mm]  && \nn \mbox{\hspn} \phantom{+}
   \colourBlue{ \cft\,\*(\ca-2\,\*\cf) }\*\biggl(
 - \frct{27536848}{2187}
 + \frct{3080828}{243}\,\*\z2
 - \frct{1572998}{81}\,\*\z3
 + \frct{68689}{3}\,\*\z4
 - \frct{19688}{9}\,\*\z2\*\z3
\nn \\ && \mbox{}
 + \frct{11012}{3}\,\*\z5
 - \frct{8728}{3}\,\*\zts
 - \frct{100496}{9}\,\*\z6
 - 896\,\*\z2\*\z5
 + 4256\,\*\z3\*\z4
 + 2448\,\*\z7
\biggr)
\nn \\ && \mbox{\hspn} 
 + \colourBlue{ \cfs\,\*(\ca-2\,\*\cf)^2 }\*\biggl(
 - \frct{57350323}{8748}
 + \frct{3490163}{243}\,\*\z2
 - \frct{2305534}{81}\,\*\z3
 + \frct{723791}{27}\,\*\z4
 + \frct{8804}{9}\,\*\z2\*\z3
\nn \\ && \mbox{}
 - \frct{23788}{9}\,\*\z5
 - \frct{17572}{3}\,\*\zts
 - \frct{100897}{9}\,\*\z6
 - 2400\,\*\z2\*\z5
 + 7112\,\*\z3\*\z4
 + 3756\,\*\z7
\biggr)
\nn \\ && \mbox{\hspn}
 + \colourBlue{ \cf\,\*(\ca-2\,\*\cf)^3 }\*\biggl(
 - \frct{14818117}{8748}
 + \frct{1487071}{243}\,\*\z2
 - \frct{280204}{27}\,\*\z3
 + \frct{59852}{9}\,\*\z4
 - \frct{1828}{9}\,\*\z2\*\z3
\nn \\
&& \mbox{}
 + 62\,\*\z5
 - \frct{7192}{3}\,\*\zts
 - \frct{6311}{9}\,\*\z6
 - 1376\,\*\z2\*\z5
 + 2776\,\*\z3\*\z4
 + 204\,\*\z7
\biggr)
\nn \\ && \mbox{\hspn}
 + \colourBlue{ \left(\dfRAnr - \frct{1}{24}\,\*\cf\*\cat\right) }\*\biggl(
   \frct{90542}{27}
 - \frct{277664}{81}\,\*\z2
 + \frct{29324}{9}\,\*\z3
 + \frct{105578}{27}\,\*\z4
\nn \\ && \mbox{}
 - 6576\,\*\z2\*\z3
 - \frct{18376}{3}\,\*\z5
 + 5312\,\*\zts
 + \frct{126328}{9}\,\*\z6
 + 1536\,\*\z2\*\z5
 - 6528\,\*\z3\*\z4
 - 2112\,\*\z7
\biggr)
\nn \\ && \mbox{\hspn}
 + \colourBlue{ \nf\,\*\cfs\,\*(\ca-2\,\*\cf) }\*\biggl(
  \frct{814505}{2187}
 - \frct{262324}{243}\,\*\z2
 + \frct{108596}{27}\,\*\z3
 - \frct{105056}{27}\,\*\z4
 - \frct{80}{3}\,\*\z2\*\z3
\nn \\ && \mbox{}
 + \frct{6640}{9}\,\*\z5
 + 48\,\*\zts
 - \frct{290}{9}\,\*\z6
\biggr)
\nn \\ && \mbox{\hspn}
 + \colourBlue{ \nf\,\*\cf\,\*(\ca-2\,\*\cf)^2 }\*\biggl(
 - \frct{71977}{2187}
 - \frct{16222}{27}\,\*\z2
 + \frct{206368}{81}\,\*\z3
 - \frct{72364}{27}\,\*\z4
 - \frct{800}{3}\,\*\z2\*\z3
\nn \\ && \mbox{}
 + \frct{15208}{9}\,\*\z5
 + \frct{176}{3}\,\*\zts
 - \frct{4178}{9}\,\*\z6
\biggr)
\nn \\ && \mbox{\hspn}
 + \colourBlue{ \nf\*\left(\dfRRnr - \frct{1}{48}\,\*\cf\*\cas\right) }\*\biggl(
 - \frct{12160}{9}
 + \frct{3328}{9}\,\*\z2
 + \frct{1216}{9}\,\*\z3
 + \frct{1072}{3}\,\*\z4
\nn \\ && \mbox{}
 + \frct{5504}{9}\,\*\z2\*\z3
 - 1088\,\*\z5
 + 128\,\*\zts
 - \frct{640}{9}\,\*\z6
\biggr)
\nn \\ && \mbox{\hspn}
 + \colourBlue{ \nfs\,\*\cf\,\*(\ca-2\,\*\cf) }\*\biggl(
   \frct{13817}{2187}
 + \frct{2344}{243}\,\*\z2
 - \frct{8384}{81}\,\*\z3
 + \frct{1156}{9}\,\*\z4
 - \frct{320}{9}\,\*\z5
\biggr)
\, , 
\eea

As mentioned above, the (space-like) results of ref.~\cite{Gehrmann:2026qbl} 
also fully determine the four-loop final-state (time-like) non-singlet
splitting functions, as their difference was already known 
\cite{Mitov:2006ic,Moch:2017uml}. 

The $N=3$ moments of the $n$-loop time-like splitting functions are an input
to the N$^n$LL resummation in the small-$z$ region of the energy-energy 
correlation (EEC), see appendix A of \cite{Dixon:2019uzg}, 
where $z = 1/2\, (1-\cos\,\chi_{ik})$ and $\chi_{ik}^{}$ denotes the angular 
separation of two final-state partons $i$ and $k$ in the center-of-mass frame.
The new $N=3$ four-loop non-singlet contribution is 

\vspace*{-5mm}
{\small{
\bea 
&& \hspn \gamma_{\,\rm ns}^{\:{\rm T}\,(3)+\!}(N\!=\!3) \;=\;
      \colourBlue{ \cff } \* \biggl(
       - \frct{20031839497}{2239488}
       + \frct{1259837}{54}\,\*\z2
       + \frct{241793}{81}\,\*\z3
       - \frct{761653}{27}\,\*\z4
\nn \\
&&\qquad
       - 15720\,\*\z2\*\z3
       - \frct{8960}{9}\,\*\z5
       - 3240\,\*\zts
       + \frct{492806}{9}\,\*\z6 
       - 6528\,\*\z2\*\z5
       - 1440\,\*\z3\*\z4
       - 8496\,\*\z7
       \biggr)
\nn \\
&&\quad \mbox{}
    + \colourBlue{ \cft \*\ca} \* \biggl(
         \frct{6815930083}{559872}
       - \frct{82466237}{1944}\,\*\z2
       + \frct{7598609}{324}\,\*\z3
       + \frct{424531}{36}\,\*\z4
       + \frct{261478}{9}\,\*\z2\*\z3
\nn \\
&&\qquad
       + \frct{14018}{9}\,\*\z5
       + \frct{27700}{3}\,\*\zts
       - \frct{678803}{9}\,\*\z6
       + 11264\*\z2\*\z5
       - 5728\*\z3\*\z4
       + 10128\*\z7
       \biggr)
\nn \\
&&\quad \mbox{}
    + \colourBlue{ \cfs \*\cas }\* \biggl(
       - \frct{6808564657}{1119744}
       + \frct{48345383}{1728}\,\*\z2
       - \frct{56825635}{1296}\,\*\z3
       + \frct{5308031}{216}\,\*\z4
       - \frct{120311}{9}\,\*\z2\*\z3
\nn \\
&&\qquad
       + \frct{58828}{9}\,\*\z5
       - \frct{26884}{3}\,\*\zts
        + \frct{174826}{9}\,\*\z6
       - 6752\,\*\z2\*\z5
       + 8776\,\*\z3\*\z4
       - 2532\,\*\z7
       \biggr)
\nn \\
&&\quad \mbox{}
    + \colourBlue{ \cf \*\cat }\* \biggl(
         \frct{3844628563}{1119744}
       - \frct{42787157}{3888}\,\*\z2
       + \frct{49360325}{2592}\,\*\z3
       - \frct{16523545}{2592}\,\*\z4
       + \frct{10757}{18}\,\*\z2\*\z3
\nn \\
&&\qquad
       - \frct{603271}{108}\,\*\z5
       + \frct{8408}{3}\,\*\zts
       + \frct{52247}{27}\,\*\z6
       + 1440\,\*\z2\*\z5
       - 3048\,\*\z3\*\z4
       - 292\,\*\z7
       \biggr)
\nn \\
&&\quad \mbox{}
    + \colourBlue{ \dfRAnr } \* \biggl(
       - \frct{135617623}{15552}
       + \frct{1760179}{162}\,\*\z2
       - \frct{391975}{36}\,\*\z3
       - \frct{790763}{108}\,\*\z4
       + 7980\,\*\z2\*\z3
\nn \\
&&\qquad
       + \frct{124778}{9}\,\*\z5
       - 5408\,\*\zts
       - \frct{160888}{9}\,\*\z6
       - 1536\,\*\z2\*\z5
       + 6528\,\*\z3\*\z4
       + 2112\,\*\z7
       \biggr)
\nn \\
&&\quad \mbox{}
    + \colourBlue{ \nf\,\*\cft } \* \biggl(
       - \frct{92045971}{279936}
       + \frct{103487}{486}\,\*\z2
       - \frct{92576}{81}\,\*\z3
       + \frct{187642}{27}\,\*\z4
       - \frct{3536}{3}\,\*\z2\*\z3
       - \frct{19040}{3}\,\*\z5
\nn \\
&&\qquad
       + 32\,\*\zts
       + \frct{21844}{9}\,\*\z6
       \biggr)
    + \colourBlue{ \nf\,\*\cfs\*\ca } \* \biggl(
       - \frct{15949219}{34992}
       - \frct{523541}{1944}\,\*\z2
       + \frct{165743}{27}\,\*\z3
       - \frct{297398}{27}\,\*\z4
\nn \\
&&\qquad
       + \frct{1192}{3}\,\*\z2\*\z3
       + 7120\,\*\z5
       + \frct{304}{3}\,\*\zts
       - 2142\*\z6
       \biggr)
    + \colourBlue{ \nf\*\dfRRnr } \* \biggl(
        \frct{10247}{9}
       - \frct{1312}{9}\,\*\z2
\nn \\
&&\qquad
       - \frct{208}{9}\,\*\z3
       - \frct{1024}{3}\,\*\z4
       - \frct{6656}{9}\,\*\z2\*\z3
       + \frct{8384}{9}\,\*\z5
       - 128\,\*\zts
       + \frct{640}{9}\,\*\z6
       \biggr)
    + \colourBlue{ \nf\,\*\cf\*\cas } \* \biggl(
       - \frct{66844303}{279936}
\nn \\
&&\qquad
       + \frct{287569}{324}\,\*\z2
       - \frct{667703}{162}\,\*\z3
       + \frct{187361}{54}\,\*\z4
       + \frct{10208}{27}\,\*\z2\*\z3
       - \frct{45554}{27}\,\*\z5
       - 56\,\*\zts
       + \frct{12494}{27}\,\*\z6
       \biggr)
\nn \\
&&\quad \mbox{}
    + \colourBlue{ \nfs\,\*\cfs } \* \biggl(
         \frct{413665}{69984}
       + \frct{4297}{243}\,\*\z2
       - \frct{20362}{81}\,\*\z3
       + \frct{2584}{9}\,\*\z4
       - \frct{640}{9}\,\*\z5
       \biggr)
    + \colourBlue{ \nft\,\*\cf } \* \biggl(
       - \frct{23587}{34992}
       + \frct{200}{81}\,\*\z3
       \biggr)
\qquad \nn \\
&&\quad \mbox{}
    + \colourBlue{ \nfs\,\*\cf\*\ca } \* \biggl(
         \frct{236063}{69984}
       - \frct{4414}{243}\,\*\z2
       + \frct{16586}{81}\,\*\z3
       - \frct{1642}{9}\,\*\z4
       + \frct{320}{9}\,\*\z5
       \biggr)
\:\: ,
\eea
}}
where the normalization and notation is the same as in 
eqs.~(A.5) -- (A.8) of ref.~\cite{Dixon:2019uzg}.

%
\section{Reciprocity-respecting sums}
\label{sec:appB}
\renewcommand{\theequation}{\ref{sec:appB}.\arabic{equation}}
\setcounter{equation}{0}
%

As far as they have been determined so far, the anomalous dimensions
for the evolution of parton distributions can be expressed in terms
of harmonic sums \cite{Vermaseren:1998uu,Blumlein:1998if}. These are
recursively defined~by
\bea
\label{eq:Hsums}
  S_{\pm m}(N) &\!=\!& \sum_{n=1}^{N}\; (\pm 1)^n \, n^{\, -m}
\:\: , \nn \\
  S_{\pm m_1^{},\,m_2^{},\,\ldots,\,m_d^{}}(N) &\!=\!& \sum_{n=1}^{N}\:
  (\pm 1)^{n} \; n^{\, -m_1^{}}\; S_{m_2^{},\,\ldots,\,m_d^{}}(n)
\:\: .
\eea
The weight $w$ of the sums is defined by the sum of the absolute
values of their indices~$m_i$. 

The inverse Mellin transforms of the sums (\ref{eq:Hsums}) and their
products with powers of $1/(N+a)$ can be expressed in terms of
harmonic polylogarithms (HPLs) \cite{Remiddi:1999ew,Moch:1999eb}. 
This means that some properties of the sums and the anomalous dimensions,
including differentiations with respect to $N$ 
-- see  eq.~(\ref{eq:gu-exp}) above -- 
can be determined via {\sc Form}-coded algorithms for the HPLs 
\cite{Remiddi:1999ew,Moch:1999eb,Vermaseren:2000nd}.

Sums up to $w = 2\:\!n-1$ contribute to the $n$-loop anomalous dimensions.
Sums with an index $-1$ do not occur, leaving 
$1,\, 3,\,7,\,17,\,41,\,99,\,239,\,\dots $ sums at 
$w = 1,\,2,\,3,\,4,\,5,\,6,\,7,\dots$. 
These are the Pell-Lucas numbers \cite{oeis-A001333} which can be evaluated 
by $\,((1 - \sqrt{2}\,)^w + (1 + \sqrt{2}\,)^w)/2$.

Reciprocity-respecting (RR) expressions in $N$-space correspond to 
$x$-space expressions with 
\beq
\label{eq:RR-x}
      f_{\mbox{\scriptsize \sc rr}}^{}(x) \;=\; 
  - x f_{\mbox{\scriptsize \sc rr}}^{}(1/x)
\eeq 
implying an invariance under $N \ra -\,N-1$ in the Mellin transform of 
$f(x)$.  Hence the RR sums can be constructed recursively via products 
of lower-weight RR sums supplemented by linear combinations with 
parameters fixed by imposing eq.~(\ref{eq:RR-x}). For an example at $w=4$
see eqs.~(2.16) and (2.17) of ref.~\cite{Davies:2016jie}. We extended
this construction to higher weights for the non-alternating cases 
(no negative index) for ref.~\cite{Moch:2017uml}.

A simpler alternative to this procedure is to make use of the binomial sums 
\beq
\label{eq:Bsums}
  {\mathbb S}_{m_1,\dots,m_d^{}} \:\:=\:\:
  (-1)^N \sum_{k=1}^{N}(-1)^k \binomial{N}{k}\binomial{N+k}{k}\,
  S_{m_1^{},...,m_d^{}}(k)
\eeq
already considered at $w \leq 4$ in ref.~\cite{Vermaseren:1998uu}, 
see eqs.~(106)--(109), (167), (168)--(171). 
Here all indices are positive. It turns out eq.~(\ref{eq:Bsums}) provides 
a basis of all reciprocity respecting linear combinations of the harmonic 
sums (\ref{eq:Hsums}). 
Combinations of non-alternating sums in (\ref{eq:Hsums}) lead to 
binomial sums (\ref{eq:Bsums}) with indices 1 and 2, where the last index
is 1. 
The number of binomial sums at weight $w$ is $2^{w-1\!}$, of which 
1, 1, 2, 3, 5, 8, 13, \dots, i.e., Fibonacci$(w)$, are non-alternating at
$w = 1,\,2,\,3,\,4,\,5,\,6,\,7,\dots\;$.

A noteworthy feature, already alluded to above, is that the large-$N$
expansion of eq.~(\ref{eq:Bsums}) does not include any
terms linear in $1/N$, if the large-$N$ logarithms are written in terms of 
$\mbox{Ln} N \,=\, \ln N + \Ge + 1/(2\,N)$, i.e., the terms including 
$N^{-1}$ of the asymptotic expansion of the analytic continuation of 
$S_1(N)$ in terms of the logarithmic derivative of the Gamma function,
$S_1(N) \Rightarrow \psi (N\!+\!1) + \Ge\,$. 

For the convenience of the reader we provide (without any claim of 
originality) a set of equations 
{for converting a function given in terms of
standard harmonic sums up to weight 7 to binomial sums.
These should be applied in order of descending weight, as 
lower-weight sums appear on the right-hand sides. E.g., 
the first equation includes $\S(1,-6)$, which is further processed in 
eq.~(\ref{eq:B9}).}
If~any non-binomial sums remain afterwards, then that function is not 
reciprocity respecting.

\noindent
Weight 7, alternating sums, depth 6:
\bea
  \S(1,1,1,1,1,-2) &\!=\!& 1/64\,\* ( -  \BS(1,1,1,1,1,2)
   + 2\,\* \S(-7) - 4\,\* \S(1,-6) - 4\,\* \S(2,-5) - 4\,\* \S(3,-4)
   - 4\,\* \S(4,-3) - 4\,\* \S(5,-2) 
\nn \\ && \mbox{\hspm} 
   + 8\,\* \S(1,1,-5) + 8\,\* \S(1,2,-4)
   + 8\,\* \S(1,3,-3) + 8\,\* \S(1,4,-2) + 8\,\* \S(2,1,-4)
   + 8\,\* \S(2,2,-3) + 8\,\* \S(2,3,-2) 
\nn \\ && \mbox{\hspm} 
   + 8\,\* \S(3,1,-3)
   + 8\,\* \S(3,2,-2) + 8\,\* \S(4,1,-2) - 16\,\* \S(1,1,1,-4)
   - 16\,\* \S(1,1,2,-3) - 16\,\* \S(1,1,3,-2) - 16\,\* \S(1,2,1,-3)
\nn \\ && \mbox{\hspm} 
   - 16\,\* \S(1,2,2,-2) - 16\,\* \S(1,3,1,-2) - 16\,\* \S(2,1,1,-3)
   - 16\,\* \S(2,1,2,-2) - 16\,\* \S(2,2,1,-2) - 16\,\* \S(3,1,1,-2)
\nn \\ && \mbox{\hspm} 
   + 32\,\* \S(1,1,1,1,-3) + 32\,\* \S(1,1,1,2,-2) + 32\,\* \S(1,1,2,1,-2)
   + 32\,\* \S(1,2,1,1,-2) + 32\,\* \S(2,1,1,1,-2) )
\nn \\[1mm] 
  \S(1,1,1,1,-2,1) &\!=\!& 1/64\,\* ( -  \BS(1,1,1,1,3)
   + 2\,\* \S(-7) - 4\,\* \S(-6,1) - 4\,\* \S(1,-6) - 4\,\* \S(2,-5)
   - 4\,\* \S(3,-4) - 4\,\* \S(4,-3) 
\nn \\ && \mbox{\hspm} 
   + 8\,\* \S(1,-5,1) + 8\,\* \S(1,1,-5)
   + 8\,\* \S(1,2,-4) + 8\,\* \S(1,3,-3) + 8\,\* \S(2,-4,1)
   + 8\,\* \S(2,1,-4) + 8\,\* \S(2,2,-3) + 8\,\* \S(3,-3,1)
\nn \\ && \mbox{\hspm} 
   + 8\,\* \S(3,1,-3) 
   + 8\,\* \S(4,-2,1) - 16\,\* \S(1,1,-4,1)
   - 16\,\* \S(1,1,1,-4) - 16\,\* \S(1,1,2,-3) - 16\,\* \S(1,2,-3,1)
\nn \\ && \mbox{\hspm} 
   - 16\,\* \S(1,2,1,-3) - 16\,\* \S(1,3,-2,1) - 16\,\* \S(2,1,-3,1)
   - 16\,\* \S(2,1,1,-3) - 16\,\* \S(2,2,-2,1) - 16\,\* \S(3,1,-2,1)
\nn \\ && \mbox{\hspm} 
   + 32\,\* \S(1,1,1,-3,1) + 32\,\* \S(1,1,1,1,-3) + 32\,\* \S(1,1,2,-2,1)
   + 32\,\* \S(1,2,1,-2,1) + 32\,\* \S(2,1,1,-2,1) )
\nn \\[1mm]
  \S(1,1,1,-2,1,1) &\!=\!& 1/64\,\* ( -  \BS(1,1,1,4)
   + 2\,\* \S(-7) - 4\,\* \S(-6,1) - 4\,\* \S(-5,2) - 4\,\* \S(1,-6)
   - 4\,\* \S(2,-5) - 4\,\* \S(3,-4) 
\nn \\ && \mbox{\hspm} 
   + 8\,\* \S(-5,1,1) + 8\,\* \S(1,-5,1)
   + 8\,\* \S(1,-4,2) + 8\,\* \S(1,1,-5) + 8\,\* \S(1,2,-4)
   + 8\,\* \S(2,-4,1) + 8\,\* \S(2,-3,2) 
\nn \\ && \mbox{\hspm} 
   + 8\,\* \S(2,1,-4)
   + 8\,\* \S(3,-3,1) + 8\,\* \S(3,-2,2) - 16\,\* \S(1,-4,1,1)
   - 16\,\* \S(1,1,-4,1) - 16\,\* \S(1,1,-3,2) - 16\,\* \S(1,1,1,-4)
\nn \\ && \mbox{\hspm} 
   - 16\,\* \S(1,2,-3,1) - 16\,\* \S(1,2,-2,2) - 16\,\* \S(2,-3,1,1)
   - 16\,\* \S(2,1,-3,1) - 16\,\* \S(2,1,-2,2) - 16\,\* \S(3,-2,1,1)
\nn \\ && \mbox{\hspm} 
   + 32\,\* \S(1,1,-3,1,1) + 32\,\* \S(1,1,1,-3,1) + 32\,\* \S(1,1,1,-2,2)
   + 32\,\* \S(1,2,-2,1,1) + 32\,\* \S(2,1,-2,1,1) )
\nn \\[1mm]
 \S(1,1,-2,1,1,1) &\!=\!& 1/64\,\* ( -  \BS(1,1,5)
   + 2\,\* \S(-7) - 4\,\* \S(-6,1) - 4\,\* \S(-5,2) - 4\,\* \S(-4,3)
   - 4\,\* \S(1,-6) - 4\,\* \S(2,-5) 
\nn \\ && \mbox{\hspm} 
   + 8\,\* \S(-5,1,1) + 8\,\* \S(-4,1,2)
   + 8\,\* \S(-4,2,1) + 8\,\* \S(1,-5,1) + 8\,\* \S(1,-4,2)
   + 8\,\* \S(1,-3,3) + 8\,\* \S(1,1,-5) 
\nn \\ && \mbox{\hspm} 
   + 8\,\* \S(2,-4,1)
   + 8\,\* \S(2,-3,2) + 8\,\* \S(2,-2,3) - 16\,\* \S(-4,1,1,1)
   - 16\,\* \S(1,-4,1,1) - 16\,\* \S(1,-3,1,2) - 16\,\* \S(1,-3,2,1)
\nn \\ && \mbox{\hspm} 
   - 16\,\* \S(1,1,-4,1) - 16\,\* \S(1,1,-3,2) - 16\,\* \S(1,1,-2,3)
   - 16\,\* \S(2,-3,1,1) - 16\,\* \S(2,-2,1,2) - 16\,\* \S(2,-2,2,1)
\nn \\ && \mbox{\hspm} 
   + 32\,\* \S(1,-3,1,1,1) + 32\,\* \S(1,1,-3,1,1) + 32\,\* \S(1,1,-2,1,2)
   + 32\,\* \S(1,1,-2,2,1) + 32\,\* \S(2,-2,1,1,1) )
\nn \\[1mm]
 \S(1,-2,1,1,1,1) &\!=\!& 1/64\,\* ( -  \BS(1,6)
   + 2\,\* \S(-7) - 4\,\* \S(-6,1) - 4\,\* \S(-5,2) - 4\,\* \S(-4,3)
   - 4\,\* \S(-3,4) - 4\,\* \S(1,-6) 
\nn \\ && \mbox{\hspm} 
   + 8\,\* \S(-5,1,1) + 8\,\* \S(-4,1,2)
   + 8\,\* \S(-4,2,1) + 8\,\* \S(-3,1,3) + 8\,\* \S(-3,2,2)
   + 8\,\* \S(-3,3,1) + 8\,\* \S(1,-5,1) 
\nn \\ && \mbox{\hspm} 
   + 8\,\* \S(1,-4,2)
   + 8\,\* \S(1,-3,3) + 8\,\* \S(1,-2,4) - 16\,\* \S(-4,1,1,1)
   - 16\,\* \S(-3,1,1,2) - 16\,\* \S(-3,1,2,1) - 16\,\* \S(-3,2,1,1)
\nn \\ && \mbox{\hspm} 
   - 16\,\* \S(1,-4,1,1) - 16\,\* \S(1,-3,1,2) - 16\,\* \S(1,-3,2,1)
   - 16\,\* \S(1,-2,1,3) - 16\,\* \S(1,-2,2,2) - 16\,\* \S(1,-2,3,1)
\nn \\ && \mbox{\hspm} 
   + 32\,\* \S(-3,1,1,1,1) + 32\,\* \S(1,-3,1,1,1) + 32\,\* \S(1,-2,1,1,2)
   + 32\,\* \S(1,-2,1,2,1) + 32\,\* \S(1,-2,2,1,1) )
\nn \\[1mm]
  \S(-2,1,1,1,1,1) &\!=\!& 1/64\,\* ( -  \BS(7)
   + 2\,\* \S(-7) - 4\,\* \S(-6,1) - 4\,\* \S(-5,2) - 4\,\* \S(-4,3)
   - 4\,\* \S(-3,4) - 4\,\* \S(-2,5) 
\nn \\ && \mbox{\hspm} 
   + 8\,\* \S(-5,1,1) + 8\,\* \S(-4,1,2)
   + 8\,\* \S(-4,2,1) + 8\,\* \S(-3,1,3) + 8\,\* \S(-3,2,2)
   + 8\,\* \S(-3,3,1) + 8\,\* \S(-2,1,4) 
\nn \\ && \mbox{\hspm} 
   + 8\,\* \S(-2,2,3)
   + 8\,\* \S(-2,3,2) + 8\,\* \S(-2,4,1) - 16\,\* \S(-4,1,1,1)
   - 16\,\* \S(-3,1,1,2) - 16\,\* \S(-3,1,2,1) - 16\,\* \S(-3,2,1,1)
\nn \\ && \mbox{\hspm} 
   - 16\,\* \S(-2,1,1,3) - 16\,\* \S(-2,1,2,2) - 16\,\* \S(-2,1,3,1)
   - 16\,\* \S(-2,2,1,2) - 16\,\* \S(-2,2,2,1) - 16\,\* \S(-2,3,1,1)
\nn \\ && \mbox{\hspm} 
   + 32\,\* \S(-3,1,1,1,1) + 32\,\* \S(-2,1,1,1,2) + 32\,\* \S(-2,1,1,2,1)
   + 32\,\* \S(-2,1,2,1,1) + 32\,\* \S(-2,2,1,1,1) )
\eea
\noindent
Depth 5:
\bea
 \S(1,1,1,-2,-2) &\!=\!& -1/32\,\* ( -  \BS(1,1,1,3,1)
   +  2\,\* \S(7) - 4\,\* \S(-5,-2) - 4\,\* \S(1,6) - 4\,\* \S(2,5)
   - 4\,\* \S(3,4) 
\nn \\ && \mbox{\hspm} 
   + 8\,\* \S(1,-4,-2) 
   + 8\,\* \S(1,1,5) + 8\,\* \S(1,2,4)
   + 8\,\* \S(2,-3,-2) + 8\,\* \S(2,1,4) + 8\,\* \S(3,-2,-2)
\nn \\ && \mbox{\hspm} 
   - 16\,\* \S(1,1,-3,-2) - 16\,\* \S(1,1,1,4) 
   - 16\,\* \S(1,2,-2,-2) - 16\,\* \S(2,1,-2,-2) )
\nn \\[1mm]
 \S(1,1,-2,1,-2) &\!=\!& -1/32\,\* ( -  \BS(1,1,4,1)
   + 2\,\* \S(7) - 4\,\* \S(-5,-2) - 4\,\* \S(-4,-3) - 4\,\* \S(1,6)
   - 4\,\* \S(2,5) 
\nn \\ && \mbox{\hspm} 
   + 8\,\* \S(-4,1,-2) + 8\,\* \S(1,-4,-2)
   + 8\,\* \S(1,-3,-3) + 8\,\* \S(1,1,5) + 8\,\* \S(2,-3,-2)
   + 8\,\* \S(2,-2,-3) 
\nn \\ && \mbox{\hspm} 
   - 16\,\* \S(1,-3,1,-2) - 16\,\* \S(1,1,-3,-2)
   - 16\,\* \S(1,1,-2,-3) - 16\,\* \S(2,-2,1,-2) )
\nn \\[1mm]
 \S(1,-2,1,1,-2) &\!=\!& -1/32\,\* ( -  \BS(1,5,1)
   + 2\,\* \S(7) - 4\,\* \S(-5,-2) - 4\,\* \S(-4,-3) - 4\,\* \S(-3,-4)
   - 4\,\* \S(1,6) 
\nn \\ && \mbox{\hspm} 
   + 8\,\* \S(-4,1,-2) + 8\,\* \S(-3,1,-3)
   + 8\,\* \S(-3,2,-2) + 8\,\* \S(1,-4,-2) + 8\,\* \S(1,-3,-3)
   + 8\,\* \S(1,-2,-4) 
\nn \\ && \mbox{\hspm} 
   - 16\,\* \S(-3,1,1,-2) - 16\,\* \S(1,-3,1,-2)
   - 16\,\* \S(1,-2,1,-3) - 16\,\* \S(1,-2,2,-2) )
\nn \\[1mm]
 \S(-2,1,1,1,-2) &\!=\!& -1/32\,\* ( -  \BS(6,1)
   + 2\,\* \S(7) - 4\,\* \S(-5,-2) - 4\,\* \S(-4,-3) - 4\,\* \S(-3,-4)
   - 4\,\* \S(-2,-5) 
\nn \\ && \mbox{\hspm} 
   + 8\,\* \S(-4,1,-2) + 8\,\* \S(-3,1,-3)
   + 8\,\* \S(-3,2,-2) + 8\,\* \S(-2,1,-4) + 8\,\* \S(-2,2,-3)
   + 8\,\* \S(-2,3,-2) 
\nn \\ && \mbox{\hspm} 
   - 16\,\* \S(-3,1,1,-2) - 16\,\* \S(-2,1,1,-3)
   - 16\,\* \S(-2,1,2,-2) - 16\,\* \S(-2,2,1,-2) )
\nn \\[1mm]
 \S(1,1,-2,-2,1) &\!=\!& -1/32\,\* ( -  \BS(1,1,3,1,1)
   +  2\,\* \S(7) - 4\,\* \S(-4,-3) - 4\,\* \S(1,6) - 4\,\* \S(2,5)
   - 4\,\* \S(6,1) 
\nn \\ && \mbox{\hspm} 
   + 8\,\* \S(-4,-2,1) + 8\,\* \S(1,-3,-3)
   + 8\,\* \S(1,1,5) + 8\,\* \S(1,5,1) + 8\,\* \S(2,-2,-3)
   + 8\,\* \S(2,4,1) 
\nn \\ && \mbox{\hspm} 
   - 16\,\* \S(1,-3,-2,1) - 16\,\* \S(1,1,-2,-3)
   - 16\,\* \S(1,1,4,1) - 16\,\* \S(2,-2,-2,1) )
\nn \\[1mm]
 \S(1,-2,1,-2,1) &\!=\!& -1/32\,\* ( -  \BS(1,4,1,1)
   +  2\,\* \S(7) - 4\,\* \S(-4,-3) - 4\,\* \S(-3,-4) - 4\,\* \S(1,6)
   - 4\,\* \S(6,1) 
\nn \\ && \mbox{\hspm} 
   + 8\,\* \S(-4,-2,1) + 8\,\* \S(-3,-3,1)
   + 8\,\* \S(-3,1,-3) + 8\,\* \S(1,-3,-3) + 8\,\* \S(1,-2,-4)
   + 8\,\* \S(1,5,1) 
\nn \\ && \mbox{\hspm} 
   - 16\,\* \S(-3,1,-2,1) - 16\,\* \S(1,-3,-2,1)
   - 16\,\* \S(1,-2,-3,1) - 16\,\* \S(1,-2,1,-3) )
\nn \\[1mm]
 \S(-2,1,1,-2,1) &\!=\!& -1/32\,\* ( -  \BS(5,1,1)
   + 2\,\* \S(7) - 4\,\* \S(-4,-3) - 4\,\* \S(-3,-4) - 4\,\* \S(-2,-5)
   - 4\,\* \S(6,1) 
\nn \\ && \mbox{\hspm} 
   + 8\,\* \S(-4,-2,1) + 8\,\* \S(-3,-3,1)
   + 8\,\* \S(-3,1,-3) + 8\,\* \S(-2,-4,1) + 8\,\* \S(-2,1,-4)
   + 8\,\* \S(-2,2,-3) 
\nn \\ && \mbox{\hspm} 
   - 16\,\* \S(-3,1,-2,1) - 16\,\* \S(-2,1,-3,1)
   - 16\,\* \S(-2,1,1,-3) - 16\,\* \S(-2,2,-2,1) )
\nn \\[1mm]
 \S(1,-2,-2,1,1) &\!=\!& -1/32\,\* ( -  \BS(1,3,1,1,1)
   +  2\,\* \S(7) - 4\,\* \S(-3,-4) - 4\,\* \S(1,6) - 4\,\* \S(5,2)
   - 4\,\* \S(6,1) 
\nn \\ && \mbox{\hspm} 
   + 8\,\* \S(-3,-3,1) + 8\,\* \S(-3,-2,2)
   + 8\,\* \S(1,-2,-4) + 8\,\* \S(1,4,2) + 8\,\* \S(1,5,1)
   + 8\,\* \S(5,1,1) 
\nn \\ && \mbox{\hspm} 
   - 16\,\* \S(-3,-2,1,1) - 16\,\* \S(1,-2,-3,1)
   - 16\,\* \S(1,-2,-2,2) - 16\,\* \S(1,4,1,1) )
\nn \\[1mm]
 \S(-2,1,-2,1,1) &\!=\!& -1/32\,\* ( -  \BS(4,1,1,1)
   + 2\,\* \S(7) - 4\,\* \S(-3,-4) - 4\,\* \S(-2,-5) - 4\,\* \S(5,2)
   - 4\,\* \S(6,1) 
\nn \\ && \mbox{\hspm} 
   + 8\,\* \S(-3,-3,1) + 8\,\* \S(-3,-2,2)
   + 8\,\* \S(-2,-4,1) + 8\,\* \S(-2,-3,2) + 8\,\* \S(-2,1,-4)
   + 8\,\* \S(5,1,1) 
\nn \\ && \mbox{\hspm} 
   - 16\,\* \S(-3,-2,1,1) - 16\,\* \S(-2,-3,1,1)
   - 16\,\* \S(-2,1,-3,1) - 16\,\* \S(-2,1,-2,2) )
\nn \\[1mm]
 \S(-2,-2,1,1,1) &\!=\!& -1/32\,\* ( -  \BS(3,1,1,1,1)
   + 2\,\* \S(7) - 4\,\* \S(-2,-5) - 4\,\* \S(4,3) - 4\,\* \S(5,2)
   - 4\,\* \S(6,1) 
\nn \\ && \mbox{\hspm} 
   + 8\,\* \S(-2,-4,1) + 8\,\* \S(-2,-3,2)
   + 8\,\* \S(-2,-2,3) + 8\,\* \S(4,1,2) + 8\,\* \S(4,2,1)
   + 8\,\* \S(5,1,1) 
\nn \\ && \mbox{\hspm} 
   - 16\,\* \S(-2,-3,1,1) - 16\,\* \S(-2,-2,1,2)
   - 16\,\* \S(-2,-2,2,1) - 16\,\* \S(4,1,1,1) )
\eea
\noindent
Depth 4, ordered by number of indices `1':
\bea
  \S(1,1,1,-4) &\!=\!& 1/16\,\* ( -  \BS(1,1,1,2,2)
   + 2\,\* \S(-7) - 4\,\* \S(1,-6) - 4\,\* \S(2,-5) - 4\,\* \S(3,-4)
\nn \\ && \mbox{\hspq} 
   + 8\,\* \S(1,1,-5) + 8\,\* \S(1,2,-4) + 8\,\* \S(2,1,-4) );
\nn \\[1mm]
  \S(1,1,-4,1) &\!=\!& 1/16\,\* ( -  \BS(1,1,2,3)
   + 2\,\* \S(-7) - 4\,\* \S(-6,1) - 4\,\* \S(1,-6) - 4\,\* \S(2,-5)
\nn \\ && \mbox{\hspq} 
   + 8\,\* \S(1,-5,1) + 8\,\* \S(1,1,-5) + 8\,\* \S(2,-4,1) );
\nn \\[1mm]
  \S(1,-4,1,1) &\!=\!& 1/16\,\* ( -  \BS(1,2,4)
   + 2\,\* \S(-7) - 4\,\* \S(-6,1) - 4\,\* \S(-5,2) - 4\,\* \S(1,-6)
\nn \\ && \mbox{\hspq} 
   + 8\,\* \S(-5,1,1) + 8\,\* \S(1,-5,1) + 8\,\* \S(1,-4,2) );
\nn \\[1mm]
  \S(-4,1,1,1) &\!=\!& 1/16\,\* ( -  \BS(2,5)
   +  2\,\* \S(-7) - 4\,\* \S(-6,1) - 4\,\* \S(-5,2) - 4\,\* \S(-4,3)
\nn \\ && \mbox{\hspq} 
   + 8\,\* \S(-5,1,1) + 8\,\* \S(-4,1,2) + 8\,\* \S(-4,2,1) );
\\[1mm]
  \S(1,1,-2,3) &\!=\!& 1/16\,\* ( -  \BS(1,1,3,2)
   + 2\,\* \S(-7) - 4\,\* \S(-4,3) - 4\,\* \S(1,-6) - 4\,\* \S(2,-5)
\nn \\ && \mbox{\hspq} 
   + 8\,\* \S(1,-3,3) + 8\,\* \S(1,1,-5) + 8\,\* \S(2,-2,3) )
\nn\\[1mm]
  \S(1,-2,1,3) &\!=\!& 1/16\,\* ( -  \BS(1,4,2)
   + 2\,\* \S(-7) - 4\,\* \S(-4,3) - 4\,\* \S(-3,4) - 4\,\* \S(1,-6)
\nn \\ && \mbox{\hspq} 
   + 8\,\* \S(-3,1,3) + 8\,\* \S(1,-3,3) + 8\,\* \S(1,-2,4) )
\nn\\[1mm]
  \S(-2,1,1,3) &\!=\!& 1/16\,\* ( -   \BS(5,2)
   + 2\,\* \S(-7) - 4\,\* \S(-4,3) - 4\,\* \S(-3,4) - 4\,\* \S(-2,5)
\nn \\ && \mbox{\hspq} 
   + 8\,\* \S(-3,1,3) + 8\,\* \S(-2,1,4) + 8\,\* \S(-2,2,3) )
\nn\\[1mm]
  \S(1,-2,3,1) &\!=\!& 1/16\,\* ( -  \BS(1,3,3)
   + 2\,\* \S(-7) - 4\,\* \S(-6,1) - 4\,\* \S(-3,4) - 4\,\* \S(1,-6)
\nn \\ && \mbox{\hspq} 
   + 8\,\* \S(-3,3,1) + 8\,\* \S(1,-5,1) + 8\,\* \S(1,-2,4) )
\nn\\[1mm]
  \S(-2,1,3,1) &\!=\!& 1/16\,\* ( -  \BS(4,3)
   +  2\,\* \S(-7) - 4\,\* \S(-6,1) - 4\,\* \S(-3,4) - 4\,\* \S(-2,5)
\nn \\ && \mbox{\hspq} 
   + 8\,\* \S(-3,3,1) + 8\,\* \S(-2,1,4) + 8\,\* \S(-2,4,1) )
\nn\\[1mm]
  \S(-2,3,1,1) &\!=\!& 1/16\,\* ( -  \BS(3,4)
   + 2\,\* \S(-7) - 4\,\* \S(-6,1) - 4\,\* \S(-5,2) - 4\,\* \S(-2,5)
\nn \\ && \mbox{\hspq} 
   + 8\,\* \S(-5,1,1) + 8\,\* \S(-2,3,2) + 8\,\* \S(-2,4,1) )
\nn\\[1mm]
  \S(1,1,3,-2) &\!=\!& 1/16\,\* ( -  \BS(1,1,2,1,2)
   + 2\,\* \S(-7) - 4\,\* \S(1,-6) - 4\,\* \S(2,-5) - 4\,\* \S(5,-2)
\nn \\ && \mbox{\hspq} 
   + 8\,\* \S(1,1,-5) + 8\,\* \S(1,4,-2) + 8\,\* \S(2,3,-2) )
\nn\\[1mm]
  \S(1,3,1,-2) &\!=\!& 1/16\,\* ( -  \BS(1,2,1,1,2)
   + 2\,\* \S(-7) - 4\,\* \S(1,-6) - 4\,\* \S(4,-3) - 4\,\* \S(5,-2)
\nn \\ && \mbox{\hspq} 
   + 8\,\* \S(1,3,-3) + 8\,\* \S(1,4,-2) + 8\,\* \S(4,1,-2) )
\nn\\[1mm]
  \S(3,1,1,-2) &\!=\!& 1/16\,\* ( -  \BS(2,1,1,1,2)
   + 2\,\* \S(-7) - 4\,\* \S(3,-4) - 4\,\* \S(4,-3) - 4\,\* \S(5,-2)
\nn \\ && \mbox{\hspq} 
   + 8\,\* \S(3,1,-3) + 8\,\* \S(3,2,-2) + 8\,\* \S(4,1,-2) )
\nn\\[1mm]
  \S(1,3,-2,1) &\!=\!& 1/16\,\* ( -  \BS(1,2,1,3)
   + 2\,\* \S(-7) - 4\,\* \S(-6,1) - 4\,\* \S(1,-6) - 4\,\* \S(4,-3)
\nn \\ && \mbox{\hspq} 
   + 8\,\* \S(1,-5,1) + 8\,\* \S(1,3,-3) + 8\,\* \S(4,-2,1) )
\nn\\[1mm]
  \S(3,1,-2,1) &\!=\!& 1/16\,\* ( -  \BS(2,1,1,3)
   + 2\,\* \S(-7) - 4\,\* \S(-6,1) - 4\,\* \S(3,-4) - 4\,\* \S(4,-3)
\nn \\ && \mbox{\hspq} 
   + 8\,\* \S(3,-3,1) + 8\,\* \S(3,1,-3) + 8\,\* \S(4,-2,1) )
\nn\\[1mm]
  \S(3,-2,1,1) &\!=\!& 1/16\,\* ( -  \BS(2,1,4) + 2\,\* \S(-7)
   - 4\,\* \S(-6,1) - 4\,\* \S(-5,2) - 4\,\* \S(3,-4)
\nn \\ && \mbox{\hspq} 
   + 8\,\* \S(-5,1,1) + 8\,\* \S(3,-3,1) + 8\,\* \S(3,-2,2) )
\\[1mm]
  \S(1,-2,-2,-2) &\!=\!& 1/16\,\* ( -  \BS(1,3,1,2)
   + 2\,\* \S(-7) - 4\,\* \S(-3,4) - 4\,\* \S(1,-6) - 4\,\* \S(5,-2)
\nn \\ && \mbox{\hspq} 
   + 8\,\* \S(-3,-2,-2) + 8\,\* \S(1,-2,4) + 8\,\* \S(1,4,-2) )
\nn \\[1mm]
  \S(-2,1,-2,-2) &\!=\!& 1/16\,\* ( -  \BS(4,1,2)
   + 2\,\* \S(-7) - 4\,\* \S(-3,4) - 4\,\* \S(-2,5) - 4\,\* \S(5,-2)
\nn \\ && \mbox{\hspq} 
   + 8\,\* \S(-3,-2,-2) + 8\,\* \S(-2,-3,-2) + 8\,\* \S(-2,1,4) )
\nn \\[1mm]
  \S(-2,-2,1,-2) &\!=\!& 1/16\,\* ( -  \BS(3,1,1,2)
   + 2\,\* \S(-7) - 4\,\* \S(-2,5) - 4\,\* \S(4,-3) - 4\,\* \S(5,-2)
\nn \\ && \mbox{\hspq} 
   + 8\,\* \S(-2,-3,-2) + 8\,\* \S(-2,-2,-3) + 8\,\* \S(4,1,-2) )
\nn \\[1mm]
  \S(-2,-2,-2,1) &\!=\!& 1/16\,\* ( -  \BS(3,1,3)
   + 2\,\* \S(-7) - 4\,\* \S(-6,1) - 4\,\* \S(-2,5) - 4\,\* \S(4,-3)
\nn \\ && \mbox{\hspq} 
   + 8\,\* \S(-2,-2,-3) + 8\,\* \S(-2,4,1) + 8\,\* \S(4,-2,1) )
\eea
\noindent
Depth 3 and below:
\bea
\label{eq:B9}
  \S(1,-4,-2) &\!=\!& -1/8\,\* ( -  \BS(1,2,3,1)
   +  2\,\* \S(7) - 4\,\* \S(-5,-2) - 4\,\* \S(1,6) )
\nn \\[1mm]
  \S(-4,1,-2) &\!=\!& -1/8\,\* ( -  \BS(2,4,1)
   + 2\,\* \S(7) - 4\,\* \S(-5,-2) - 4\,\* \S(-4,-3) )
\nn \\[1mm]
  \S(-4,-2,1) &\!=\!& -1/8\,\* ( -  \BS(2,3,1,1)
   + 2\,\* \S(7) - 4\,\* \S(-4,-3) - 4\,\* \S(6,1) )
\nn \\[1mm]
  \S(1,-2,-4) &\!=\!& -1/8\,\* ( -  \BS(1,3,2,1)
   + 2\,\* \S(7) - 4\,\* \S(-3,-4) - 4\,\* \S(1,6) )
\nn \\[1mm]
  \S(-2,1,-4) &\!=\!& -1/8\,\* ( -  \BS(4,2,1)
   + 2\,\* \S(7) - 4\,\* \S(-3,-4) - 4\,\* \S(-2,-5) )
\nn \\[1mm]
  \S(-2,-4,1) &\!=\!& -1/8\,\* ( -  \BS(3,2,1,1)
   + 2\,\* \S(7) - 4\,\* \S(-2,-5) - 4\,\* \S(6,1) )
\nn \\[1mm]
  \S(-2,-2,3) &\!=\!& -1/8\,\* ( -  \BS(3,1,2,1)
   + 2\,\* \S(7) - 4\,\* \S(-2,-5) - 4\,\* \S(4,3) )
\nn \\[1mm]
  \S(-2,3,-2) &\!=\!& -1/8\,\* ( -  \BS(3,3,1)
   +  2\,\* \S(7) - 4\,\* \S(-5,-2) - 4\,\* \S(-2,-5) )
\nn \\[1mm]
  \S(3,-2,-2) &\!=\!& -1/8\,\*( -  \BS(2,1,3,1)
   + 2\,\* \S(7) - 4\,\* \S(-5,-2) - 4\,\* \S(3,4) )
\\[1mm]
  \S(1,-6) &\!=\!& -1/4\,\* (  \BS(1,2,2,2) - 2\,\* \S(-7) )
\nn \\[1mm]
  \S(-6,1) &\!=\!& -1/4\,\* (  \BS(2,2,3) - 2\,\* \S(-7) )
\nn \\[1mm]
  \S(-2,5) &\!=\!& -1/4\,\* (  \BS(3,2,2) - 2\,\* \S(-7) )
\nn \\[1mm]
  \S(5,-2) &\!=\!& -1/4\,\* (  \BS(2,2,1,2) - 2\,\* \S(-7) )
\nn \\[1mm]
  \S(-4,3) &\!=\!& -1/4\,\* (  \BS(2,3,2) - 2\,\* \S(-7) )
\nn \\[1mm]
  \S(3,-4) &\!=\!& -1/4\,\* (  \BS(2,1,2,2) - 2\,\* \S(-7) )
\eea
Weight 7, non-alternating sums, depths 7:
\bea
 \S(1,1,1,1,1,1,1) &\!=\!& -1/128\,\* ( -  \BS(1,1,1,1,1,1,1)
   + 2\,\* \S(7)   - 4\,\* \S(1,6) - 4\,\* \S(2,5)
   - 4\,\* \S(3,4) - 4\,\* \S(4,3) - 4\,\* \S(5,2) 
\nn \\ && \mbox{\hspm} 
   - 4\,\* \S(6,1)
   + 8\,\* \S(1,1,5) + 8\,\* \S(1,2,4) + 8\,\* \S(1,3,3)
   + 8\,\* \S(1,4,2) + 8\,\* \S(1,5,1) + 8\,\* \S(2,1,4)
   + 8\,\* \S(2,2,3) + 8\,\* \S(2,3,2) 
\nn \\ && \mbox{\hspm} 
   + 8\,\* \S(2,4,1)
   + 8\,\* \S(3,1,3) + 8\,\* \S(3,2,2) + 8\,\* \S(3,3,1)
   + 8\,\* \S(4,1,2) + 8\,\* \S(4,2,1) + 8\,\* \S(5,1,1)
   - 16\,\* \S(1,1,1,4) 
\nn \\ && \mbox{\hspm} 
   - 16\,\* \S(1,1,2,3) - 16\,\* \S(1,1,3,2)
   - 16\,\* \S(1,1,4,1) - 16\,\* \S(1,2,1,3) - 16\,\* \S(1,2,2,2)
   - 16\,\* \S(1,2,3,1) - 16\,\* \S(1,3,1,2) 
\nn \\ && \mbox{\hspm} 
   - 16\,\* \S(1,3,2,1)
   - 16\,\* \S(1,4,1,1) - 16\,\* \S(2,1,1,3) - 16\,\* \S(2,1,2,2)
   - 16\,\* \S(2,1,3,1) - 16\,\* \S(2,2,1,2) - 16\,\* \S(2,2,2,1)
\nn \\ && \mbox{\hspm} 
   - 16\,\* \S(2,3,1,1) - 16\,\* \S(3,1,1,2) - 16\,\* \S(3,1,2,1)
   - 16\,\* \S(3,2,1,1) - 16\,\* \S(4,1,1,1)
   + 32\,\* \S(1,1,1,1,3) 
\nn \\ && \mbox{\hspm} 
   + 32\,\* \S(1,1,1,3,1) + 32\,\* \S(1,1,3,1,1)
   + 32\,\* \S(1,3,1,1,1) + 32\,\* \S(3,1,1,1,1) + 32\,\* \S(1,1,1,2,2)
   + 32\,\* \S(1,1,2,1,2) 
\nn \\ && \mbox{\hspm} 
   + 32\,\* \S(1,2,1,1,2) + 32\,\* \S(2,1,1,1,2)
   + 32\,\* \S(1,1,2,2,1) + 32\,\* \S(1,2,1,2,1) + 32\,\* \S(2,1,1,2,1)
   + 32\,\* \S(1,2,2,1,1) 
\nn \\ && \mbox{\hspm} 
   + 32\,\* \S(2,1,2,1,1) + 32\,\* \S(2,2,1,1,1)
   - 64\,\* \S(1,1,1,1,1,2) - 64\,\* \S(1,1,1,1,2,1)
   - 64\,\* \S(1,1,1,2,1,1) 
\nn \\ && \mbox{\hspm} 
   - 64\,\* \S(1,1,2,1,1,1) - 64\,\* \S(1,2,1,1,1,1) - 64\,\* \S(2,1,1,1,1,1) )
\eea
Depth 5:
\bea
 \S(1,1,1,1,3) &\!=\!& -1/32\,\* ( -  \BS(1,1,1,1,2,1) + 2\,\* \S(7)
   - 4\,\* \S(1,6) - 4\,\* \S(2,5) - 4\,\* \S(3,4) - 4\,\* \S(4,3)
\nn \\ && \mbox{\hspq} 
   + 8\,\* \S(1,1,5) + 8\,\* \S(1,2,4) 
   + 8\,\* \S(1,3,3) + 8\,\* \S(2,1,4)
   + 8\,\* \S(2,2,3) + 8\,\* \S(3,1,3) 
\nn \\ && \mbox{\hspq} 
   - 16\,\* \S(1,1,1,4)
   - 16\,\* \S(1,1,2,3) - 16\,\* \S(1,2,1,3) - 16\,\* \S(2,1,1,3) )
\nn \\[1mm]
 \S(1,1,1,3,1) &\!=\!& -1/32\,\* ( -  \BS(1,1,1,2,1,1) + 2\,\* \S(7)
   - 4\,\* \S(1,6) - 4\,\* \S(2,5) - 4\,\* \S(3,4) - 4\,\* \S(6,1)
\nn \\ && \mbox{\hspq} 
   + 8\,\* \S(1,1,5) + 8\,\* \S(1,2,4) + 8\,\* \S(1,5,1) + 8\,\* \S(2,1,4)
   + 8\,\* \S(2,4,1) + 8\,\* \S(3,3,1) 
\nn \\ && \mbox{\hspq} 
   - 16\,\* \S(1,1,1,4)
   - 16\,\* \S(1,1,4,1) - 16\,\* \S(1,2,3,1) - 16\,\* \S(2,1,3,1) );
\nn \\[1mm]
 \S(1,1,3,1,1) &\!=\!& -1/32\,\* ( -  \BS(1,1,2,1,1,1) + 2\,\* \S(7)
   - 4\,\* \S(1,6) - 4\,\* \S(2,5) - 4\,\* \S(5,2) - 4\,\* \S(6,1)
\nn \\ && \mbox{\hspq} 
   + 8\,\* \S(1,1,5) + 8\,\* \S(1,4,2) + 8\,\* \S(1,5,1) + 8\,\* \S(2,3,2)
   + 8\,\* \S(2,4,1) + 8\,\* \S(5,1,1) 
\nn \\ && \mbox{\hspq} 
   - 16\,\* \S(1,1,3,2)
   - 16\,\* \S(1,1,4,1) - 16\,\* \S(1,4,1,1) - 16\,\* \S(2,3,1,1) );
\nn \\[1mm]
 \S(1,3,1,1,1) &\!=\!& -1/32\,\* ( -  \BS(1,2,1,1,1,1) + 2\,\* \S(7)
   - 4\,\* \S(1,6) - 4\,\* \S(4,3) - 4\,\* \S(5,2) - 4\,\* \S(6,1)
\nn \\ && \mbox{\hspq} 
   + 8\,\* \S(1,3,3) + 8\,\* \S(1,4,2) + 8\,\* \S(1,5,1) + 8\,\* \S(4,1,2)
   + 8\,\* \S(4,2,1) + 8\,\* \S(5,1,1) 
\nn \\ && \mbox{\hspq} 
   - 16\,\* \S(1,3,1,2)
   - 16\,\* \S(1,3,2,1) - 16\,\* \S(1,4,1,1) - 16\,\* \S(4,1,1,1) );
\nn \\[1mm]
 \S(3,1,1,1,1) &\!=\!& -1/32\,\* ( -  \BS(2,1,1,1,1,1) + 2\,\* \S(7)
   - 4\,\* \S(3,4) - 4\,\* \S(4,3) - 4\,\* \S(5,2) - 4\,\* \S(6,1)
\nn \\ && \mbox{\hspq} 
   + 8\,\* \S(3,1,3) + 8\,\* \S(3,2,2) + 8\,\* \S(3,3,1) + 8\,\* \S(4,1,2)
   + 8\,\* \S(4,2,1) + 8\,\* \S(5,1,1) 
\nn \\ && \mbox{\hspq} 
   - 16\,\* \S(3,1,1,2)
   - 16\,\* \S(3,1,2,1) - 16\,\* \S(3,2,1,1) - 16\,\* \S(4,1,1,1) );
\eea
Depths 3 and below:
\bea
  \S(1,1,5) &\!=\!& -1/8\,\* ( -  \BS(1,1,2,2,1)
   + 2\,\* \S(7) - 4\,\* \S(1,6) - 4\,\* \S(2,5) )
\nn \\[1mm]
 \S(1,3,3) &\!=\!& -1/8\,\* ( -  \BS(1,2,1,2,1)
   + 2\,\* \S(7) - 4\,\* \S(1,6) - 4\,\* \S(4,3) )
\nn \\[1mm]
 \S(1,5,1) &\!=\!& -1/8\,\* ( -  \BS(1,2,2,1,1)
   + 2\,\* \S(7) - 4\,\* \S(1,6) - 4\,\* \S(6,1) )
\nn \\[1mm]
 \S(3,1,3) &\!=\!& -1/8\,\* ( -  \BS(2,1,1,2,1)
   + 2\,\* \S(7) - 4\,\* \S(3,4) - 4\,\* \S(4,3) )
\nn \\[1mm]
 \S(3,3,1) &\!=\!& -1/8\,\* ( -  \BS(2,1,2,1,1)
   + 2\,\* \S(7) - 4\,\* \S(3,4) - 4\,\* \S(6,1) )
\nn \\[1mm]
 \S(5,1,1) &\!=\!& -1/8\,\* ( -  \BS(2,2,1,1,1)
   + 2\,\* \S(7) - 4\,\* \S(5,2) - 4\,\* \S(6,1) )
\nn \\[1mm]
 \S(7) &\!=\!& 1/2\,\*  \BS(2,2,2,1)
\eea
\noindent
Weight 6, alternating sums, depth 5:
\bea
 \S(-2,1,1,1,1) &\!=\!& 1/32\,\* ( -  \BS(6)
   - 2\,\* \S(-6) + 4\,\* \S(-5,1) + 4\,\* \S(-4,2) + 4\,\* \S(-3,3)
   + 4\,\* \S(-2,4) - 8\,\* \S(-4,1,1) 
\nn \\ && \mbox{\hspq}
   - 8\,\* \S(-3,1,2)
   - 8\,\* \S(-3,2,1) - 8\,\* \S(-2,1,3) - 8\,\* \S(-2,2,2)
   - 8\,\* \S(-2,3,1) 
\nn \\ && \mbox{\hspq}
   + 16\,\* \S(-3,1,1,1) + 16\,\* \S(-2,1,1,2)
   + 16\,\* \S(-2,1,2,1) + 16\,\* \S(-2,2,1,1) )
\nn \\[1mm]
 \S(1,-2,1,1,1) &\!=\!& 1/32\,\* ( -  \BS(1,5)
   - 2\,\* \S(-6) + 4\,\* \S(-5,1) + 4\,\* \S(-4,2) + 4\,\* \S(-3,3)
   + 4\,\* \S(1,-5) - 8\,\* \S(-4,1,1) 
\nn \\ && \mbox{\hspq}
   - 8\,\* \S(-3,1,2)
   - 8\,\* \S(-3,2,1) - 8\,\* \S(1,-4,1) - 8\,\* \S(1,-3,2)
   - 8\,\* \S(1,-2,3) 
\nn \\ && \mbox{\hspq}
   + 16\,\* \S(-3,1,1,1) + 16\,\* \S(1,-3,1,1)
   + 16\,\* \S(1,-2,1,2) + 16\,\* \S(1,-2,2,1) )
\nn \\[1mm]
 \S(1,1,-2,1,1) &\!=\!& 1/32\,\* ( -  \BS(1,1,4)
   - 2\,\* \S(-6) + 4\,\* \S(-5,1) + 4\,\* \S(-4,2) + 4\,\* \S(1,-5)
   + 4\,\* \S(2,-4) - 8\,\* \S(-4,1,1) 
\nn \\ && \mbox{\hspq}
   - 8\,\* \S(1,-4,1)
   - 8\,\* \S(1,-3,2) - 8\,\* \S(1,1,-4) - 8\,\* \S(2,-3,1)
   - 8\,\* \S(2,-2,2) 
\nn \\ && \mbox{\hspq}
   + 16\,\* \S(1,-3,1,1) + 16\,\* \S(1,1,-3,1)
   + 16\,\* \S(1,1,-2,2) + 16\,\* \S(2,-2,1,1) )
\nn \\[1mm]
 \S(1,1,1,-2,1) &\!=\!& 1/32\,\* ( -  \BS(1,1,1,3)
   - 2\,\* \S(-6) + 4\,\* \S(-5,1) + 4\,\* \S(1,-5) + 4\,\* \S(2,-4)
   + 4\,\* \S(3,-3) - 8\,\* \S(1,-4,1) 
\nn \\ && \mbox{\hspq}
   - 8\,\* \S(1,1,-4)
   - 8\,\* \S(1,2,-3) - 8\,\* \S(2,-3,1) - 8\,\* \S(2,1,-3)
   - 8\,\* \S(3,-2,1) 
\nn \\ && \mbox{\hspq}
   + 16\,\* \S(1,1,-3,1) + 16\,\* \S(1,1,1,-3)
   + 16\,\* \S(1,2,-2,1) + 16\,\* \S(2,1,-2,1) )
\nn \\[1mm]
 \S(1,1,1,1,-2) &\!=\!& 1/32\,\* ( -  \BS(1,1,1,1,2)
   - 2\,\* \S(-6) + 4\,\* \S(1,-5) + 4\,\* \S(2,-4) + 4\,\* \S(3,-3)
   + 4\,\* \S(4,-2) - 8\,\* \S(1,1,-4) 
\nn \\ && \mbox{\hspq}
   - 8\,\* \S(1,2,-3)
   - 8\,\* \S(1,3,-2) - 8\,\* \S(2,1,-3) - 8\,\* \S(2,2,-2)
   - 8\,\* \S(3,1,-2) 
\nn \\ && \mbox{\hspq}
   + 16\,\* \S(1,1,1,-3) + 16\,\* \S(1,1,2,-2)
   + 16\,\* \S(1,2,1,-2) + 16\,\* \S(2,1,1,-2) )
\eea
Depth 4:
\bea
 \S(-2,-2,1,1) &\!=\!& -1/16\,\* ( -  \BS(3,1,1,1)
   - 2\,\* \S(6) + 4\,\* \S(-2,-4) + 4\,\* \S(4,2) + 4\,\* \S(5,1)
\nn \\ && \mbox{\hspq}
   - 8\,\* \S(-2,-3,1) - 8\,\* \S(-2,-2,2) - 8\,\* \S(4,1,1) )
\nn \\[1mm]
 \S(-2,1,-2,1) &\!=\!& -1/16\,\* ( -  \BS(4,1,1)
   - 2\,\* \S(6) + 4\,\* \S(-3,-3) + 4\,\* \S(-2,-4) + 4\,\* \S(5,1)
\nn \\ && \mbox{\hspq}
   - 8\,\* \S(-3,-2,1) - 8\,\* \S(-2,-3,1) - 8\,\* \S(-2,1,-3) )
\nn \\[1mm]
 \S(-2,1,1,-2) &\!=\!& -1/16\,\* ( -  \BS(5,1)
   - 2\,\* \S(6) + 4\,\* \S(-4,-2) + 4\,\* \S(-3,-3) + 4\,\* \S(-2,-4)
\nn \\ && \mbox{\hspq}
   - 8\,\* \S(-3,1,-2) - 8\,\* \S(-2,1,-3) - 8\,\* \S(-2,2,-2) )
\nn \\[1mm]
 \S(1,-2,-2,1) &\!=\!& -1/16\,\* ( -  \BS(1,3,1,1)
   - 2\,\* \S(6) + 4\,\* \S(-3,-3) + 4\,\* \S(1,5) + 4\,\* \S(5,1)
\nn \\ && \mbox{\hspq}
   - 8\,\* \S(-3,-2,1) - 8\,\* \S(1,-2,-3) - 8\,\* \S(1,4,1) )
\nn \\[1mm]
 \S(1,-2,1,-2) &\!=\!& -1/16\,\* ( -  \BS(1,4,1)
   - 2\,\* \S(6) + 4\,\* \S(-4,-2) + 4\,\* \S(-3,-3) + 4\,\* \S(1,5)
\nn \\ && \mbox{\hspq}
   - 8\,\* \S(-3,1,-2) - 8\,\* \S(1,-3,-2) - 8\,\* \S(1,-2,-3) )
\nn \\[1mm]
 \S(1,1,-2,-2) &\!=\!& -1/16\,\* ( -  \BS(1,1,3,1)
   - 2\,\* \S(6) + 4\,\* \S(-4,-2) + 4\,\* \S(1,5) + 4\,\* \S(2,4)
\nn \\ && \mbox{\hspq}
   - 8\,\* \S(1,-3,-2) - 8\,\* \S(1,1,4) - 8\,\* \S(2,-2,-2)
\eea
Depth 3 and below:
\bea
 \S(1,1,-4) &\!=\!& 1/8\,\* ( -  \BS(1,1,2,2)
   - 2\,\* \S(-6) + 4\,\* \S(1,-5) + 4\,\* \S(2,-4) )
\nn \\[1mm]
 \S(1,-4,1) &\!=\!& 1/8\,\* ( -  \BS(1,2,3)
   - 2\,\* \S(-6) + 4\,\* \S(-5,1) + 4\,\* \S(1,-5) )
\nn \\[1mm]
 \S(-4,1,1) &\!=\!& 1/8\,\* ( -  \BS(2,4)
   - 2\,\* \S(-6) + 4\,\* \S(-5,1) + 4\,\* \S(-4,2) )
\nn \\[1mm]
 \S(-2,3,1) &\!=\!& 1/8\,\* ( -  \BS(3,3)
   - 2\,\* \S(-6) + 4\,\* \S(-5,1) + 4\,\* \S(-2,4) )
\nn \\[1mm]
 \S(-2,1,3) &\!=\!& 1/8\,\* ( -  \BS(4,2)
   - 2\,\* \S(-6) + 4\,\* \S(-3,3) + 4\,\* \S(-2,4) )
\nn \\[1mm]
 \S(1,-2,3) &\!=\!& 1/8\,\* ( -  \BS(1,3,2)
   - 2\,\* \S(-6) + 4\,\* \S(-3,3) + 4\,\* \S(1,-5) )
\nn \\[1mm]
 \S(3,-2,1) &\!=\!& 1/8\,\* ( -  \BS(2,1,3)
   - 2\,\* \S(-6) + 4\,\* \S(-5,1) + 4\,\* \S(3,-3) )
\nn \\[1mm]
 \S(1,3,-2) &\!=\!& 1/8\,\* ( -  \BS(1,2,1,2)
   - 2\,\* \S(-6) + 4\,\* \S(1,-5) + 4\,\* \S(4,-2) )
\nn \\[1mm]
 \S(3,1,-2) &\!=\!& 1/8\,\* ( -  \BS(2,1,1,2)
   - 2\,\* \S(-6) + 4\,\* \S(3,-3) + 4\,\* \S(4,-2) )
\nn \\[1mm]
 \S(-2,-2,-2) &\!=\!& 1/8\,\* ( -  \BS(3,1,2)
   - 2\,\* \S(-6) + 4\,\* \S(-2,4) + 4\,\* \S(4,-2) )
\\[1mm]
 \S(-4,-2) &\!=\!& -1/4\,\* ( -  \BS(2,3,1) - 2\,\* \S(6) )
\nn \\[1mm]
 \S(-2,-4) &\!=\!& -1/4\,\* ( -  \BS(3,2,1) - 2\,\* \S(6) )
\nn \\[1mm]
 \S(-6)    &\!=\!& -1/2\,\*  \BS(2,2,2)
\eea
Weight 6, non-alternating sums, depth 6:
\bea
 \S(1,1,1,1,1,1) &\!=\!& -1/64\,\* ( -  \BS(1,1,1,1,1,1)
   - 2\,\* \S(6) + 4\,\* \S(1,5) + 4\,\* \S(2,4) + 4\,\* \S(3,3)
   + 4\,\* \S(4,2) + 4\,\* \S(5,1) 
\nn \\ && \mbox{\hspm}
   - 8\,\* \S(1,1,4) - 8\,\* \S(1,2,3)
   - 8\,\* \S(1,3,2) - 8\,\* \S(1,4,1) - 8\,\* \S(2,1,3) - 8\,\* \S(2,2,2)
   - 8\,\* \S(2,3,1) - 8\,\* \S(3,1,2) 
\nn \\ && \mbox{\hspm}
   - 8\,\* \S(3,2,1) - 8\,\* \S(4,1,1)
   + 16\,\* \S(1,1,1,3) + 16\,\* \S(1,1,2,2) + 16\,\* \S(1,1,3,1)
   + 16\,\* \S(1,2,1,2) + 16\,\* \S(1,2,2,1) 
\nn \\ && \mbox{\hspm}
   + 16\,\* \S(1,3,1,1)
   + 16\,\* \S(2,1,1,2) + 16\,\* \S(2,1,2,1) + 16\,\* \S(2,2,1,1)
   + 16\,\* \S(3,1,1,1)   - 32\,\* \S(1,1,1,1,2) 
\nn \\ && \mbox{\hspm}
   - 32\,\* \S(1,1,1,2,1)
   - 32\,\* \S(1,1,2,1,1) - 32\,\* \S(1,2,1,1,1) - 32\,\* \S(2,1,1,1,1) )
\eea
Depth 4 and below:
\bea
 \S(1,1,1,3) &\!=\!& -1/16\,\* ( -  \BS(1,1,1,2,1)
   - 2\,\* \S(6) + 4\,\* \S(1,5) + 4\,\* \S(2,4) + 4\,\* \S(3,3)
   - 8\,\* \S(1,1,4) - 8\,\* \S(1,2,3) - 8\,\* \S(2,1,3) )
\nn \\[1mm]
 \S(1,1,3,1) &\!=\!& -1/16\,\* ( -  \BS(1,1,2,1,1)
   - 2\,\* \S(6) + 4\,\* \S(1,5) + 4\,\* \S(2,4) + 4\,\* \S(5,1)
   - 8\,\* \S(1,1,4) - 8\,\* \S(1,4,1) - 8\,\* \S(2,3,1) )
\nn \\[1mm]
 \S(1,3,1,1) &\!=\!& -1/16\,\* ( -  \BS(1,2,1,1,1) - 2\,\* \S(6)
   + 4\,\* \S(1,5) + 4\,\* \S(4,2) + 4\,\* \S(5,1) - 8\,\* \S(1,3,2)
   - 8\,\* \S(1,4,1) - 8\,\* \S(4,1,1) )
\nn \\[1mm]
 \S(3,1,1,1) &\!=\!& -1/16\,\* ( -  \BS(2,1,1,1,1) - 2\,\* \S(6)
   + 4\,\* \S(3,3) + 4\,\* \S(4,2) + 4\,\* \S(5,1) - 8\,\* \S(3,1,2)
   - 8\,\* \S(3,2,1) - 8\,\* \S(4,1,1) )
\nn \\
\\[-2mm]
 \S(1,5) &\!=\!& -1/4\,\* ( -  \BS(1,2,2,1) - 2\,\* \S(6) )
\nn \\[1mm]
 \S(3,3) &\!=\!& -1/4\,\* ( -  \BS(2,1,2,1) - 2\,\* \S(6) )
\nn \\[1mm]
 \S(5,1) &\!=\!& -1/4\,\* ( -  \BS(2,2,1,1) - 2\,\* \S(6) )
\eea
\noindent
Weight 5, alternating sums, depth 4:
\bea
 \S(-2,1,1,1) &\!\!=\!\!& 1/16\,\* ( -  \BS(5)
   + 2\,\* \S(-5) - 4\,\* \S(-4,1) - 4\,\* \S(-3,2) - 4\,\* \S(-2,3)
   + 8\,\* \S(-3,1,1) + 8\,\* \S(-2,1,2) + 8\,\* \S(-2,2,1) )
\nn \\[1mm]
 \S(1,-2,1,1) &\!\!=\!\!& 1/16\,\* ( -  \BS(1,4)
   + 2\,\* \S(-5) - 4\,\* \S(-4,1) - 4\,\* \S(-3,2) - 4\,\* \S(1,-4)
   + 8\,\* \S(-3,1,1) + 8\,\* \S(1,-3,1) + 8\,\* \S(1,-2,2) )
\nn \\[1mm]
 \S(1,1,-2,1) &\!\!=\!\!& 1/16\,\* ( -  \BS(1,1,3)
   + 2\,\* \S(-5) - 4\,\* \S(-4,1) - 4\,\* \S(1,-4) - 4\,\* \S(2,-3)
   + 8\,\* \S(1,-3,1) + 8\,\* \S(1,1,-3) + 8\,\* \S(2,-2,1) )
\nn \\[1mm]
 \S(1,1,1,-2) &\!\!=\!\!& 1/16\,\* ( -  \BS(1,1,1,2)
   + 2\,\* \S(-5) - 4\,\* \S(1,-4) - 4\,\* \S(2,-3) - 4\,\* \S(3,-2)
   + 8\,\* \S(1,1,-3) + 8\,\* \S(1,2,-2) + 8\,\* \S(2,1,-2) )
\nn \\
\eea
Depth 3 and below:
\bea
 \S(-2,-2,1) &\!=\!& -1/8\,\* ( -  \BS(3,1,1)
   + 2\,\* \S(5) - 4\,\* \S(-2,-3) - 4\,\* \S(4,1) )
\nn \\[1mm]
 \S(-2,1,-2) &\!=\!& -1/8\,\* ( -  \BS(4,1)
   + 2\,\* \S(5) - 4\,\* \S(-3,-2) - 4\,\* \S(-2,-3) )
\nn \\[1mm]
 \S(1,-2,-2) &\!=\!& -1/8\,\* ( -  \BS(1,3,1)
   + 2\,\* \S(5) - 4\,\* \S(-3,-2) - 4\,\* \S(1,4) )
\nn \\[1mm]
 \S(-4,1) &\!=\!& 1/4\,\* ( -  \BS(2,3) + 2\,\* \S(-5) )
\nn \\[1mm]
 \S(-2,3) &\!=\!& 1/4\,\* ( -  \BS(3,2) + 2\,\* \S(-5) )
\nn \\[1mm]
 \S(1,-4) &\!=\!& 1/4\,\* ( -  \BS(1,2,2) + 2\,\* \S(-5) )
\nn \\[1mm]
 \S(3,-2) &\!=\!& 1/4\,\* ( -  \BS(2,1,2) + 2\,\* \S(-5) )
\eea
Non-alternating sums, depth 5:
\bea
 \S(1,1,1,1,1) &\!=\!& -1/32\,\* ( -  \BS(1,1,1,1,1)
   + 2\,\* \S(5) - 4\,\* \S(1,4) - 4\,\* \S(2,3) - 4\,\* \S(3,2)
   - 4\,\* \S(4,1) 
\nn \\ && \mbox{\hspq}
   + 8\,\* \S(1,1,3) + 8\,\* \S(1,2,2)
   + 8\,\* \S(1,3,1) + 8\,\* \S(2,1,2) + 8\,\* \S(2,2,1)
   + 8\,\* \S(3,1,1) 
\nn \\ && \mbox{\hspq}
   - 16\,\* \S(1,1,1,2) - 16\,\* \S(1,1,2,1)
   - 16\,\* \S(1,2,1,1) - 16\,\* \S(2,1,1,1) )
\eea
Depth 3 and lower:
\bea
 \S(1,1,3) &\!=\!& -1/8\,\* ( -  \BS(1,1,2,1)
   + 2\,\* \S(5) - 4\,\* \S(1,4) - 4\,\* \S(2,3) )
\nn \\[1mm]
 \S(1,3,1) &\!=\!& -1/8\,\* ( -  \BS(1,2,1,1)
   + 2\,\* \S(5) - 4\,\* \S(1,4) - 4\,\* \S(4,1) )
\nn \\[1mm]
 \S(3,1,1) &\!=\!& -1/8\,\* ( -  \BS(2,1,1,1)
   + 2\,\* \S(5) - 4\,\* \S(3,2) - 4\,\* \S(4,1) )
\nn \\[1mm]
 \S(5)  &\!=\!& 1/2\,\*  \BS(2,2,1)
\eea
Weight 4, alternating sums:
\bea
 \S(-2,1,1) &\!=\!& 1/8\,\* ( -  \BS(4)
   - 2\,\* \S(-4) + 4\,\* \S(-3,1) + 4\,\* \S(-2,2) )
\nn \\[1mm]
 \S(1,-2,1) &\!=\!& 1/8\,\* ( -  \BS(1,3)
   - 2\,\* \S(-4) + 4\,\* \S(-3,1) + 4\,\* \S(1,-3) )
\nn \\[1mm]
 \S(1,1,-2) &\!=\!& 1/8\,\* ( -  \BS(1,1,2)
   - 2\,\* \S(-4) + 4\,\* \S(1,-3) + 4\,\* \S(2,-2) )
\nn \\[1mm]
 \S(-2,-2) &\!=\!& -1/4\,\* ( -  \BS(3,1) - 2\,\*  \S(4) )
\nn \\[1mm]
 \S(-4)    &\!=\!& -1/2\,\*  \BS(2,2)
\eea
Non-alternating sums:
\bea
 \S(1,1,1,1) &\!=\!& -1/16\,\* ( -  \BS(1,1,1,1)
   - 2\,\* \S(4) + 4\,\* \S(1,3) + 4\,\* \S(2,2) + 4\,\* \S(3,1)
   - 8\,\* \S(1,1,2) - 8\,\* \S(1,2,1) - 8\,\* \S(2,1,1) )
\nn \\[1mm]
 \S(1,3)  &\!=\!& -1/4\,\* ( -  \BS(1,2,1) - 2\,\* \S(4) )
\nn \\[1mm]
 \S(3,1)  &\!=\!& -1/4\,\* ( -  \BS(2,1,1) - 2\,\* \S(4) )
\eea
Weight 3 and below:
\bea
 \S(-2,1)  &\!=\!& 1/4\,\* ( -  \BS(3) + 2\,\* \S(-3) )
\nn \\[0.5mm]
 \S(1,-2)  &\!=\!& 1/4\,\* ( -  \BS(1,2) + 2\,\* \S(-3) )
\nn \\[0.5mm]
 \S(1,1,1) &\!=\!& -1/8\,\* ( -  \BS(1,1,1)
   + 2\,\* \S(3) - 4\,\* \S(1,2) - 4\,\* \S(2,1))
\nn \\[0.5mm]
 \S(3)   &\!=\!& 1/2\,\*  \BS(2,1)
\\[1mm]
 \S(1,1) &\!=\!& -1/4\,\* ( -  \BS(1,1) - 2\,\* \S(2) )
\nn \\[0.5mm]
 \S(-2)  &\!=\!& -1/2\,\*  \BS(2)
\nn \\[0.5mm]
 \S(1)   &\!=\!&  1/2\,\*  \BS(1)
\eea
The inverse conversion is obtained by simply exchanging the roles of 
the harmonic sums on the left and the binomial sums on the right.
In this direction the order of the statements in irrelevant.

%
\section{The quartic-Casimir part of the anomalous dimensions}
\label{sec:appC}
\renewcommand{\theequation}{\ref{sec:appC}.\arabic{equation}}
\setcounter{equation}{0}
%

Finally we write down the contributions of the quartic group 
invariants to the four-loop non-singlet anomalous dimensions.
Since these terms cannot receive any lower-order contributions
in eq.~(\ref{eq:gu-exp}), they must be reciprocity respecting.
Hence they can be written in terms of $\eta = 1/N \,-\, 1/(N+1)$
and binomial sums. 

\pagebreak

Without the terms with weight-7 sums already given in 
eq.~(\ref{eq:gns3W7}), the $d_{FA}^{\,(4)}/\nc \equiv \dfRAnc$
contribution reads
\bea
\label{eq:g3d4FA}
&& \hspn\hspn
 \frct{1}{16}\: \gamma_{\,\rm ns}^{\:(3)+}
 \Big|_{w\,<\,7,\,d_{FA}^{\,(4)}/\nc} 
 \;=\; 
\nn
\\[0.5mm]  && \nn \mbox{} 
   \eta \,\* (
       - 16\,\* \BS(1,1,2,2) 
       + 48\,\* \BS(1,1,3,1) 
       - 32\,\* \BS(1,1,4) 
       + 16\,\* \BS(1,2,2,1) 
       - 16\,\* \BS(1,2,3) 
       + 32\,\* \BS(1,3,2) 
       + 16\,\* \BS(1,4,1) 
   )
\nn \\[0.8mm] && \mbox{}
       + \BS(1,2,1,2) \,\* (
          - 94/3
          - 16\,\*\eta
          )
       + \BS(1,3,1,1) \,\* (
            94/3
          - 32\,\*\eta
          )
       + \BS(2,1,1,2) \,\* (
          - 140/3
          - 32\,\*\eta
          )
\nn \\[0.8mm] && \mbox{}
       + \BS(2,1,2,1) \,\* (
          - 24
          - 8\,\*\eta
          )
       + \BS(2,1,3) \,\* (
            42
          + 4\,\*\eta
          )
       + \BS(2,2,1,1) \,\* (
            140/3
          + 56\,\*\eta
          )
\nn \\[0.8mm] && \mbox{}
       + \BS(2,2,2) \,\* (
          - 12
          - 12\,\*\eta
          )
       + \BS(2,3,1) \,\* (
          - 10
          - 28\,\*\eta
          )
       + \BS(2,4) \,\* (
            4
          + 20\,\*\eta
          )
       + \BS(3,1,1,1) \,\* (
            24
          - 16\,\*\eta
          )
\nn \\[0.8mm] && \mbox{}
       + \BS(3,1,2) \,\* (
            40
          + 60\,\*\eta
          )
       + \BS(3,2,1) \,\* (
            18
          + 12\,\*\eta
          )
       + \BS(3,3) \,\* (
          - 16
          + 4\,\*\eta
          )
       + \BS(4,1,1) \,\* (
          - 78
          - 36\,\*\eta
          )
\nn \\[0.8mm] && \mbox{}
       + \BS(4,2) \,\* (
            8
          - 12\,\*\eta
          )
       + \BS(5,1) \,\* (
            4
          - 12\,\*\eta
          )
       + \BS(1,1,2,1) \,\* (
            14\,\*\eta
          + 17\,\*\eta^2
          )
       + \BS(1,1,3) \,\* (
          - 88\,\*\eta
          - 68\,\*\eta^2
          )
\nn \\[0.8mm] && \mbox{}
       + \BS(1,2,1,1) \,\* (
            2\,\*\eta
          - 21\,\*\eta^2
          )
       + \BS(1,2,2) \,\* (
            48\,\*\eta
          + 38\,\*\eta^2
          )
       + \BS(1,3,1) \,\* (
          - 24\,\*\eta
          - 14\,\*\eta^2
          )
\nn \\[0.8mm] && \mbox{}
       + \BS(1,4) \,\* (
            48\,\*\eta
          + 48\,\*\eta^2
          )
       + \BS(2,1,1,1) \,\* (
          - 16\,\*\eta
          + 4\,\*\eta^2
          )
       + \BS(2,1,2) \,\* (
            196/3
          + 20/3\,\*\eta
          - 4\,\*\eta^2
          )
\nn \\[0.8mm] && \mbox{}
       + \BS(2,2,1) \,\* (
            14
          - 52\,\*\eta
          - 40\,\*\eta^2
          )
       + \BS(2,3) \,\* (
          - 18
          + 64\,\*\eta
          + 88\,\*\eta^2
          )
\nn \\[0.8mm] && \mbox{}
       + \BS(3,1,1) \,\* (
          - 238/3
          + 328/3\,\*\eta
          + 88\,\*\eta^2
          )
       + \BS(3,2) \,\* (
            6
          - 76\,\*\eta
          - 76\,\*\eta^2
          )
\nn \\[0.8mm] && \mbox{}
       + \BS(4,1) \,\* (
            12
          - 36\,\*\eta
          - 60\,\*\eta^2
          )
       + \BS(1,1,2) \,\* (
            14\,\*\eta
          - 52\,\*\eta^2
          - 28\,\*\eta^3
          + 32\,\*\z3
          )
\nn \\[0.8mm] && \mbox{}
       + \BS(1,2,1) \,\* (
            35\,\*\eta
          - 79\,\*\eta^2
          - 57\,\*\eta^3
          + 40\,\*\z3
          )
       + \BS(1,3) \,\* (
          - 108\,\*\eta
          + 162\,\*\eta^2
          + 58\,\*\eta^3
          - 72\,\*\z3
          )
\nn \\[0.8mm] && \mbox{}
       + \BS(2,1,1) \,\* (
          - 10/3
          - 21\,\*\eta
          + 67\,\*\eta^2
          + 53\,\*\eta^3
          - 8\,\*\z3
          )
       + \BS(2,2) \,\* (
            31/3
          - 42\,\*\eta^2
          - 24\,\*\eta^3
          - 24\,\*\z3
          )
\nn \\[0.8mm] && \mbox{}
       + \BS(3,1) \,\* (
          - 7
          + 56\,\*\eta
          + 36\,\*\eta^2
          + 50\,\*\eta^3
          - 32\,\*\z3
          )
       + \BS(4) \,\* (
            24\,\*\eta
          - 92\,\*\eta^2
          - 52\,\*\eta^3
          + 64\,\*\z3
          )
\nn \\[0.8mm] && \mbox{}
       + \BS(1,2) \,\* (
          - 4
          - 188/3\,\*\eta
          + 148/3\,\*\eta^2
          + 58/3\,\*\eta^3
          + 14\,\*\eta^4
          - 188/3\,\*\z3
          - 32\,\*\z3\,\*\eta
          )
\nn \\[0.8mm] && \mbox{}
       + \BS(2,1) \,\* (
            5/3
          - 10\,\*\eta
          - 20\,\*\eta^2
          - 24\,\*\eta^3
          - 10\,\*\eta^4
          - 280/3\,\*\z3
          - 52\,\*\z3\,\*\eta
          )
\nn \\[0.8mm] && \mbox{}
       + \BS(3) \,\* (
          - 184/3\,\*\eta
          + 268/3\,\*\eta^2
          + 274/3\,\*\eta^3
          + 40\,\*\eta^4
          + 132\,\*\z3
          + 96\,\*\z3\,\*\eta
          )
\nn \\[0.8mm] && \mbox{}
       + \BS(1,1) \,\* (
          - 26\,\*\eta
          + 18\,\*\eta^2
          - 40\,\*\eta^3
          - 68\,\*\eta^4
          - 20\,\*\eta^5
          + 120\,\*\z3\,\*\eta
          + 68\,\*\z3\,\*\eta^2
          - 80\,\*\z5
          )
\nn \\[0.8mm] && \mbox{}
       + \BS(2) \,\* (
            59/3
          + \eta
          + 178/3\,\*\eta^2
          + 40\,\*\eta^3
          + 88\,\*\eta^4
          + 24\,\*\eta^5
          + 476/3\,\*\z3
          + 40/3\,\*\z3\,\*\eta
          + 8\,\*\z3\,\*\eta^2
          )
\quad \nn \\[0.8mm] && \mbox{}
       + \BS(1) \,\* (
          - 31/3\,\*\eta
          + 218/3\,\*\eta^2
          + 48\,\*\eta^3
          + 216\,\*\eta^4
          + 168\,\*\eta^5
          + 36\,\*\eta^6
          + 4/3\,\*\z3
          + 104\,\*\z3\,\*\eta
\nn \\[0.8mm] && \mbox{\hspp}
          - 112\,\*\z3\,\*\eta^2
          + 56\,\*\z3\,\*\eta^3
          + 580/3\,\*\z5
          + 80\,\*\z5\,\*\eta
          )
\nn \\[0.8mm] && \mbox{}
       - 6
          + 4\,\*\eta
          + 55/3\,\*\z3
          - 132\,\*\z3\,\*\eta
          + 120\,\*\z3\,\*\eta^2
          - 740/3\,\*\z5
          + 210\,\*\z5\,\*\eta
          - 460\,\*\z5\,\*\eta^2
\eea

\noindent
Its $d_{FF}^{\,(4)}/\nc \equiv \dfRRnc$ counterpart already
determined in ref.~\cite{Kniehl:2025ttz} is much simpler with
\bea
\label{eq:g3d4FF}
&& \hspn\hspn
 \frct{3}{16}\: \gamma_{\,\rm ns}^{\:(3)+}
 \Big|_{\,\nf\,d_{FF}^{\,(4)}/\nc}
 \;=\;
\nn
\\[0.5mm]  && \nn \mbox{} 
       + 8\,\* \BS(1,2,1,2) 
       - 8\,\* \BS(1,3,1,1) 
       - 8\,\* \BS(2,1,1,2) 
       + 8\,\* \BS(2,2,1,1) 
       + \BS(1,1,2,1) \,\* (
            24\,\*\eta
          - 24\,\*\eta^2
          )
\nn \\[0.5mm] && \mbox{}
       + \BS(1,2,1,1) \,\* (
          - 24\,\*\eta
          + 24\,\*\eta^2
          )
       + \BS(2,1,2) \,\* (
          - 8\,\*\eta
          )
       + \BS(2,2,1) \,\* (
          - 12
          )
       + \BS(3,1,1) \,\* (
            12
          + 8\,\*\eta
          )
\nn \\[0.5mm] && \mbox{}
       + \BS(1,2,1) \,\* (
          - 24\,\*\eta
          + 24\,\*\eta^3
          )
       + \BS(2,1,1) \,\* (
            20
          + 24\,\*\eta
          - 24\,\*\eta^3
          )
       + \BS(2,2) \,\* (
          - 26
          + 24\,\*\eta
          )
\nn \\[0.5mm] && \mbox{}
       + \BS(3,1) \,\* (
            6
          - 24\,\*\eta
          )
       + \BS(1,2) \,\* (
            24
          - 8\,\*\eta
          + 4\,\*\eta^2
          + 8\,\*\eta^3
          + 16\,\*\z3
          )
\nn \\[0.5mm] && \mbox{}
       + \BS(2,1) \,\* (
          - 10
          + 78\,\*\eta
          - 16\,\*\z3
          )
       + \BS(3) \,\* (
            12\,\*\eta
          + 4\,\*\eta^2
          + 8\,\*\eta^3
          )
\nn \\[0.5mm] && \mbox{}
       + \BS(1,1) \,\* (
            48\,\*\eta
          - 78\,\*\eta^2
          - 96\,\*\z3\,\*\eta
          + 96\,\*\z3\,\*\eta^2
          )
\nn \\[0.5mm] && \mbox{}
       + \BS(2) \,\* (
          - 118
          + 62\,\*\eta
          - 40\,\*\eta^2
          - 24\,\*\eta^3
          - 24\,\*\z3
          - 16\,\*\z3\,\*\eta
          )
\nn \\[0.5mm] && \mbox{}
       + \BS(1) \,\* (
            86\,\*\eta
          - 170\,\*\eta^2
          - 8\,\*\z3
          + 96\,\*\z3\,\*\eta
          - 96\,\*\z3\,\*\eta^3
          - 80\,\*\z5
          )
\nn \\[0.5mm] && \mbox{}
       + 36
          - 24\,\*\eta
          + 106\,\*\z3
          - 200\,\*\z3\,\*\eta
          + 72\,\*\z3\,\*\eta^2
          + 240\,\*\z5
          - 400\,\*\z5\,\*\eta
          + 480\,\*\z5\,\*\eta^2
\eea

The difference of the `plus' and `minus' cases turns out the be
rather complicated for the new $\nfz$ part \cite{Gehrmann:2026qbl}.
It includes the function $f(N)$ of eq.~(\ref{eq:fwrap}) as well as
weight-6 objects (two with $\z3$) without $\eta\,$ which also 
combine to vanish in the large-$N$ limit. 
Both structures, the $\zeta$-function parts of which were seen 
before in refs.~\cite{Moch:2017uml,Vogt:2023unp,Kniehl:2025jfs},
occur also in other $\nfz$ contributions not shown here.
\bea
\label{eq:dgns3d4FA}
&& \hspn\hspn
 \frct{1}{16}\: \Big[ 
  \gamma_{\,\rm ns}^{\:(3)+} \,-\, \gamma_{\,\rm ns}^{\:(3)-}
 \Big]_{d_{FA}^{\,(4)}/\nc}
 \;=\;
\nn
\\[0.5mm]  && \nn \mbox{} \phantom{+}
    8\,\* (1 + \eta) \,\* 
    (
         \BS(2,3,1) 
       - \BS(2,4) 
       - \BS(3,1,2) 
       + \BS(3,3) 
       + \BS(4,2) 
       - \BS(5,1) 
    )
\nn \\[0.5mm] && \mbox{}
    + 16\,\* (\eta + \eta^2) \,\* 
    (
       - \BS(1,1,2,1) 
       + 6\,\* \BS(1,1,3) 
       + \BS(1,2,1,1) 
       - 3\,\* \BS(1,2,2) 
       - \BS(1,3,1) 
       - 2\,\*\BS(1,4)
    )
\nn \\[0.5mm] && \mbox{}
       + \BS(2,1,2) \,\* (
            4/3
          + 100/3\,\*\eta
          + 16\,\*\eta^2
          )
       + \BS(2,2,1) \,\* (
            48\,\*\eta
          + 48\,\*\eta^2
          )
       + \BS(2,3) \,\* (
          - 96\,\*\eta
          - 96\,\*\eta^2
          )
\nn \\[0.5mm] && \mbox{}
       + \BS(3,1,1) \,\* (
          - 4/3
          - 340/3\,\*\eta
          - 96\,\*\eta^2
          )
       + \BS(3,2) \,\* (
            64\,\*\eta
          + 64\,\*\eta^2
          )
       + \BS(4,1) \,\* (
            64\,\*\eta
          + 64\,\*\eta^2
          )
\nn \\[0.5mm] && \mbox{}
       + \BS(1,1,2) \,\* (
            64\,\*\eta^2
          + 32\,\*\eta^3
          )
       + \BS(1,2,1) \,\* (
          - 24\,\*\eta
          + 88\,\*\eta^2
          + 56\,\*\eta^3
          )
\nn \\[0.5mm] && \mbox{}
       + \BS(1,3) \,\* (
            156\,\*\eta
          - 120\,\*\eta^2
          - 52\,\*\eta^3
          )
       + \BS(2,1,1) \,\* (
            24\,\*\eta
          - 88\,\*\eta^2
          - 56\,\*\eta^3
          )
\nn \\[0.5mm] && \mbox{}
       + \BS(2,2) \,\* (
          - 36\,\*\eta
          + 48\,\*\eta^2
          + 28\,\*\eta^3
          )
       + \BS(3,1) \,\* (
          - 72\,\*\eta
          - 48\,\*\eta^2
          - 48\,\*\eta^3
          )
\nn \\[0.5mm] && \mbox{}
       + \BS(4) \,\* (
          - 48\,\*\eta
          + 56\,\*\eta^2
          + 40\,\*\eta^3
          )
       + \BS(1,2) \,\* (
          - 24\,\*\eta
          - 52\,\*\eta^2
          + 92/3\,\*\eta^3
          - 8\,\*\eta^4
          )
\nn \\[0.5mm] && \mbox{}
       + \BS(2,1) \,\* (
          - 12\,\*\eta
          + 20\,\*\eta^2
          + 48\,\*\eta^3
          + 16\,\*\eta^4
          + 24\,\*\z3
          + 24\,\*\z3\,\*\eta
          )
\nn \\[0.5mm] && \mbox{}
       + \BS(3) \,\* (
            160\,\*\eta
          - 260/3\,\*\eta^2
          - 496/3\,\*\eta^3
          - 52\,\*\eta^4
          - 48\,\*\z3
          - 48\,\*\z3\,\*\eta
          )
\nn \\[0.5mm] && \mbox{}
       + \BS(1,1) \,\* (
            96\,\*\eta^3
          + 128\,\*\eta^4
          + 32\,\*\eta^5
          - 128\,\*\z3\,\*\eta
          - 128\,\*\z3\,\*\eta^2
          )
\nn \\[0.5mm] && \mbox{}
       + \BS(2) \,\* (
            46/3\,\*\eta
          - 202/3\,\*\eta^2
          - 96\,\*\eta^3
          - 148\,\*\eta^4
          - 36\,\*\eta^5
          + 8/3\,\*\z3
          + 296/3\,\*\z3\,\*\eta
          + 64\,\*\z3\,\*\eta^2
          )
\nn \\[0.5mm] && \mbox{}
       + \BS(1) \,\* (
          - 152/3\,\*\eta^2
          - 48\,\*\eta^3
          - 216\,\*\eta^4
          - 168\,\*\eta^5
          - 36\,\*\eta^6
          - 216\,\*\z3\,\*\eta
          + 16\,\*\z3\,\*\eta^2
          - 56\,\*\z3\,\*\eta^3
          )
\quad \nn \\[0.5mm] && \mbox{}
       - 118/3\,\*\eta
          + 12\,\*\eta^2
          - 296\,\*\z3\,\*\eta
          - 176/3\,\*\z3\,\*\eta^2
          + 104\,\*\z3\,\*\eta^3
          - 20/3\,\*\z5
          - 260/3\,\*\z5\,\*\eta 
\:\: ,\\[3mm] 
\label{eq:dg3d4FF}
&& \hspn\hspn
 \frct{3}{16}\: \Big[
  \gamma_{\,\rm ns}^{\:(3)+} \,-\, \gamma_{\,\rm ns}^{\:(3)-}
 \Big]_{\nf\,d_{FF}^{\,(4)}/\nc}
 \;=\;
\nn
\\[0.5mm]  && \nn \mbox{} \phantom{+}
         \BS(3) \,\* (
          - 24\,\*\eta
          - 8\,\*\eta^2
          - 16\,\*\eta^3
          )
       + \BS(2) \,\* (
          - 12\,\*\eta
          + 60\,\*\eta^2
          + 24\,\*\eta^3
          )
\nn \\[0.5mm] && \mbox{}
       + 236\,\*\eta
          - 248\,\*\eta^2
          + 48\,\*\z3\,\*\eta
          + 16\,\*\z3\,\*\eta^2
          + 32\,\*\z3\,\*\eta^3
\:\: .
\eea

\vspace*{4mm}
\noindent
{\sc Form} files with the main results of the article and the
conversions between harmonic sums and binomial sums in this appendix
can be obtained from the preprint server http://arXiv.org.
They are also available from the authors upon request.

\end{document}